%% file: main.tex
\documentclass[nonatbib]{elsarticle}
\usepackage[utf8]{inputenc}
\usepackage[english]{babel}
\usepackage{subfiles}
\usepackage{booktabs}
\usepackage{graphicx}
\usepackage{multirow}
\usepackage{lscape}
\usepackage[table,xcdraw]{xcolor}
\usepackage[hidelinks]{hyperref}
\usepackage{geometry}
\makeatletter
\let\c@author\relax
\makeatother
\usepackage[backend=biber, style=authoryear, maxcitenames=1, isbn=false, date=year, url=false]{biblatex}
\AtEveryBibitem{\clearfield{note} \clearlist{language}}
\makeatletter
\def\ps@pprintTitle{%
  \let\@oddhead\@empty
  \let\@evenhead\@empty
  \let\@oddfoot\@empty
  \let\@evenfoot\@oddfoot}
\makeatother

\title{Impacts of electric carsharing on a power sector with variable renewables\tnoteref{t1}}

\author[1,2,3]{Adeline Guéret\corref{cor1}}

\author[1]{Wolf-Peter Schill}

\author[1]{Carlos Gaete-Morales}

\cortext[cor1]{Corresponding author: \texttt{agueret@diw.de}.}
\affiliation[1]{organization={DIW Berlin},
addressline={Mohrenstraße 58},
city={10117 Berlin},
country={Germany}}

\affiliation[2]{organization={Technische Universität Berlin},
addressline={Straße des 17.\ Juni 135},
city={10623 Berlin},
country={Germany}}

\affiliation[3]{organization={OFCE - Sciences Po Paris},
addressline={10, Place de Catalogne},
city={75014 Paris},
country={France}}

\begin{document}


\begin{abstract}
Electrifying the car fleet is a major strategy for mitigating emissions in the transport sector. As electrification cannot solve all negative externalities associated with cars, reducing the size of the car fleet would be beneficial. Electric carsharing could allow to reconcile current car usage habits with a smaller fleet, but this may reduce the potential of electric cars to align their grid interactions with variable renewable electricity generation. We investigate how electric carsharing may impact the power sector, combining three methods: sequence clustering of car travel diaries, generation of synthetic electric vehicle time series, and power sector modelling. We show that switching to electric carsharing only moderately increases power sector costs, less than 110 euros per substituted car in our main setting. This effect is largest with bidirectional charging. We conclude that the power sector interactions of shared electric car fleets could still be aligned with variable renewable electricity generation. 
\end{abstract}

\maketitle

\section{Introduction}
Electrifying the vehicle fleet is a major strategy for reducing greenhouse gas (GHG) emissions in the passenger road transport sector, which represents the largest source of GHG emissions in the transport sector (\cite{noauthor_global_2023, ipcc2023}). This is because battery electric vehicles (BEVs) often have lower life-cycle emissions than internal combustion engine vehicles (ICEVs) already today, and potentially much more so in the future (\cite{sacchi_when_2022, cox_life_2020, knobloch_net_2020, hoekstra2019}). As policymakers increasingly support the electric vehicle industry, the electrification of passenger cars is also becoming more and more widespread. This trend is expected to continue and accelerate as bans on the sale of new non-zero emission cars are announced in various parts of the world. In the European Union, the ban will take effect in 2035. This will also be the case in some states of the United States such as California. In China, at least half of all new cars sold after 2035 will have to be electric, fuel cell or plug-in hybrid.

At the same time, there is a growing interest in carsharing, especially in urban areas (\cite{creutzig2024, fulton2018, kortum2016free}), where cars are responsible for environmental externalities (\cite{leroutier2023tackling}). By reducing the overall fleet size, carsharing has the potential to increase available street space in cities (\cite{creutzig2020fair, glotz2016reclaim}). Carsharing and car electrification are interrelated, as the former could also foster consumer uptake of electric cars. Shared cars could thus play a role in bringing the electrification of the car fleet to scale (\cite{hoerler2021carsharing, shaheen2020zero, schluter2019car, mccollum2018interaction}). Besides, as light-duty vehicles represent the main mode of transport for private citizens (\cite{ipcc2023}), carsharing can be perceived as a more appealing transport mode than other mobility options such as public transport or active mobility, as it could require smaller behavioral adaptation compared to a so-called ``modal shift''. However, while electric carsharing offers numerous benefits, it also poses some challenges to the power sector. One challenge is the potential loss of flexibility. privately owned BEVs with system-oriented charging capabilities can provide flexibility to the power sector. This facilitates the integration of variable renewable energy sources (vRES) such as solar and wind power (\cite{liu2024, xu2023, taljegard2019impact, hanemann2017effects}) and can help to shave peak loads (\cite{xu2018planning}). This way, BEVs can also reduce stationary electricity storage needs (\cite{xu2023storage}) and respective use of mineral resources (\cite{deetman2021}). Electric carsharing may reduce the flexibility potential of BEVs due to their increased driving frequency, which translates into a lower grid availability, as well as to the lower aggregate battery capacity of the fleet. 

Taking into account both the trend of electrification of passenger transport on the one hand and the development of sharing options on the other, we investigate the impacts of a switch from private BEVs to electric carsharing on a power sector dominated by variable renewable energy sources. In our study, carsharing refers to a free-floating fleet of shared vehicles that can be rented on a per-trip basis. Carsharing differs from ride-sharing or share-pooling in that individuals do not share a vehicle during a journey. In Germany, carsharing has been in use for longer than ride-pooling and is therefore more established (\cite{burghard2022sharing}). Carsharing might contribute to reducing car ownership (\cite{jochem_does_2020, mounce_potential_2019}) and hence the car fleet size (\cite{spieser2014toward}). Therefore, carsharing can also be relevant in the transport decarbonization strategy, as a reduction in the size of the car fleet may be desirable beyond the mere electrification of the passenger road transport sector in order to achieve decarbonization targets (\cite{dirnaichner_life-cycle_2022, milovanoff2020electrification}). Case studies have shown that carsharing can also help reduce other negative externalities and, for instance, increase the availability of public space in central city areas (\cite{diana_spatial_2022}). Finally, carsharing could help mitigate negative impacts of cars on marine and freshwater ecotoxicity as well as on mineral resource scarcity (\cite{vilaca_life_2022}). Hence, electric carsharing could also help tackle the potential trade-off between decarbonizing transport and the need for critical metals (\cite{zhang_trade-off_2023, zeng2022}).

In addition to the challenge of electrifying the passenger car fleet, meeting decarbonization targets also requires an increasing use of electricity generated from renewable energy sources (RES). In most countries, the renewable energy sources with the largest potential are wind and solar power, which have variable generation patterns (\cite{lopezprol2021}). This requires future power sectors to be more flexible. Various options for temporal and spatial flexibility exist, including various types of energy storage (\cite{schill2020}). While electrifying passenger road transport increases electricity demand, and thus the need for integrating variable renewables, it may also increase the power sector flexibility if BEVs are charged smartly or bidirectionally using vehicle-to-grid (V2G) technologies (\cite{syla2024, luca2018impact, schill2015power, loisel2014large}). In particular, it has been found that BEVs may help balance daily variations of solar power (\cite{brown2018synergies}). One important factor enabling this flexibility is the fact that many privately owned vehicles remain idle for extended periods of time. Assuming sufficient grid connection, electric vehicles could charge during times of abundant renewable power and, if bidirectional charging is possible, feed electricity back into the grid during times of scarcity. The actual flexibility potential that BEVs can provide to the power sector depends on various factors, such as driving profiles and plug-in behaviours (\cite{gschwendtner_impact_2023}). Furthermore, the overall battery capacity of the BEV fleet that is equipped with smart or bidirectional charging is also a significant factor.

A few analyses focus specifically on the potential of carsharing fleets for grid congestion management (\cite{brinkel2022grid}) or take the perspective of operators of shared BEV fleets (\cite{wiedemann2024vehicle, kahlen2018electric}) or the end users' perspective to investigate whether carsharing users are willing to offer temporal flexibility when using shared BEVs (\cite{suel2024vehicle}). To the best of our knowledge, the effects on the power sector's optimal investment and dispatch related to switching from privately owned BEVs to electric carsharing have not been studied yet. We aim to address this research gap with a quantitative model-based study. We do so for 2030 scenarios of the German power sector with renewable energy shares of around 85\%. We use a combination of three open-source quantitative methods: (i) sequence clustering techniques on car travel diaries from a national German mobility survey, (ii) \textit{emobpy}, a probabilistic tool for generating synthetic BEV time series, and (iii) DIETER, a capacity expansion and dispatch model of the power sector. For clarity, we assume no changes in demand for car mobility, and consider three different BEV charging strategies which apply to the whole vehicle fleet. 

We show that a switch to electric carsharing decreases the flexibility potential of BEVs and thus increases power sector costs. This cost increase however remains moderate in most settings analyzed, and remains below 110~euros per substituted car and year in our main scenario. Costs effects of carsharing are driven by changes in the optimal power plant portfolio and its hourly use. The nature and magnitude of impacts of switching to carsharing depend on the charging strategy considered. Costs increase the most when all electric vehicles are assumed to be charged bidirectionally, i.e.~make full use of their flexibility potential. A switch to carsharing still appears to be compatible with a large share of renewable energy in the electricity mix of about 85\%. Overall, our research demonstrates that future BEV fleets that include relevant amounts of shared cars can still be operated largely in line with variable renewable energy sources and contribute to their system integration. 

\section{A combination of three quantitative methods}
\label{sec:method}

\begin{figure}[!ht]
    \centering
    \includegraphics[width=17.5cm]{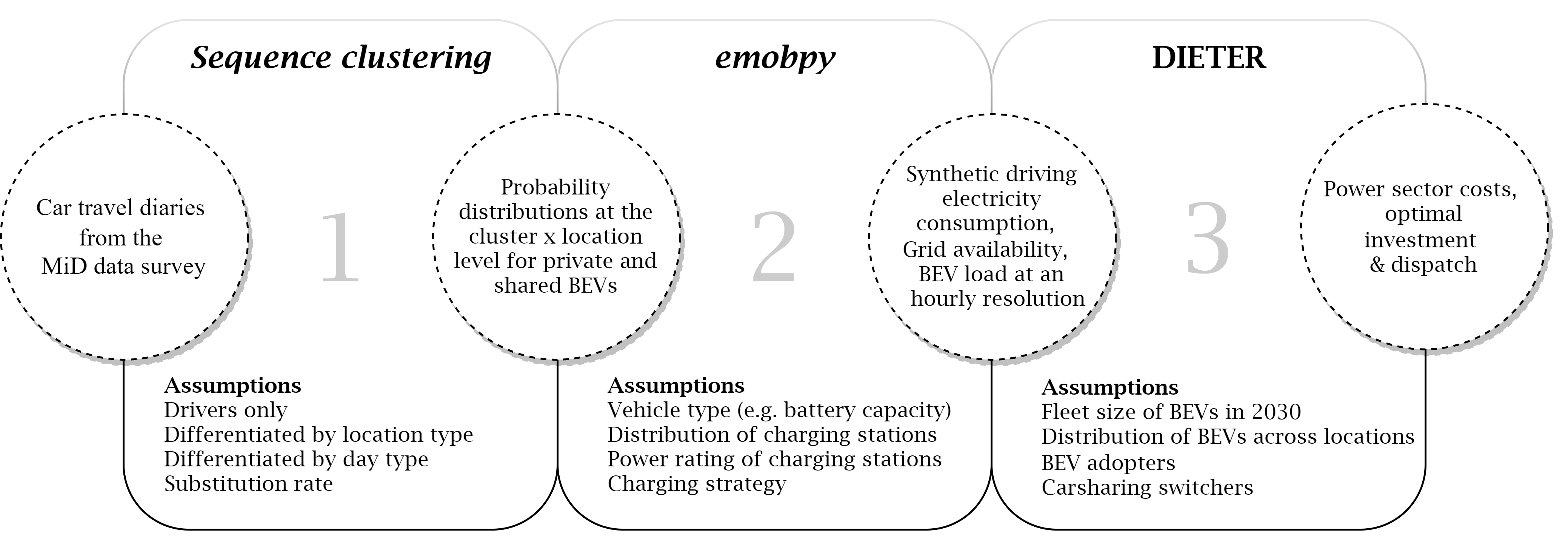}
    \caption{Flowchart of methods, main assumptions, inputs and outputs}
    \label{fig:workflow}
\end{figure}

Our methodological approach is based on a three-step procedure (Figure~\ref{fig:workflow}). First, probability distributions for car mobility are computed based on empirical data. These distributions are generated for both private and shared BEVs, assuming a constant substitution rate. Secondly, the open-source tool \textit{emobpy} is used to generate time series for the electric consumption and grid availability of BEVs at an hourly resolution. Thirdly, these time series serve as input for a linear optimisation open-source power sector model called DIETER. The first two steps are carried out using Python exclusively, while the third step also involves GAMS programming.

\subsection{Sequence clustering}

To generate mobility probability distributions, we use travel diaries from the representative survey \textit{Mobilität in Deutschland} (short MiD, see~\cite{eggs2018mobilitat}). The travel diaries are reshuffled into a sequence format, and a clustering algorithm is applied to identify distinct groups of car usage. 

\subsubsection{Travel diaries} 

GPS tracking and big data methods can provide access to highly detailed time series with high geographical and temporal resolution for existing car users. However, electric cars and carsharing are still in the early stages of development. Therefore, generalizing the behaviour of early adopters to a country-wide setting could lead to flawed conclusions (\cite{yeh2022improving}). This study departs from existing empirical data on electric car or carsharing use and instead uses travel diaries from a representative survey, which allows to consider geographical components that are relevant in this context, such as the level of urbanization. In addition, travel diaries provide detailed information on car use at a high temporal resolution. 

This study uses the B1 version of the 2017 edition of the MiD, commissioned by the German Federal Ministry for Transport and Digital Infrastructure \textit{(Bundesministerium für Digitales und Verkehr}, BMDV). The data collection took place in Germany from May 2016 to September 2017. A total of 316,361 individuals from 156,420 households were surveyed, representing the German population at the time (around 41 million households and 82 million individuals). For the analysis, we only consider car trips with available departure and arrival times. We exclude passenger trips to avoid double counting, as most car trips taken as a passenger are with another household member as the driver. We also exclude car trips undertaken during one's professional activity. Individuals whose diary implied overlapping trips, i.e. at least two trips taking place at least partially at the same time, and trips that span two days were also removed. The final database comprises 355,136 car trips.

\subsubsection{Sequencing}

Travel diaries contain a wealth of information, including the timing of departure and arrival, trip duration, distance, and destination. Additionally, the serial aspect of the data can provide further insight into individuals' car use. The sequence format allows for a large amount of information to be encapsulated into a one-dimensional object, making comparisons across individuals easier without losing any information. Hence, we recode travel diaries into sequences, which results in a sample of 119,519 sequences.

Sequence formatting requires two essential features: a fixed time interval and a taxonomy of mutually exclusive possible states. The time interval decomposes a given time frame into blocks of equal duration. In this case, the time frame is a full day from midnight until 23:59 o'clock.  The time interval is five minutes, meaning that each day is broken into 288 blocks. By ``sequence'', we refer to the succession of these 288 blocks that map the clock time. The following step in sequence formatting involves assigning each block to a single ``state''. These states correspond to the variable of interest, which in this case is car mobility. We define two states: ``idle'' when the car is not moving and ``on the move'' when a car trip takes place. 

\subsubsection{Clustering}

We implement the Hierarchical Agglomerative Clustering (HAC) algorithm using the Levenshtein distance as a distance measure and Ward's method as the linkage criterion. The clustering is performed on subdatasets that partition the overall sequence dataset by location type (metropolis, big city, middle-size city, small city, and rural area). The location typology corresponds to the variable RegioStatR Gem5 provided in the MiD dataset. Additional information is available on the BMDV website at \href{https://bmdv.bund.de/SharedDocs/DE/Artikel/G/regionalstatistische-raumtypologie.html}{\textcolor{blue}{this link}}. The final number of clusters is selected using dendrograms (Figure \ref{fig:dend}).

The computed descriptive statistics for the clusters (Table \ref{tab:clust_stat}) indicate that car usage can be characterised by the daily distance travelled by car, which correlates well with the average trip distance and duration. For example, individuals in cluster 6 travel approximately 11 kilometers by car, with an average trip distance of 5.5 kilometers and an average trip duration of 14 minutes. In contrast, individuals in cluster 4 travel an average of 50 kilometers, with an average trip distance of 18.5 kilometers and an average trip duration of 34.2 minutes. To visualise differences in car use, we refer to sequence index plots (Figure \ref{fig:sequences_16}).

This applies to all locations, even though there may be some differences in the characteristics of the clusters across locations. For example, in the cluster of users who drive short distances overall during the day, the total distance travelled as well as the average trip distance are greater in less urbanized areas. Conversely, the average trip duration is bigger in more urbanised areas, which is in line with congestion being more prevalent in big cities than rural areas.

\subsection{Synthetic BEV time series}

Using the open-source tool \textit{emobpy}, we generate synthetic BEV time series for private and shared BEVs based on clustering results. Synthetic time series are computed at the cluster level for each location at a 5~-minute resolution. In all computations, we use the default values for the Volkswagen ID.3 model. For shared BEVs across all clusters and private BEVs from the cluster with longest trips, we assume a battery capacity of 100~kWh, compared to 58~kWh for other private vehicles. As shared BEVs must be able to make more trips per day, we assume that such vehicles would be equipped with bigger batteries than privately owned cars that are used only by one household. This battery capacity is slightly above the battery capacity of models currently offered by carsharing operators. Yet, it appears plausible that future models will be equipped with bigger batteries, as the battery manufacturing costs keep decreasing. Already today, carsharing operators provide models with above-average batteries, such as the Tesla model~Y, the Tesla model~3 or the VW~ID.4, with battery capacities of about 80~kWh. Probability distributions for charging station availability can be found in Table~\ref{tab:emobpy_assumptions}. The charging strategy used to compute the grid electricity time series is \textit{balanced}. For more detailed information on \textit{emobpy}, we refer to the dedicated publication (\cite{gaete-morales_open_2021}) and the appendix (\ref{sec:appendix_emobpy}).

\subsubsection{Vehicle mobility} To generate the vehicle mobility time series, several input probability distributions have to be provided: the distribution of the number of trips per day, the distribution of the departure time of the trips and the joint distribution of trip duration and trip distance. We introduce new features to \textit{emobpy} in order to generate time series based on more detailed distributions. More precisely, the distribution of the departure time is conditional on the total number of trips and the trip rank, on top of the trip destination and the day type. Similarly, the joint probability of duration and distance is also conditional on the trip destination, the day type and, as a new feature, the total number of trips. The conditionality on the total number of trips and the trip rank, where applicable, is only active for the generation of time series for private cars and not shared ones.  

\paragraph{Shared BEVs}
Assuming a substitution rate of about five between private cars and shared cars, we derive probability distributions for shared BEVs based on the probability distributions for private BEVs. This substitution rate can be seen as a rather conservative assumption since previous research has shown that free-floating carsharing could replace between 8 and 20 private cars (\cite{jochem_does_2020}). Yet, the conditions under which carsharing would substitute private car ownership are still debated (\cite{silvestri2021contribution}). In this regard, assuming a conservative substitution rate of about five by 2030 seems reasonable. The distribution of number of trips for shared BEVs is the same as for private BEVs, but centered around a new mean (see~\ref{sec:appendix_emobpy}). We assume that the distribution of trip departure times remains the same as in the private case, except that we condition only on the trip destination. For the joint distribution of trip duration and distance, we also assume the same as in the private case and remove the conditioning on the total number of trips (Table~\ref{tab:emobpy_condprob}). Thus, this modelling approach is based on the assumption that carsharing does not induce a change in the number or type of car trips made.

\paragraph{Weekly time series} 
Time series are computed at the cluster level for each location type on weekdays. On Saturdays and Sundays, time series are generated at the location level, as summarised in Table~\ref{tab:emobpy_condprob}. This approach avoids making additional assumptions on how clusters should be matched across day types. One of our working hypotheses is thus that car usage during the week drives the likelihood of switching to carsharing, regardless of weekend usage. 

\subsection{Power sector modelling}

We apply the open-source power sector model DIETER (\textit{Dispatch and Investment Evaluation Tool with Endogenous Renewables}). As a capacity expansion model, it minimizes fixed and variable power sector costs under various feasibility and policy constraints. The model is solved for a full year at an hourly resolution (8760 consecutive hours). Its main focus is the integration of variable renewable energy sources by means of various options for power sector flexibility. We apply the model to a 2030 scenario of Germany with a renewable share of around 85\%. For clarity and generalizability, we abstract from the European interconnection and model Germany as an electric island, assuming perfect expansion of the transmission and distribution grids within the country. Sector coupling with the heating sector is modelled assuming a stock of 7.5~million heat pumps, which are assumed to be operated inflexibly, i.e.~to provide heat at the time of demand. The electricity consumption time series of heat pumps has been derived from \cite{roth2023flexible}. Using around 52~TWh of electricity, these heat pumps generate around 143~TWh of heat, corresponding to abou twenty percent of the annual space heating demand from residential buildings in Germany. In additional model runs, we also include a 30~TWh flat annual demand for green hydrogen, which corresponds to an additional electricity demand of around 42~TWh. 

\paragraph{Technologies and renewable energy constraint}
We model four conventional generation technologies, including lignite, hard coal, open-cycle gas turbines (OCGT) and closed-cycle gas turbines (CCGT), as well as five renewable generation technologies, including solar photovoltaics, onshore wind, offshore wind, biomass and run-of-river hydropower (Table~\ref{tab:costs_powergeneration}). Variable costs for conventional technologies include a carbon price of 130~euros per tonne of CO$_{2}$eq. Three storage technologies are considered: lithium-ion batteries, pumped-hydro storage, and hydrogen-based long-duration storage; the latter being a combination of electrolysis, hydrogen storage, and hydrogen gas turbines (Table~\ref{tab:parameters_storage}). A policy constraint reflects the renewable energy target of the German government: at least 80\% of the electricity consumed has to come from renewable energy sources. Yet, this constraint is not binding in our parameterisation, as renewable shares around 85\% lead to the lowest power sector costs (Table~\ref{tab:res_shares}). Assuming plausible limits for expanding hydro and bio energy in Germany, the majority of this renewable electricity has to come from wind and solar power. In our parameterisation, this implies variable renewable energy shares of 80\% or slightly more in all scenarios.

\paragraph{Modelling interactions with BEVs} 
Sector coupling with BEVs is modelled using distinct vehicle profiles. These profiles represent individual vehicles which are scaled up to the overall fleet. In the case of uncontrolled charging, each profile is considered an exogenous electricity demand that is added to the overall demand. When considering smart or bidirectional charging, the model chooses when and how much electricity should flow into a vehicle, while ensuring battery levels that are compatible with the mobility constraint, i.e. all trips until the next charging episode can be undertaken. This relies on the exogenous time series ``driving electricity consumption'' and ``grid availability'' derived with \textit{emobpy}. Discharging electricity from BEV back to the grid incurs additional variable costs of 15~euros/MWh to account for battery degradation.

The model is calibrated based on a 15 million BEV fleet in 2030, which aligns with the target of the German coalition. The distribution of cars across locations is assumed to be the same as during the survey (Table~\ref{tab:dieter_cardistribution}). Additionally, it is assumed that clusters with very long trips (clusters 1 and 2 in each location with more than 250 kilometers driven on average during the day) are not regular daily patterns, but rather trips undertaken under exceptional circumstances, such as holidays. As these patterns represent a very small proportion of overall sequences (around 2.4\%), we exclude them from our analysis and assume that the 15~million BEV fleet is distributed across locations and clusters 3 to 6 (Table~\ref{tab:cars_clustloc}). This means that very long trips that would constitute daily mobility habits are assumed not to be undertaken with electric cars by 2030. Nevertheless, very long trips can still take place sporadically during the day as well as over weekends.

\subfile{car_distribution_cluster_location.tex}

\subsection{Scenarios}

We investigate a number of scenarios which differ with respect to the size of the overall vehicle fleet, the number of shared cars, and the charging strategy. Scenarios are compared to respective reference settings without carsharing, i.e.~references only include privately owned BEV profiles.

\paragraph{Frameworks} We group our analyses into two broad frameworks. The first framework only considers BEVs that switch to carsharing in the scenario (``Shared only''). The second framework additionally includes other electric cars that stay private both in the reference and the scenario (``Shared + other BEVs''). Hence, the BEV fleets considered in the latter framework are larger than the ones in the former (Table~\ref{tab:scenarios}). The former framework highlights the power sector effects of switching to carsharing, assuming away many other potential sources of flexibility. While results are instructive, they may not be policy-relevant. The latter framework is designed to mimic a more realistic setting with additional BEVs that do not switch to carsharing, but remain privately owned. Depending on the charging strategy, these additional ``non-switching'' BEVs also serve as a proxy for other flexibility options that are not included in the model. We thus consider ``Shared + other BEVs'' to be the main framework.

\subfile{scenarios.tex}

\paragraph{Carsharing uptake} 

\begin{figure}[!ht]
    \centering
    \includegraphics[width=15.5cm]{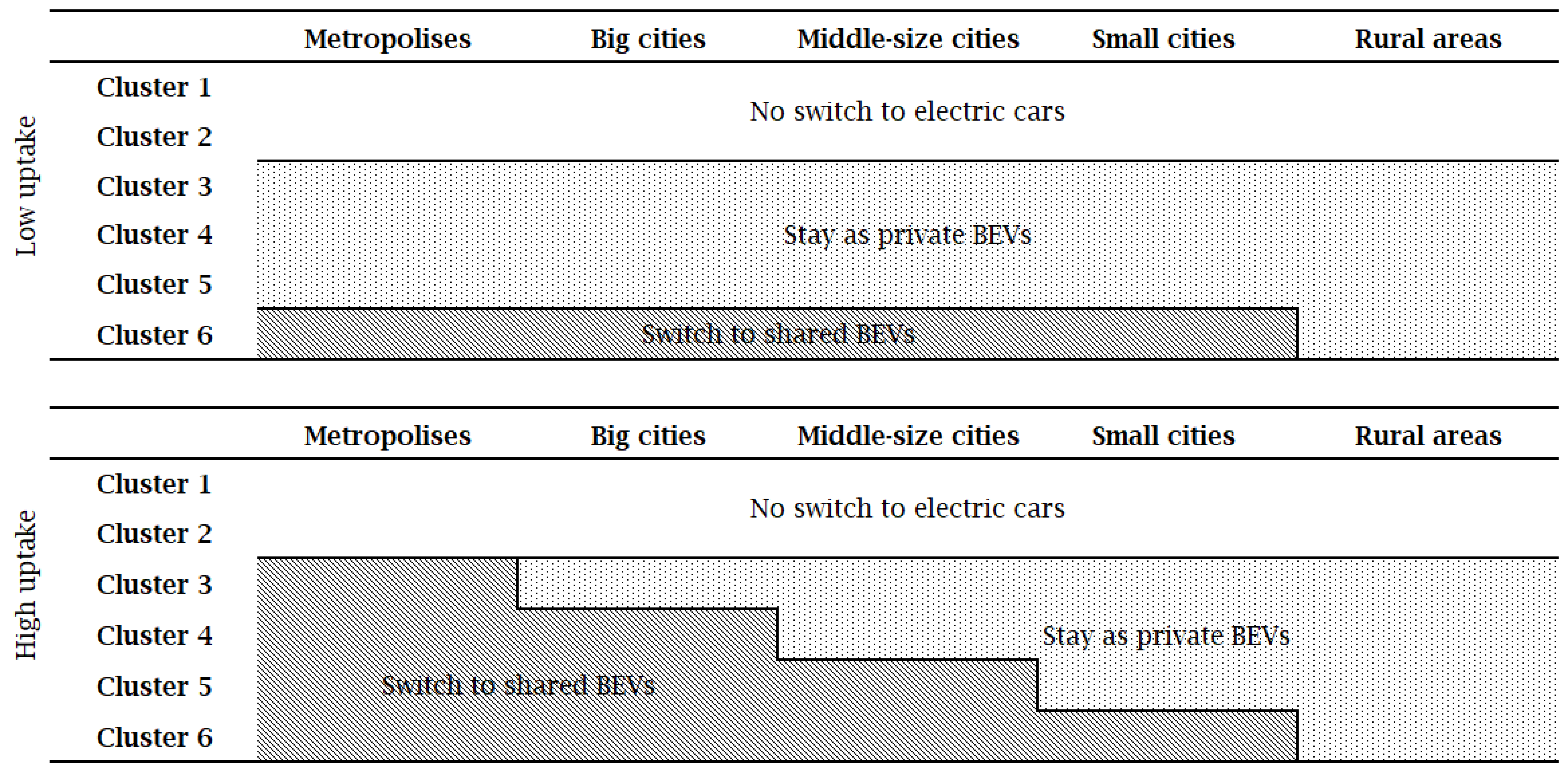}
    \caption{Definition of carsharing uptake regimes based on cluster and location types. Note that this represents the reference fleet of the ``Shared + other BEVs'' framework. For the ``Shared only'' framework, the reference fleet is only constituted of the shaded areas that switch to shared BEVs in the scenario.}
    \label{fig:carsharing_uptake}
\end{figure}

For each framework, we consider two possible carsharing uptake regimes: a low one and a high one. In the low uptake regime, only BEVs belonging to the cluster of shortest trips (cluster 6) are assumed to switch to carsharing. In the high uptake regime, we assume that all BEVs in metropolises switch to carsharing as well as most BEVs in big cities and some in the middle-size cities. In small cities, only cluster 6 is assumed to switch to carsharing. For both uptake regimes, we assume that there is no switch to carsharing in rural areas. The number of substituted BEVs is approximately 3.9~million in the low uptake scenarios compared to 6.2~million in the high uptake scenarios. Further details are available in Figure~\ref{fig:carsharing_uptake} and Table~\ref{tab:scenarios}. The low uptake regime is more conservative, as only users that rely the less on their cars switch to carsharing. The high uptake regime is more progressive and aims at illustrating a very optimistic scenario when analysing impacts on the power sector. The distribution of the number of daily trips in the overall fleet can be seen in Figure~\ref{fig:emobpy_dist_ntrips} and the distribution of where electric cars are located at each time step in Figure~\ref{fig:emobpy_freqdestination}. Average grid availability for each hour of the week is illustrated in Figure~\ref{fig:ev_gridavailability}.

\paragraph{Charging strategy} 

For each scenario, we consider various charging strategies: uncontrolled charging, smart charging, or bidirectional charging. In order not to mix up effects, we assume that both the reference and the respective scenarios always have the same charging strategy. Additionally, only one charging strategy is used for the entire fleet at a time to enable meaningful comparisons. This means that all BEVs charge smartly in the reference and the scenarios, or they all charge bidirectionally, etc. We consider different charging strategies since it is unclear which charging strategy will be mostly used by BEVs in the future. While it is unlikely that the whole BEV fleet will be operated in an uncontrolled manner (\cite{schill2015power}), it also seems unlikely that the whole fleet will be operated bidirectionally. By considering these strategies and homogeneous fleets, we analyse extreme settings which can be thought of as boundaries for outcomes of alternative scenarios with BEV fleets with mixed charging strategies.

\section{Results}
\label{sec:results}
 
In the main part of the paper, we focus on the main framework where only part of the total BEV fleet considered in the reference switches to carsharing (``Shared + other BEVs''), i.e.~the most policy-relevant framework. Results for the ``Shared only'' framework are provided in the appendix (\ref{sec:additional_results}). Qualitative results are similar across frameworks unless explicitly mentioned.

\subsection{Power sector cost increases remain moderate and are largest for bidirectional charging scenario}

\begin{figure}[!ht]
    \centering
    \includegraphics[width=17.5cm]{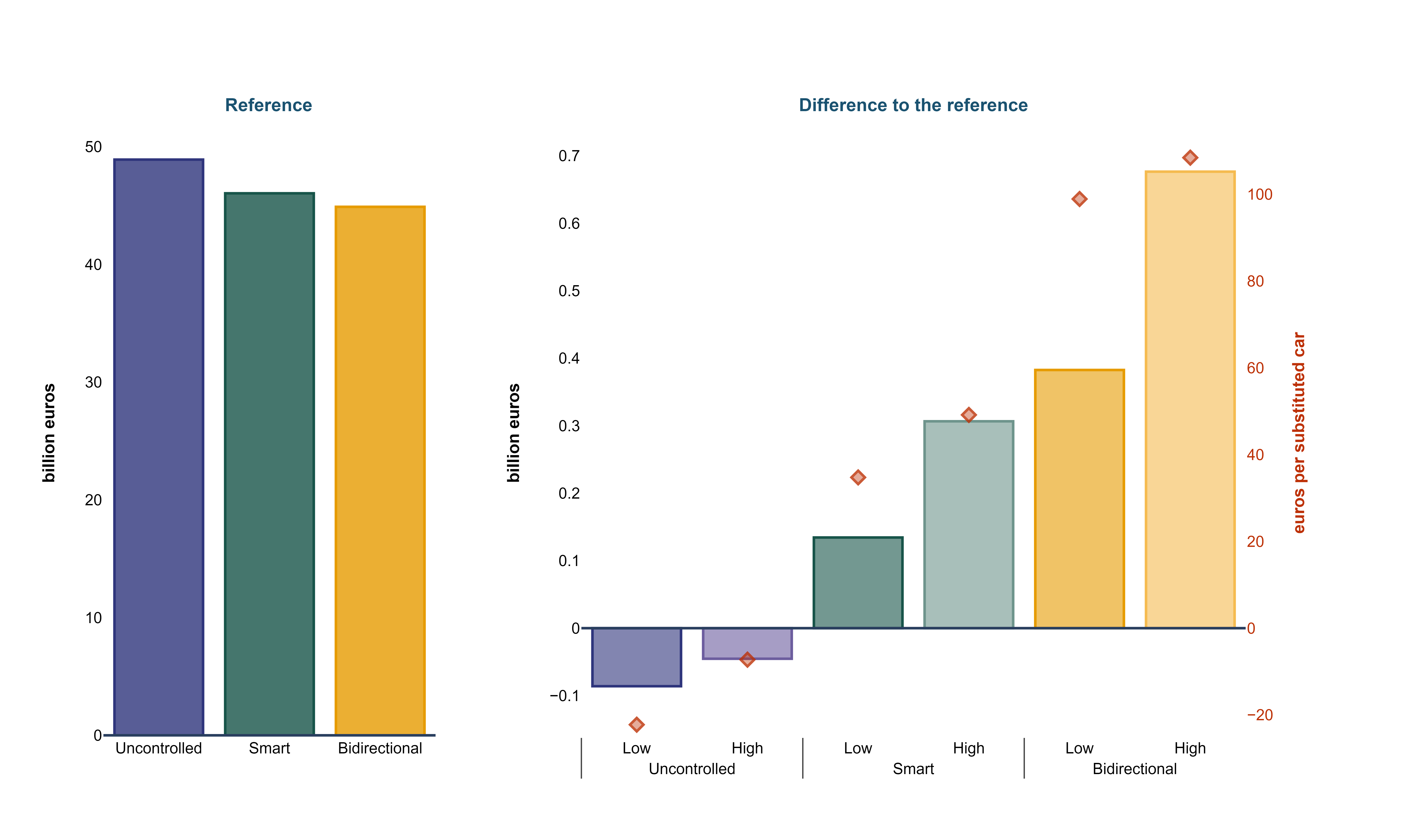}
    \caption{Power sector costs by charging strategy and carsharing uptake [``Shared + other BEVs'']}
    \label{fig:system_costs}
\end{figure}

The adoption of carsharing in the BEV fleet results in increased power sector costs across almost all frameworks, carsharing uptake regimes, and charging strategies (Figure~\ref{fig:system_costs} and Table~\ref{tab:system_costs}). This cost increase is relatively small compared to the overall sector costs, and always remains below 110~euros per substituted car per year in the main framework ``Shared + other BEVs'', even in the case with bidirectional charging. In the ``Shared + other BEVs'' framework, it represents about 1.5\% of the total sector costs and 2.2\% in the ``Shared only'' framework. Note that the load of substituted BEVs is also small compared to the overall load and ranges between 1.1\% and 2.6\% of the overall load in the ``Shared + other BEVs'' framework, depending on the uptake regime. Sensitivity runs indicate that cost effects are qualitatively robust with respect to alternative weather years (Figures~\ref{fig:SensitivityWeather_systemcosts_ref} and~\ref{fig:SensitivityWeather_systemcosts_scen}) and alternative technology cost assumptions (Figure~\ref{fig:SensitivityCosts_systemcosts_ref} and Figure~\ref{fig:SensitivityCosts_systemcosts_scen}).

In the case where all BEVs charge in an uncontrolled way, switching to electric carsharing may even lead to slight cost reductions. This is because charging a large fleet of BEVs without considering power sector dynamics leads to a very expensive power sector, as BEVs considerably increase existing system peak loads (Figures~\ref{fig:resload_summer} and~\ref{fig:resload_winter}). By switching to carsharing, the changing charging patterns may even slightly ease this burden. The cost decreases related to carsharing with the uncontrolled charging strategy should not be misinterpreted. The absolute sector costs remain the highest compared to other charging strategies, even after a switch to carsharing. Since the perspective of having a large BEV fleet charged in an uncontrolled way is neither realistic nor desirable (compare~\cite{schill2015power}), we do not focus on this charging strategy in the rest of the paper. Results can nevertheless be found in the appendix. 

Assuming that the whole BEV fleet charges smartly or bidirectionally, costs in the reference decrease compared to a setting with uncontrolled charging, as the former charging strategies help avoid costly load peaks (see section~\ref{subsec:cap_and_disp}), which reduces the need for peaker plants and storage capacity. In turn, switching to carsharing then increases costs, regardless of the uptake regime. This is because substituting private BEVs that are charged smartly or bidirectionally with shared BEVs partly takes away the flexibility that private BEVs provide to the power sector, as shared cars are driven more often. Note that we assume that private cars provide power sector flexibility without additional investment costs. We also do not include charging infrastructure costs in our model, considering that the charging infrastructure assumptions are derived from exogenous mobility requirements. Substituting these private cars away implies an increase in the investment needed in the power sector. The cost increase is about two to three times higher when the whole BEV fleet is charged bidirectionally compared to when it is charged smartly, irrespective of the modelled uptake regime. 

For both the smart and the bidirectional charging strategy, the increase in cost is higher when the carsharing uptake is higher. In the ``Shared + other BEVs'' framework, this holds not only in absolute terms, but also in relative terms: the cost increase in euros per substituted car is always greater, everything else being equal, when considering the high uptake. It ranges between 35 and 49~euros per substituted BEV per year in the smart charging case, and between 99 and 108~euros for bidirectional charging 

When considering bidirectional charging, the increase in power sector costs related to carsharing is 40\% to 75\%~higher in the ``Shared only'' framework, where no private BEV remain as a potential flexibility resource, compared to ``Shared + other BEVs'' (Table~\ref{tab:system_costs}). This is because the remaining private BEVs (which are absent in the ``Shared only'' framework) are more likely to be used flexibly since they are driven less and are thus available for grid interactions more often. Feeding back electricity from BEV batteries into the grid is valuable in power sectors with high shares of renewables, as it helps to substitute peak generation and electricity storage capacities (\cite{kern2022modeling, sovacool2017future}). Reducing the potential of a BEV fleet to engage in V2G accordingly hurts more than just reducing its potential to charge smartly. When assuming that all BEVs only charge smartly, but do not engage in V2G, the relative cost increases hardly differ between the ``Shared only'' and ``Shared + other BEVs'' frameworks.

\begin{figure}[!ht]
    \centering
    \includegraphics[width=17.5cm]{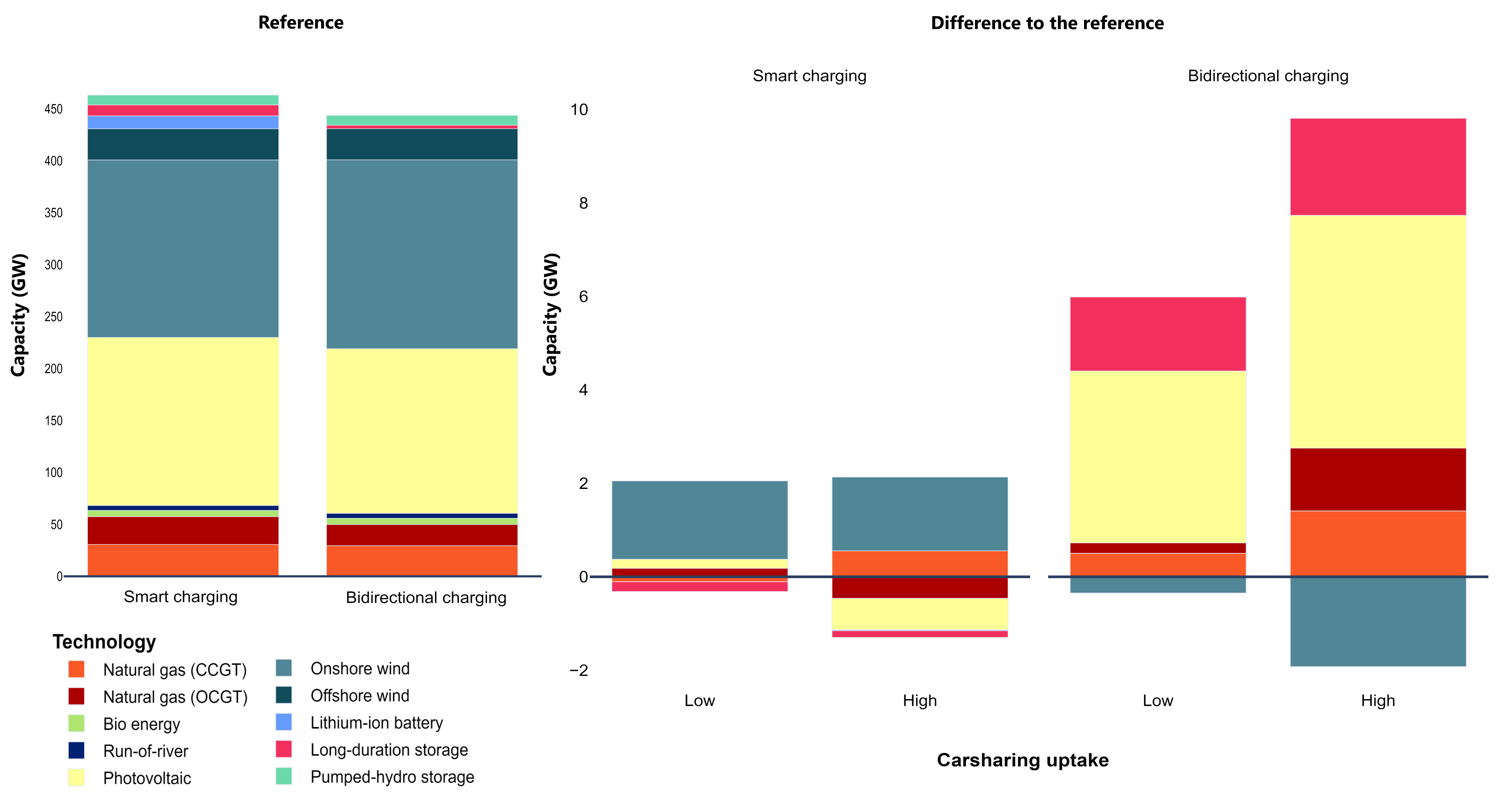}
    \caption{Generation and storage discharging capacity by technology [``Shared + other BEVs'']}
    \label{fig:power_capacity}
\end{figure}

\subsection{Changes in generation capacity and dispatch depend on charging strategy}
\label{subsec:cap_and_disp}

The switch to carsharing triggers changes in the optimal electricity generation capacity mix that are different for smart and bidirectional charging. In the case of smart charging, the switch to carsharing reduces the potential to flexibilize the demand. In the case of bidirectional charging, switching to carsharing also entails a loss in the ability of the BEV fleet to feed back electricity into the grid when the demand is relatively high compared to the available supply.

In the case of smart charging, the effects on the generation capacity generally remain small, with a moderate increase in onshore wind power investments of slightly less than 2~GW (Figure~\ref{fig:power_capacity}). This finding is qualitatively robust also for alternative weather years, where onshore wind capacity increases for most weather years by 1.7~GW on average, and only decreases in three of 22 weather years (Figure~\ref{fig:SensitivityWeather_capacity_scen}). Besides, we observe for all weather years except one additional PV investments but the magnitude of the additional capacity depends more strongly on the weather year, ranging from 0.4 to 7~GW with an average of 2.6~GW. 

For bidirectional charging, capacity effects are larger. The optimal PV generation capacity increases, by between around 1.5 and 5~GW. In contrast, optimal wind power capacities tend to decrease with carsharing (Figure~\ref{fig:power_capacity}, see also Figure~\ref{fig:gencap_allscen}). This finding holds for many of the alternative weather years (15 out of 21) and alternative cost assumptions (Figure~\ref{fig:SensitivityCosts_capacity_ref} and Figure~\ref{fig:SensitivityCosts_capacity_scen}). This is because the relatively large cumulative battery capacity of the fleet of privately owned BEV tend to foster the integration of wind power in the reference. This is visible in the cumulative BEV battery charging level time series, which does not only show diurnal patterns related to PV variability, but is also dominated by longer-duration patterns that relate to wind power variability (Figures~\ref{fig:ev_storagelevel_year}~and~\ref{fig:ev_storagelevel}). If part of the BEV fleet switches to carsharing, the potential for such longer-duration storage decreases in line with the shrinking overall battery capacity. In turn, the share of wind power in the optimal capacity mix decreases, while the use of PV with its shorter-duration diurnal variations increases. Note that this switch from wind to solar power is not driven by the changes in BEV grid availability, as the latter is not the limiting factor even in the high uptake scenario (compare Figure~\ref{fig:ev_gridavailability}).

\begin{figure}[!ht]
    \centering
    \includegraphics[width=17.5cm]{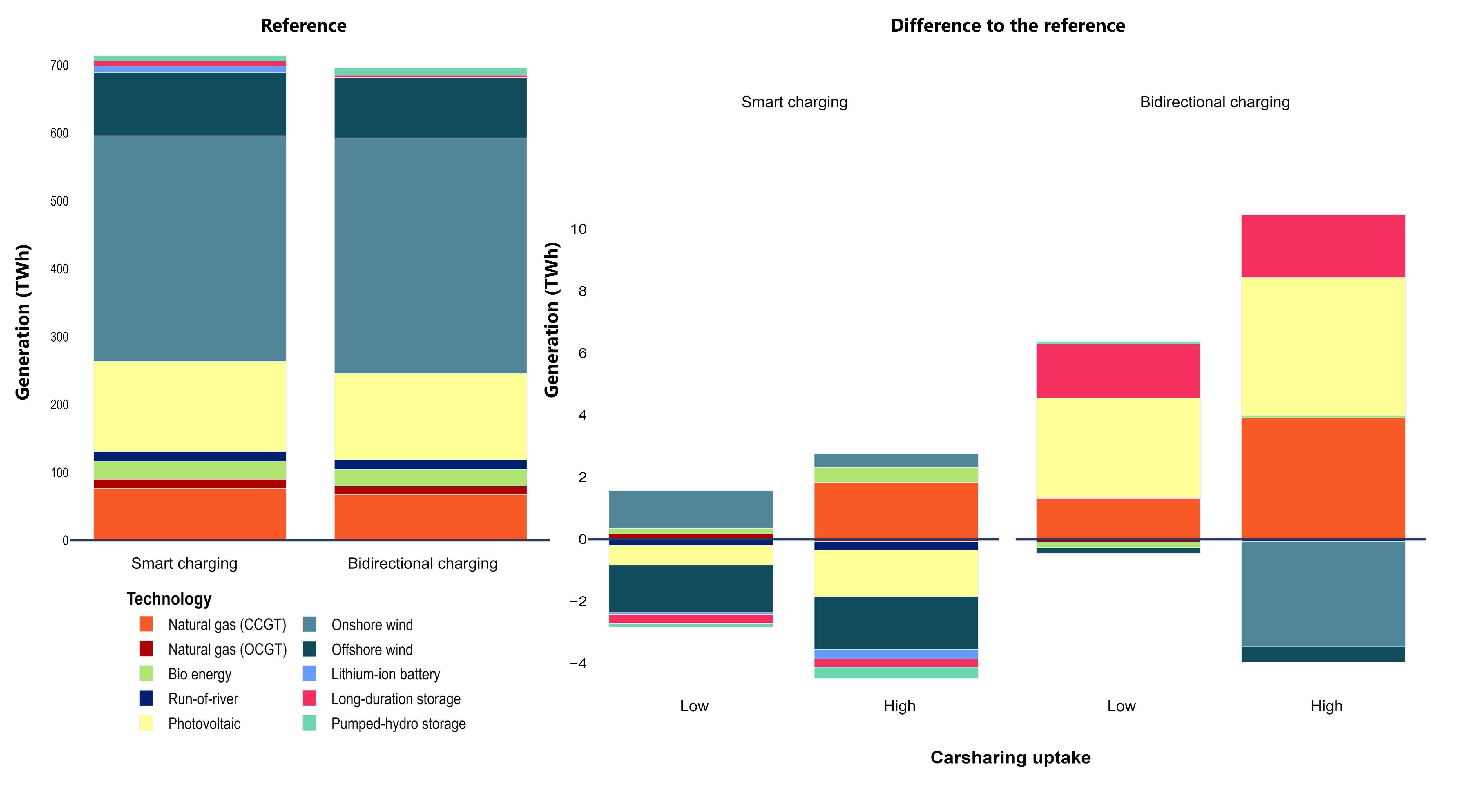}
    \caption{Yearly electricity generation by technology [``Shared + other BEVs'']}
    \label{fig:power_generation}
\end{figure}

As the longer-duration storage capabilities of the car fleet decrease when private BEVs with bidirectional charging are substituted with shared BEVs, optimal investments into stationary long-duration electricity storage increase. This is true for both charging and discharging power (Figures~\ref{fig:sto_charging_scen} and~\ref{fig:sto_discharging_scen}) as well as storage energy capacity (Figure~\ref{fig:sto_energy_scen}) of stationary long-duration storage. Depending on the weather year considered, we find substantial variations in the magnitude of the additional investment in long-duration storage. For the discharging capacity, the additional investment varies between 0.5~GW and 7.5~GW, with an average of 3~GW (Figure~\ref{fig:SensitivityWeather_capacity_scen}). For the energy capacity, the additional investment also shows important variations but is always positive, except for one year (Figures~\ref{fig:SensitivityWeather_storage_ref} and~\ref{fig:SensitivityWeather_storage_scen}). Likewise, investments in firm CCGT and OCGT capacities increase (Figure~\ref{fig:power_capacity}). This generally holds for most weather year sensitivities (Figures~\ref{fig:SensitivityWeather_capacity_ref} and~\ref{fig:SensitivityWeather_capacity_scen}). Yet, these additional capacities do not necessarily translate into greater yearly electricity generation from gas power plants (Figure~\ref{fig:power_generation} and Figures~\ref{fig:SensitivityWeather_generation_ref} and~\ref{fig:SensitivityWeather_generation_scen}). In particular, generation from OCGT plants hardly increases, as some of the additional firm capacity is only needed during very few hours in which the residual load is particularly high, and which are served by V2G of private BEVs in the reference, but to a lower extent by shared BEVs. In contrast, electricity generation from CCGT plants increases (between one and four additional~TWh for the year chosen in the main setting, see Figure~\ref{fig:powergen_allscen} and Figures~\ref{fig:SensitivityCosts_generation_ref} and~\ref{fig:SensitivityCosts_generation_scen}). This means that the renewable share slightly decreases and the electricity mix gets slightly more carbon-intensive. In the case of the ``Shared only'' setting, where no additional private BEVs are available for providing long-duration storage capabilities, we observe even higher investments into stationary storage charging and discharging capacity (Figures~\ref{fig:sto_charging_ref}-\ref{fig:sto_discharging_scen}). In contrast to the bidirectional charging case, the switch to carsharing hardly impacts stationary electricity storage needs when BEVs only charge smartly. This is due to the fact that necessary storage requirements are already realized in the reference (Figures~\ref{fig:sto_charging_ref} to ~\ref{fig:sto_energy_scen}).

\subsection{Shared BEVs can still charge and discharge in line with variable renewables, but to a lesser extent}

\begin{figure}
    \centering
    \includegraphics[width=15cm]{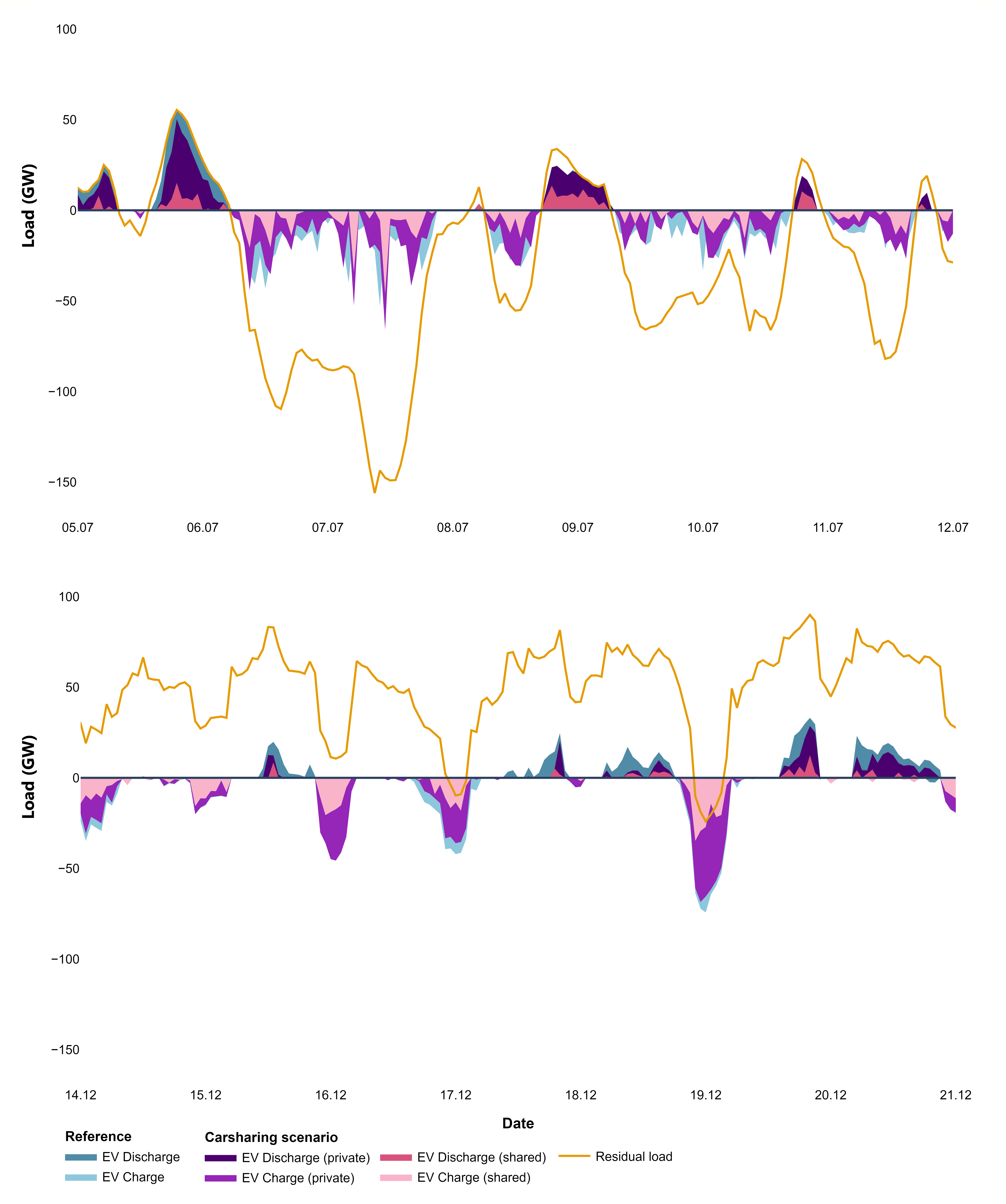}
    \caption{Residual load curve, BEV charging and discharging for a subset of summer hours (upper panel) and winter hours (lower panel) [``Shared + other BEVs'', bidirectional charging, high uptake]. Note that the area plots are partly unstacked to better illustrate the decreasing flexibility of shared BEVs. The carsharing scenario areas (purple and pink) are stacked on top of each other, while the reference areas (blue) are not.}
    \label{fig:EV_dispatch_resload_high}
\end{figure}

We observe that, with perfect foresight, BEV fleets with shared cars could still be operated largely in line with variable renewable generation patterns (Figures~\ref{fig:EV_dispatch_resload_high} and~\ref{fig:EV_dispatch_resload_smart_high}). In particular, optimised charging takes place when RES availability is high, e.g.~during negative residual load events in summer, and during periods with low positive residual load in winter (Figure~\ref{fig:EV_dispatch}). Conversely, BEVs are discharged at times when the residual load is high, meaning that the demand exceeds variable renewable electricity generation.

A fleet with shared BEVs has a lower potential to balance renewable variations, i.e.~to shift energy between negative and positive residual load periods, than a fleet of only private BEVs: light or dark blue areas on Figure~\ref{fig:EV_dispatch_resload_high} are always larger than or equal to the sum of the purple and pink areas. The difference in magnitude is slightly smaller in the low uptake case (Figure~\ref{fig:EV_dispatch_resload_low}). The charging and discharging patterns of the shared fleet (light purple and light pink) are more variable compared to those of the overall fleet. The reason is that shared vehicles are driven much more frequently and hence can only charge or discharge for shorter periods before they are used again. These results also apply to the ``Shared only'' setting (Figure~\ref{fig:EV_dispatch_resload_sharedonly_high}). 

\subsection{Green hydrogen production as another source of flexibility}

In additional model runs, we include green hydrogen production as an example of another sector coupling technology which increases the demand-side flexibility of the system. More precisely, we consider a yearly industrial demand for green hydrogen of 30~TWh, which is assumed to be produced domestically via electrolysis. This largely reflects current German hydrogen policy targets for 2030. We assume that this demand for green hydrogen comes on top of the hydrogen required for reconversion to electricity, i.e.~for long-duration electricity storage, and that the industrial hydrogen demand is equally distributed over every hour of the year. In general, green hydrogen can be flexibly produced at low costs in periods when plenty of renewable electricity is available and prices are low. Yet, shifting hydrogen production to such periods requires according electrolysis and hydrogen storage capacities. The model optimises these capacities as well as their hourly use as endogenous model variables, rendering hydrogen production a flexible sector coupling option. 

Including additional green hydrogen production in the model does not lead to major changes of the power sector cost effects of carsharing (Table~\ref{tab:system_costs}). In the ``Shared + other BEVs'' framework, the additional costs per substituted car slightly decrease by 4-6\%, compared to the setting with no green hydrogen industrial demand. The reason for this is that green hydrogen production serves as a flexible load, which makes smart and bidirectional BEV charging less beneficial than in the setting without hydrogen. Reducing the BEV fleet's flexibility potential by shifting to carsharing thus hurts less. In the ``Shared only'' framework, relative effects are even smaller. One of the reasons why the power sector cost effects of carsharing are robust against adding flexible green hydrogen production is that the renewable share slightly differs (Table~\ref{tab:res_shares}). Especially in the cases with bidirectional charging, carsharing induces a slightly higher use of non-renewable electricity generation when green hydrogen production is added compared to the baseline specification, which helps to reduce power sector costs effects. This goes along with respective changes in renewable generation capacity (Figure~\ref{fig:SensitivityH2_gencap_ref} and Figure~\ref{fig:SensitivityH2_gencap_scen}). In alternative model parameterisations where the minimum renewable share is always binding, we conversely find that green hydrogen production increases the power sector costs effects of carsharing, as electrolysis would then compete with smartly charged BEVs for cheap electricity in periods where renewable availability is high.

\section{Discussion}
\label{sec:discussion}

Our model results rely on a set of assumptions. The parameterisation is on purpose rather restrictive when it comes to the flexibility of the power sector: it does not include interconnection with other neighbouring countries, nor other flexible demand-side options that could engage in temporal load shifting. Including additional flexibility options would likely decrease the impacts of carsharing, which would translate into smaller cost increases as well as smaller deviations from the reference in terms of capacity mix and dispatch. In this respect, our main results can be interpreted as an upper bound of the power sector impacts. 

When it comes to the specific mobility time series, our framework is also rather conservative in terms of flexibility. In the ``Shared only'' framework, we consider only one cluster that switches to carsharing, namely the one with the smallest amount of trips per day and with the shortest overall travelled distance. These vehicles are idle most of the time, and thus offer the greatest flexibility potential if smart charging or vehicle-to-grid technologies are available. Assuming that such profiles switch entirely to carsharing removes the greatest source of flexibility that one could get from the sector coupling with passenger cars. The results thus provide a very conservative quantification of how much the flexibility potential of BEVs would be reduced. The same applies to the ``Shared + other BEVs'' framework, although to a smaller extent, since we also consider more intense car usage by including clusters 3 to 5. Cost increases still remain moderate.

Another key assumption of our setting is that car mobility needs remain constant, irrespective of the way they are fulfilled. This means that the switch to carsharing does not induce a modification of when and how people decide to move with a car. This is disputable as it could be the case that switching to carsharing comes along with less car trips or shorter car trips, potentially combined with other transport modes such as active modes or public transport. If carsharing induced a decrease in overall car use, this could entail a greater flexibility of the shared cars than the one we model in this paper, hence a smaller cost increase. This is confirmed by a sensitivity analysis where we assume shared cars are used less, such that their electricity consumption decreases by~10\% (see Table~\ref{tab:sensitivity_eved_systemcosts} and Figures~\ref{fig:SensitivityEved_capacity} and~\ref{fig:SensitivityEved_generation}), which corroborates our upper bound interpretation of the modelled flexibility reduction effects. 

Previous research has shown that carsharing has the potential to decrease private car ownership, and one shared car might replace up to 20 private cars in the most optimistic cases (\cite{jochem_does_2020}). Nevertheless, whether carsharing ultimately reduces car ownership and contributes to a decrease of the fleet size relies on an efficient policy design without which carsharing might even lead to an increase of the car fleet size. For example, carsharing policies should ensure that carsharing is used as a complement rather than a substitute to public transport and active mobility (\cite{silvestri2021contribution}). In this work, we assume a substitution rate of about five, which reflects a decrease in private car ownership, but remains conservative in comparison to some potential substitution rates found in the literature. Assuming a higher substitution rate might lead to higher cost increases, but it could also imply greater mobility changes, such as smaller and/or less numerous car trips. More research needs to be carried out in order to better understand how these effects interact. 

Next, car users and fleet operators are assumed to have perfect foresight when it comes to charging and discharging their cars. This means that people can perfectly predict their car trips over time. This absence of uncertainty when it comes to the timing and duration of future trips enables to optimize the grid interactions of the overall fleet to the fullest extent possible, providing a benchmark that may not be achievable with more realistic behaviours. In that sense, our modelling might overestimate the flexibility of both private and shared BEVs. Note that we focus on deviations to respective references, such that assumptions present in both settings play less of a role. Besides, we consider that the whole BEV fleet uses the same charging strategy. This might lead to an overestimation of the flexibility potential when smart or bidirectional charging is considered. Nevertheless, considering only one charging strategy across the fleet can be interpreted as a limit case. Results for mixed fleets with several charging strategies should lie in between these limit cases. Integrating mixed fleets in modelling exercises would allow to better represent the different abilities of carsharing operators and private car owners to choose flexible (dis-)charging strategies and could be done in further research. 

Some modelling features rely on stylised assumptions for tractability and numerical feasibility, especially concerning BEVs time series. For example, including more car mobility profiles and additional vehicle models would be desirable to generate an even more representative fleet, but this heavily increases the numerical burden. On the power modelling side, a potential limitation is the optimisation at a national scale and not at the substation level, which could potentially overestimate the flexibility potential of BEVs with smart charging (\cite{strobel2022joint}). Further research on how to increase the representativity of mobility patterns while not under- or overestimating the flexibility potential of BEVs in power sector models is still needed. Besides, we assume a bigger battery capacity for shared BEVs compared to private ones. Our assumption is also slightly bigger than what is available on the carsharing market today, but it appears plausible that this will not be the case in the future anymore. Assuming a smaller battery capacity for shared cars would further constrain the flexibility potential of shared BEVs, which might lead to higher cost increases. Nevertheless, too small a battery size would not allow vehicles to accommodate as many trips as required by the constant mobility need assumption. In this regard, it seems that the overestimation of the flexibility potential of shared cars remains small. Likewise, we assume that shared cars always find a charging station after a trip and that such charging stations have a power rating of 75~kW. Even if this is optimistic compared to today's situation, it appears plausible that future carsharing fleets could be managed by their operators such that users can always find a charging station, potentially a specifically designed one for carsharing users. 

While most effects determined in our analysis are relatively small in size, both in terms of cost increases and changes in the capacity mix, they should be put in perspective with potential gains from switching to electric carsharing and reducing the BEV fleet size. BEVs are also associated with negative externalities which can be thought as social costs (\cite{gossling2019social, de2018total}). Methods to derive estimates of such costs have been developed (\cite{schroder2023ending}). In a case study for Munich, the overall external cost of a BEV is estimated to be around 14~euro-cents per kilometer driven. Assuming BEVs are driven ten kilometers a day, and five days per week, the yearly external cost of a BEV is around 364~euros. Removing one BEV from the system could thus represent a gain of 364~euros per year for society. In our main scenario, these reduced negative externalities are around three times larger than power sector cost increases under the assumption of bidirectional charging, and even around a factor seven larger assuming smart charging. These estimates need to be interpreted cautiously since they depend on many underlying assumptions and are specific to each urban setting. Nevertheless, they highlight that the increase in the power sector costs is only a part of the overall impact of switching from private to shared electric cars. In addition, carsharing may also reduce the carbon footprint of BEVs because of higher overall use of cars over their lifetime (\cite{morfeldt2022}). An evaluation of the extent to which the changes in power sector carbon emissions might be counteracted by the reduction in emissions related to car and battery manufacturing could constitute future research work. From these insights, we conclude that the potential societal gains from the reduction of the car fleet size are likely to outweigh the potentially adverse electricity sector effects. Nevertheless, more research needs to be done in this direction, in order to better understand how different effects play one with another. Further interdisciplinary research on how to optimally design and operate carsharing fleets and the charging infrastructure supporting them is also required.

Overall, our research demonstrates that future BEV fleets that include relevant amounts of shared cars can still be operated largely in line with variable renewable energy sources and thus contribute to their system integration. Note that this relies on the assumption that vehicle operators, or some aggregators, have perfect foresight on mobility patterns and charging availability, both for privately owned and shared BEVs. Hence, our results imply an upper bound of the beneficial system effects of flexible BEVs. Yet, we argue that system-oriented charging of future shared car fleets appears to be plausible, given that carsharing operators are likely to have an interest in low charging costs.

We conclude that the moderate power system cost increases related to electric carsharing are only one aspect of various, and potentially beneficial, changes brought by such a structural shift in private transport. Further research has to be done to comprehensively evaluate how all these aspects interact one with another. At the same time, policymakers should strive to enhance the ability of carsharing operators to operate their fleet as flexibility as possible and design policies ensuring that carsharing enables a decrease in private car ownership.

\section*{Author contributions}
\textbf{Adeline Guéret}: Conceptualisation, Methodology, Software, Formal analysis, Investigation, Data Curation, Writing - Original Draft, Writing - Review \& Editing, Visualisation.
\textbf{Wolf-Peter Schill}: Conceptualisation, Investigation, Writing - Review \& Editing, Funding acquisition.
\textbf{Carlos Gaete-Morales}: Software, Visualisation.

\section*{Data and code availability}

\begin{itemize}
    \item All data - except the Mobility in Germany survey data - and codes are available under permissive licenses in public repositories. 
    \item The Mobility in Germany survey data (B1 data package) can be requested against a small service fee here: \url{https://daten.clearingstelle-verkehr.de/order-form.html#279}
    \item The general repository of the DIETER model can be found here: \url{https://gitlab.com/diw-evu/dieter_public}. A general model documentation is available here: \url{https://diw-evu.gitlab.io/dieter_public/dieterpy/}.
    \item A complete installation guide and documentation of \textit{emobpy} can be found here: \url{https://pypi.org/project/emobpy/}. 
    \item The particular tool and model versions used for the present study, together with the complete input dataset (except the original data from the Mobility in Germany survey), is available under permissive licenses in the following dedicated repository: \url{https://gitlab.com/diw-evu/projects/electric-carsharing}. 
\end{itemize}

\section*{Acknowledgments}
We thank the members of the research group \textit{Transformation of the Energy Economy} at DIW Berlin and participants to the ITEA Conference 2023, the European Transport Conference 2023, the British Institute of Energy Economics Research Conference 2023, the YEEES 32 Seminar 2023, the International Energy Workshop 2024, the EAERE Conference 2024, the 5$^{th}$ CIRED Summer School and the Chalmers Transport Seminar, as well as participants to the DIW EVU/KLI Brown Bag Seminar and the DIW Graduate Center Workshop 2023 for their constructive feedback and helpful comments. We acknowledge research grants by the German Federal Ministry of Education and Research via the ``Ariadne'' projects (Fkz 03SFK5NO \& 03SFK5NO-2).

\section*{Declaration of interests}
The authors declare no competing interests.

\newpage
\printbibliography

\appendix

\newpage
\clearpage
\section{Clusters}

\begin{figure}[!ht]
    \centering
    \includegraphics[width=16.5cm]{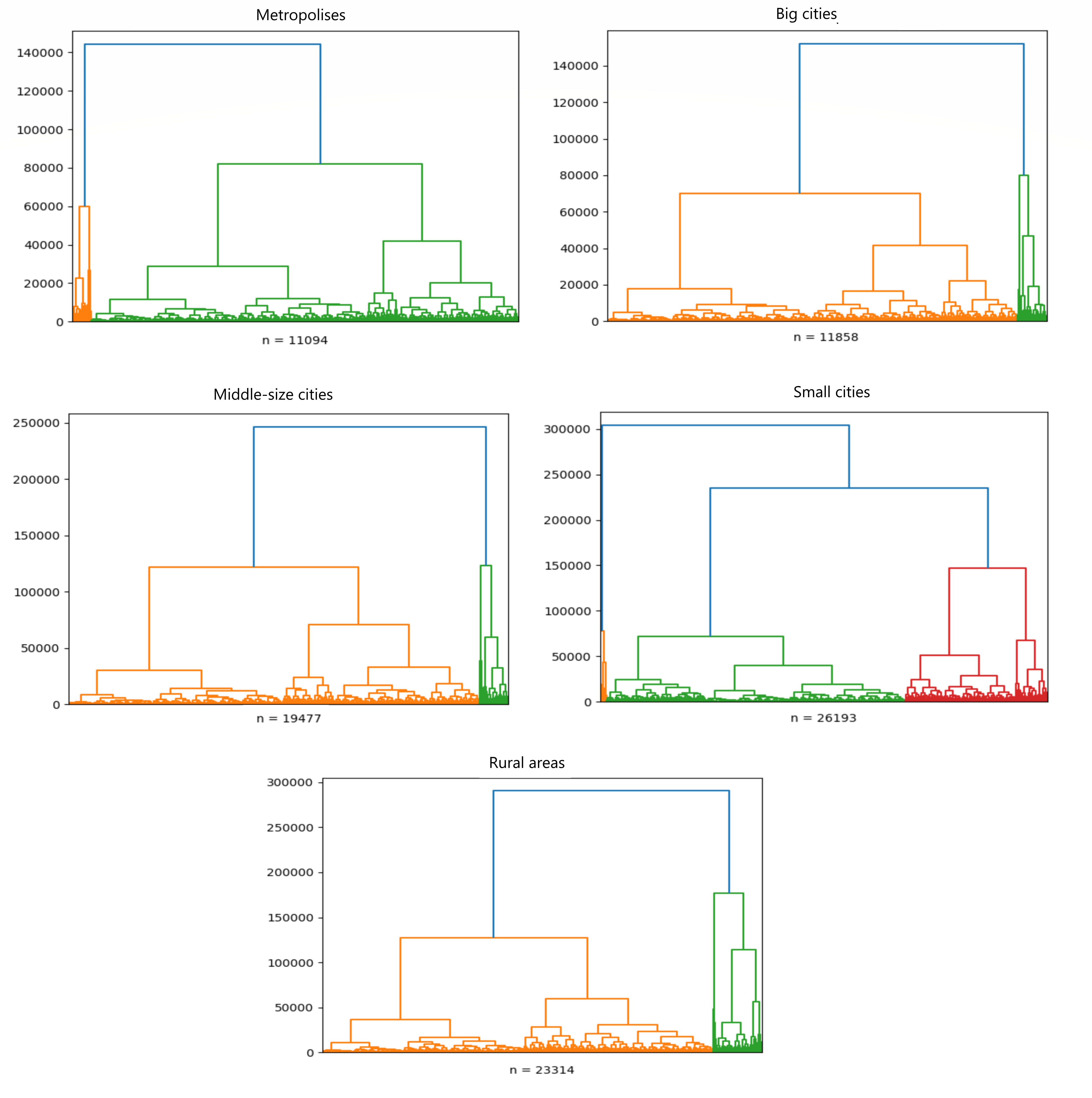}
    \caption{Dendrograms for the sequences for all location types (working days)}
    \label{fig:dend}
\end{figure}

\subfile{cluster_statistics.tex}

\begin{figure}[!ht]
    \centering
    \includegraphics[width=17.5cm]{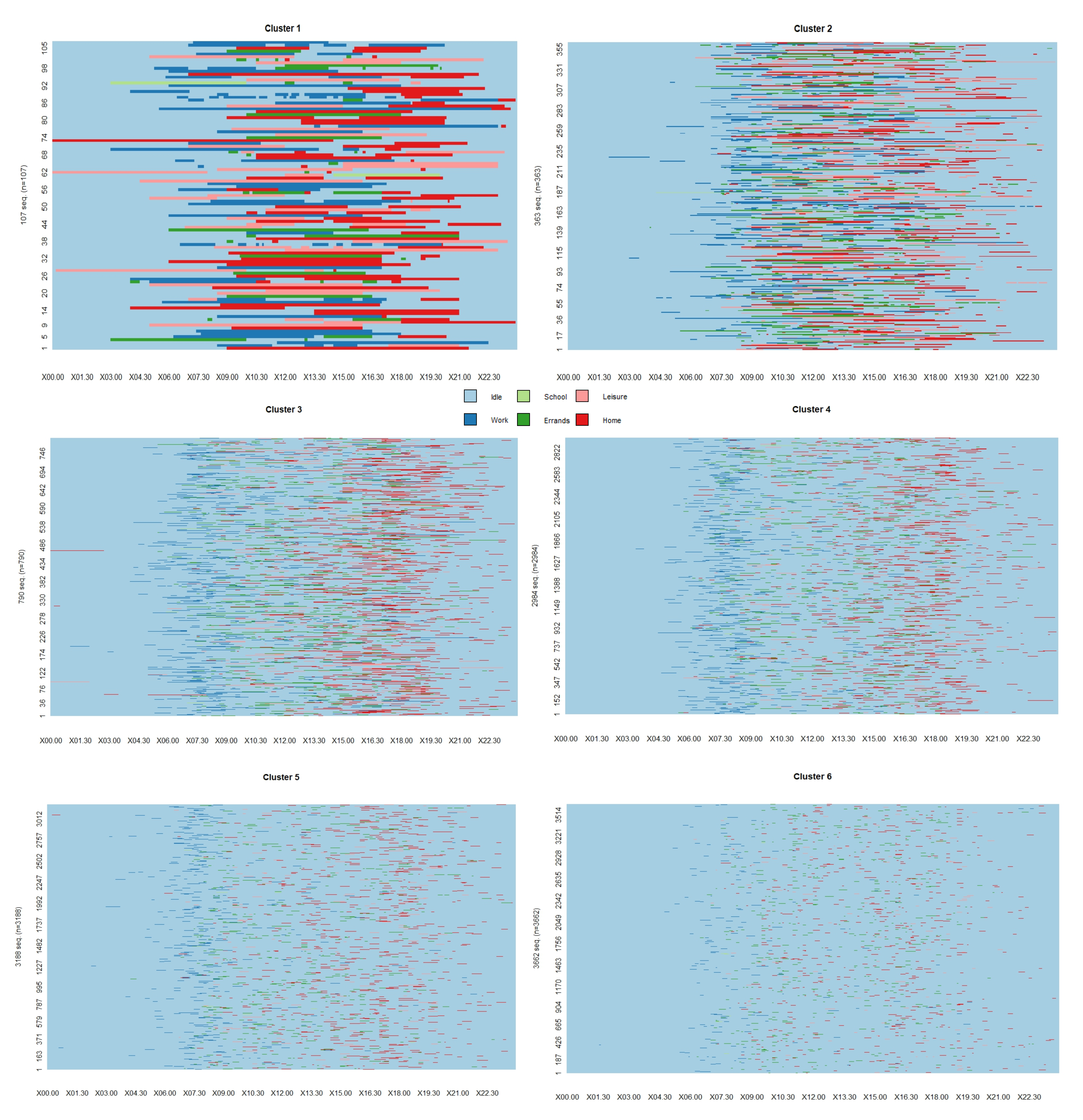}
    \caption{Sequence index plots (metropolises, working days, clusters 1 to 6)}
    \label{fig:sequences_16}
\end{figure}

\clearpage
\newpage

\section{emobpy}
\label{sec:appendix_emobpy}

\textit{emobpy} is an open-source Python tool designed to generate synthetic BEV time series. Using the tool, four different types of time series can be generated: (i) vehicle mobility, (ii) driving electricity consumption, (iii) grid availability and (iv) grid electricity demand. This order matters since each time series builds upon the previous one in order to be created. 

The \textit{vehicle mobility} time series indicates how many trips are undertaken during a day and, for each trip the destination, distance and duration. Possible destinations are work/school, leisure, home and errands. The generation of the vehicle mobility time series for shared BEVs relies on the same distributions as for private BEVs of a given cluster in a given location but centered around a new mean. To calculate the mean number of trips for shared BEVs, we first calculate the total number of private cars in a cluster, assuming that each sequence corresponds to one car. We then compute the total number of shared cars based on the assumed substitution rate. Finally, we calculate the average number of trips per shared car in the cluster. 

The \textit{driving electricity consumption} time series computes for 5-minutes time steps how much electricity is used to power the trip depending on vehicle and climatic features such as the motor or the ambient temperature as well as the speed of the trip. 

The \textit{grid availability} time series indicates whether the vehicle is plugged to a charging station and if yes, at which power rating. The grid availability time series is also computed at the 5-minute time step and depends on the assumed average charging station availability (Table~\ref{tab:emobpy_assumptions}) and the mobility pattern of a particular vehicle.  
For shared BEVs, we assume a higher charging station availability than for private BEVs, as we assume that carsharing operators are able to design optimally both users' financial incentives and their charging network so that customers can always plug the shared BEV they have used after a ride. 

Finally, the \textit{grid electricity} time series corresponds to the electricity actually taken from the grid for charging a vehicle, using the consumption and availability time series and assuming a given charging behaviour. We only compute this last time series when looking at scenarios with uncontrolled charging. 

\subfile{emobpy_condprob.tex}

\subfile{emobpy_assumptions.tex}

\begin{figure}[!ht]
    \centering
    \includegraphics[width=16cm]{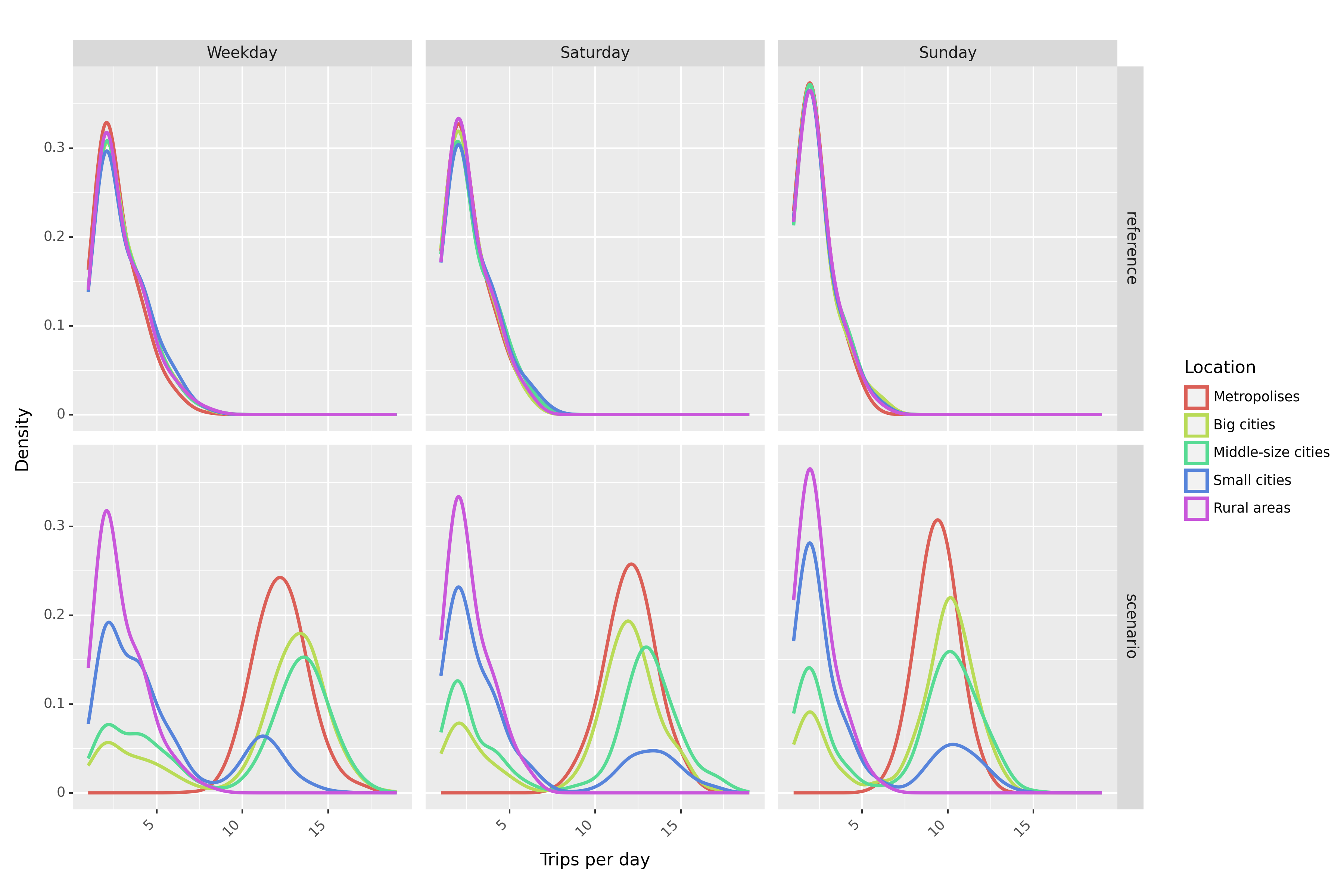}
    \caption{Density of number of daily trips by location and day type in the reference and in the scenario [``Shared + other BEVs'', high uptake]}
    \label{fig:emobpy_dist_ntrips}
\end{figure}

\begin{figure}[!ht]
    \centering
    \includegraphics[width=14cm]{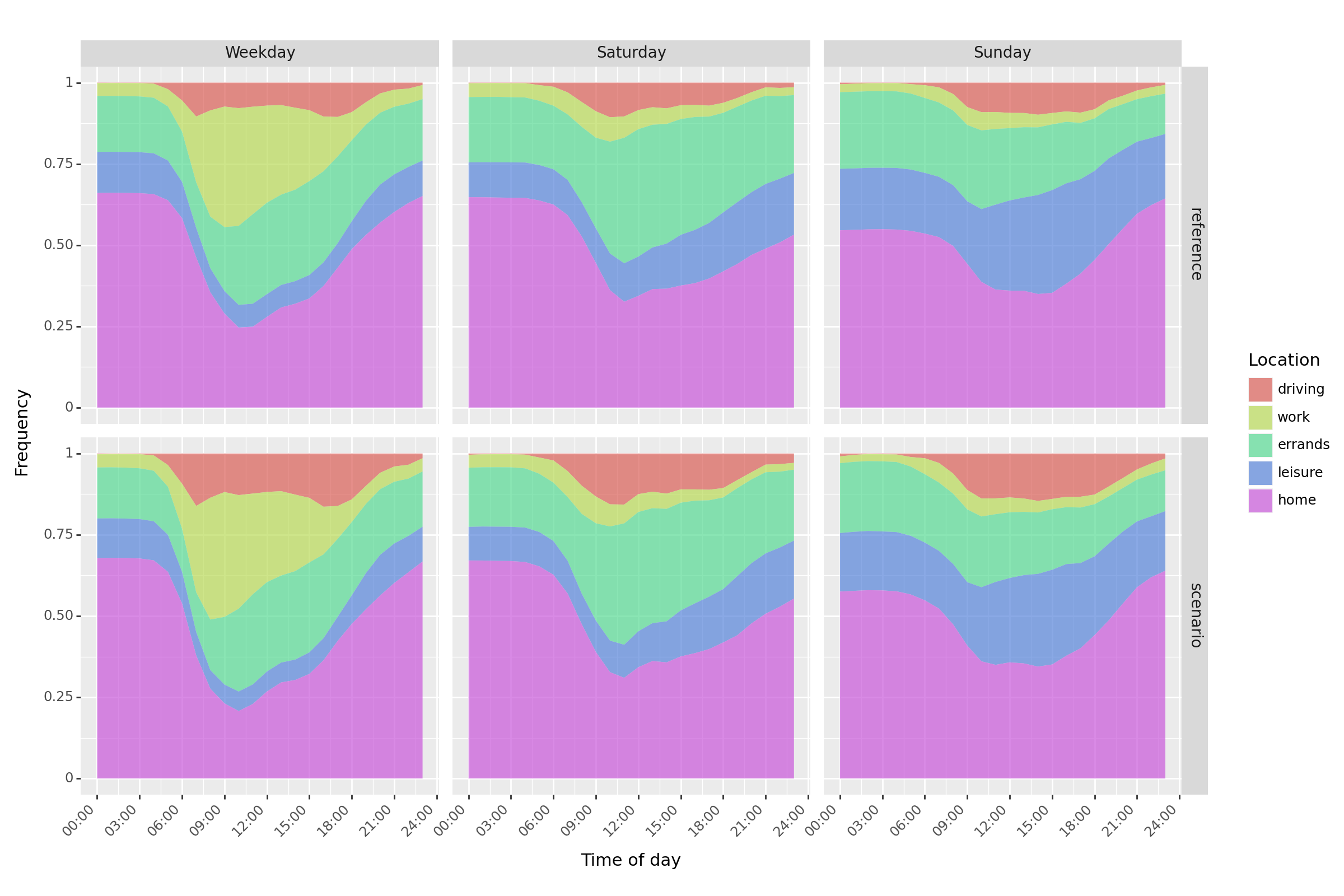}
    \caption{Distribution of BEVs's parking place for each time step by location and day type in the reference and in the scenario [``Shared + other BEVs'', high uptake]}
    \label{fig:emobpy_freqdestination}
\end{figure}

\begin{figure}[!ht]
    \centering
    \includegraphics[width=14cm]{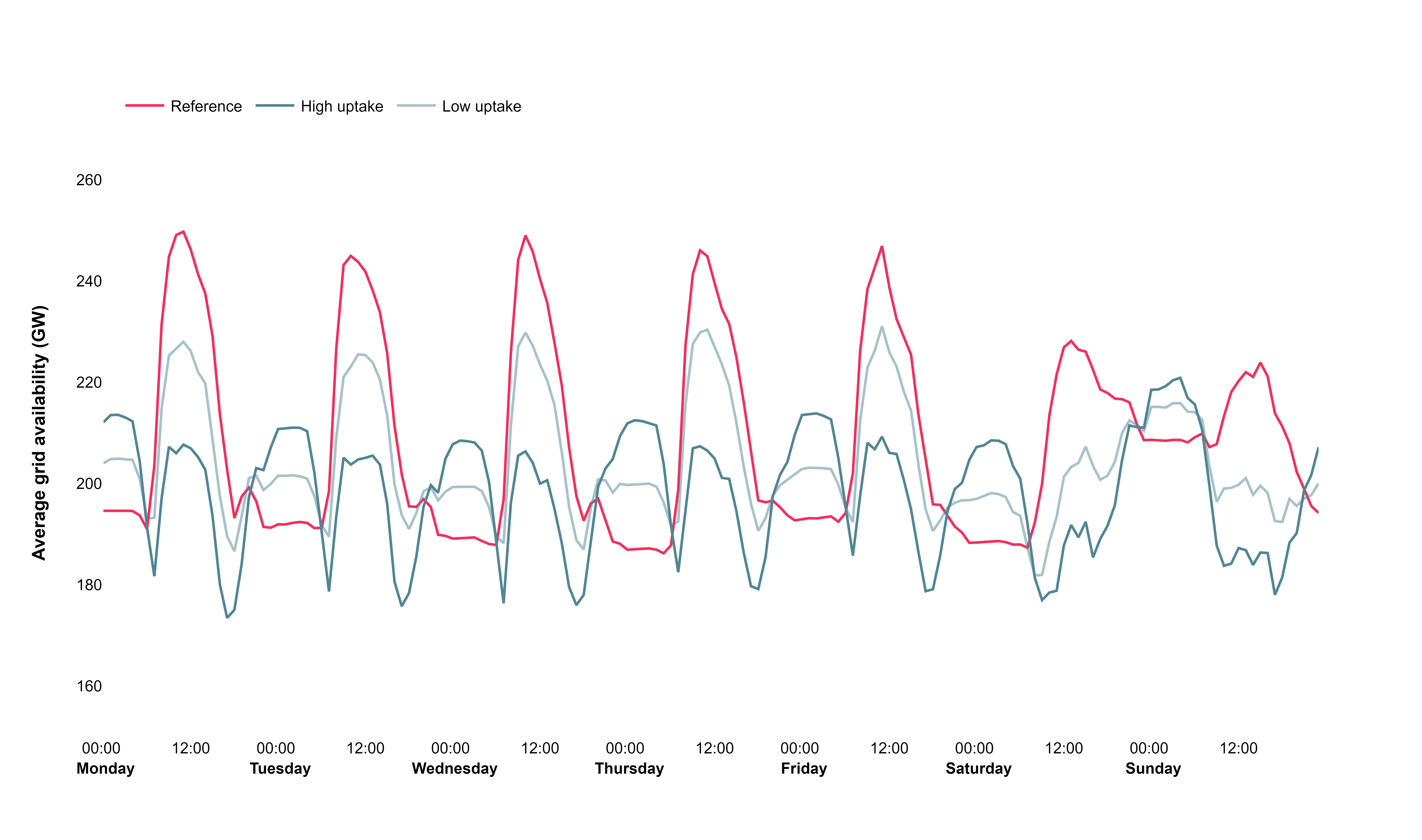}
    \caption{Battery electric vehicles' average overall grid availability for each hour of the week in the reference and carsharing scenarios [``Shared + other BEVs'']}
    \label{fig:ev_gridavailability}
\end{figure}

\clearpage
\newpage

\section{DIETER}

\subfile{dieter_cardistribution.tex}

\subfile{dieter_parametersgeneration.tex}

\subfile{dieter_parametersstorage.tex}

\clearpage
\newpage

\section{Detailed results for all scenarios} 
\label{sec:additional_results}

\subsection{Power sector costs}
\label{sec:costs}

\subfile{system_costs.tex}

\subsection{Renewable energy shares}

\subfile{res_shares.tex}

\clearpage
\newpage

\subsection{Power generation and capacity mix}
\label{sec:generation}

\begin{figure}[!ht]
    \centering
    \includegraphics[width=15.5cm]{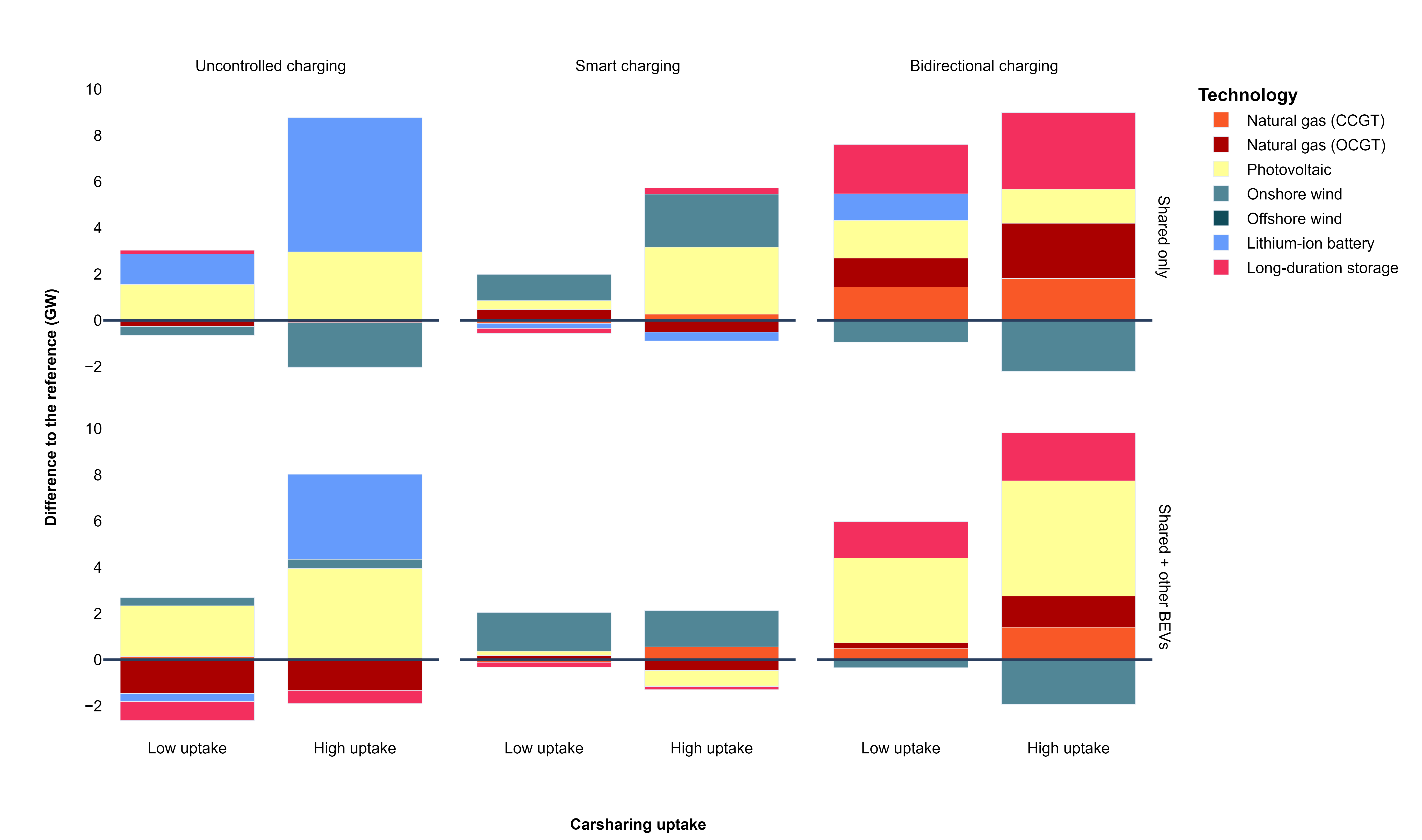}
    \caption{Effects of carsharing on optimal generation and storage discharging capacity}
    \label{fig:gencap_allscen}
\end{figure}

\begin{figure}[!ht]
    \centering
    \includegraphics[width=15.5cm]{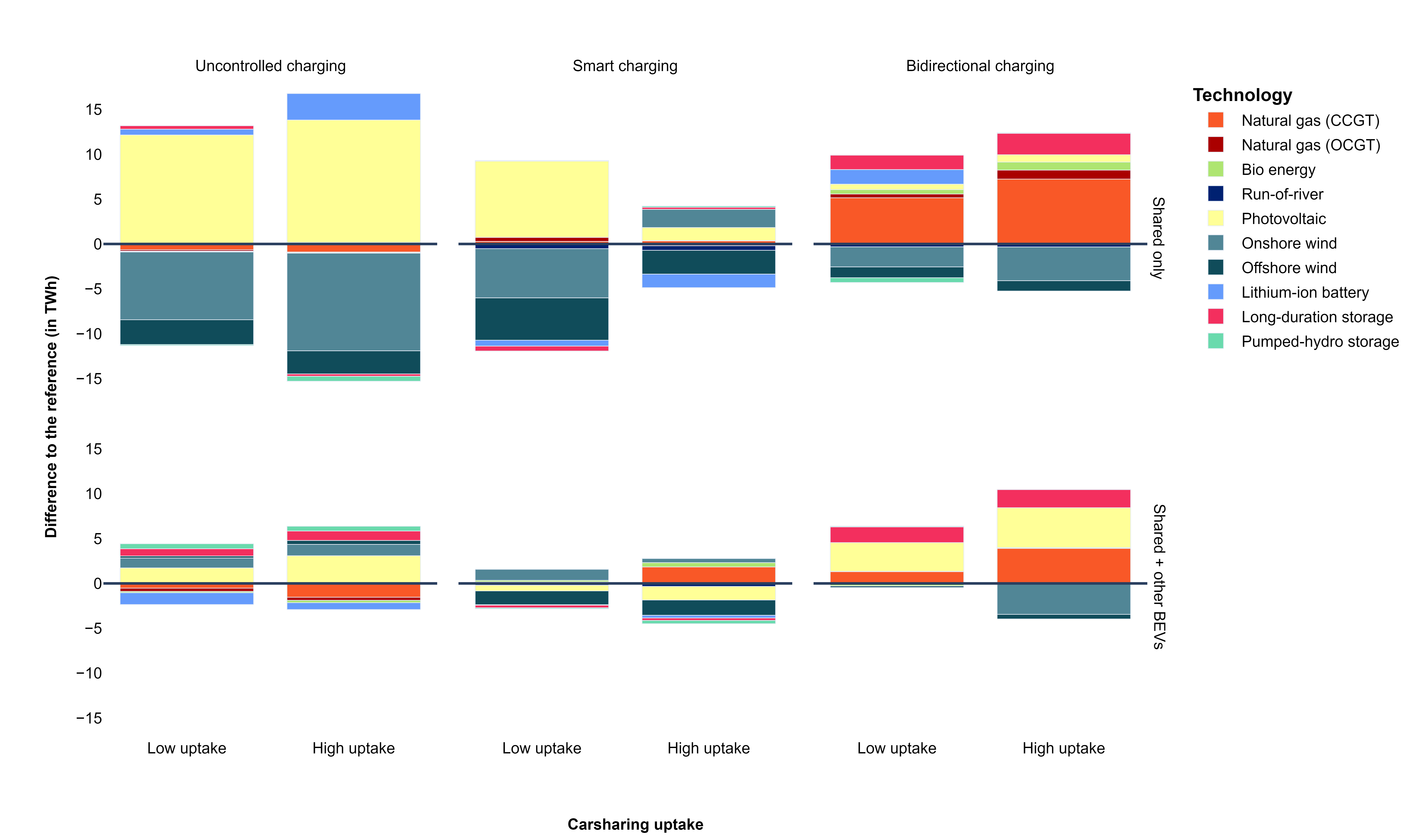}
    \caption{Effects of carsharing on yearly electricity generation}
    \label{fig:powergen_allscen}
\end{figure}

\clearpage
\newpage

\subsection{Power dispatch}
\label{sec:dispatch}

\begin{figure}[!ht]
    \centering
    \includegraphics[width=15.5cm]{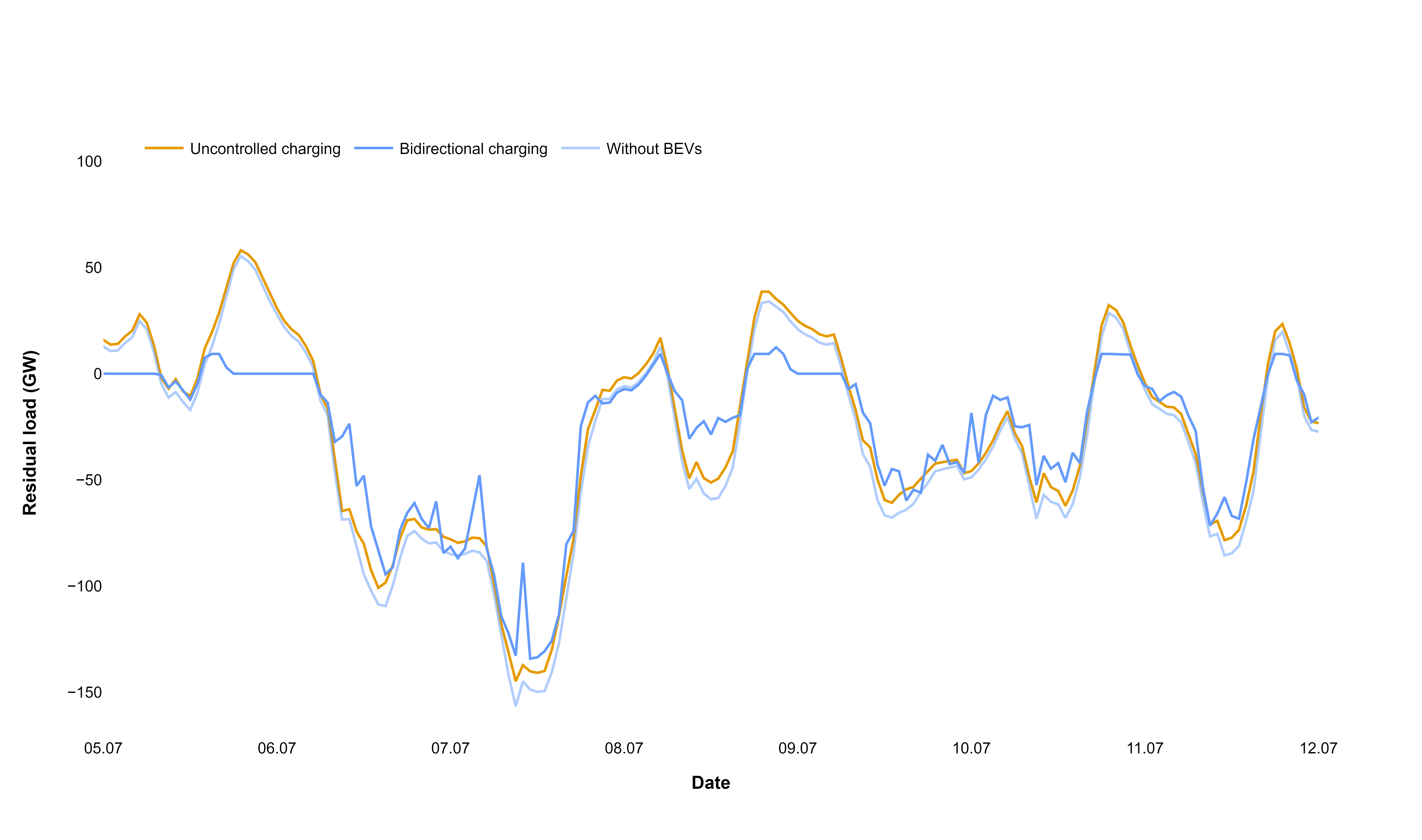}
    \caption{Residual load for different charging strategies for a subset of summer hours [``Shared + other BEVs'', reference]}
    \label{fig:resload_summer}
\end{figure}

\begin{figure}[!ht]
    \centering
    \includegraphics[width=15.5cm]{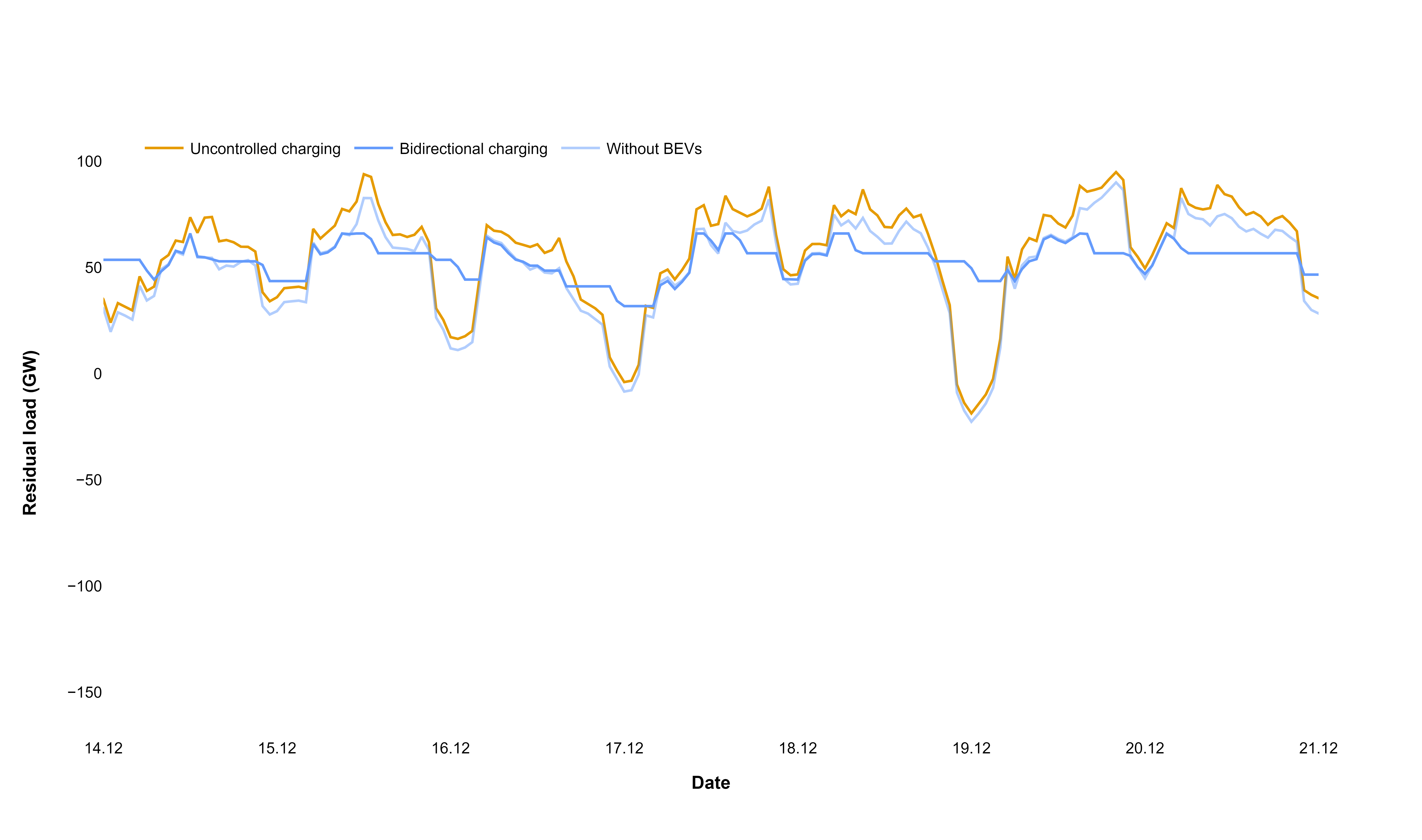}
    \caption{Residual load for different charging strategies for a subset of winter hours [``Shared + other BEVs'', reference]}
    \label{fig:resload_winter}
\end{figure}

\clearpage

\begin{figure}[!ht]
    \centering
    \includegraphics[width=17.5cm]{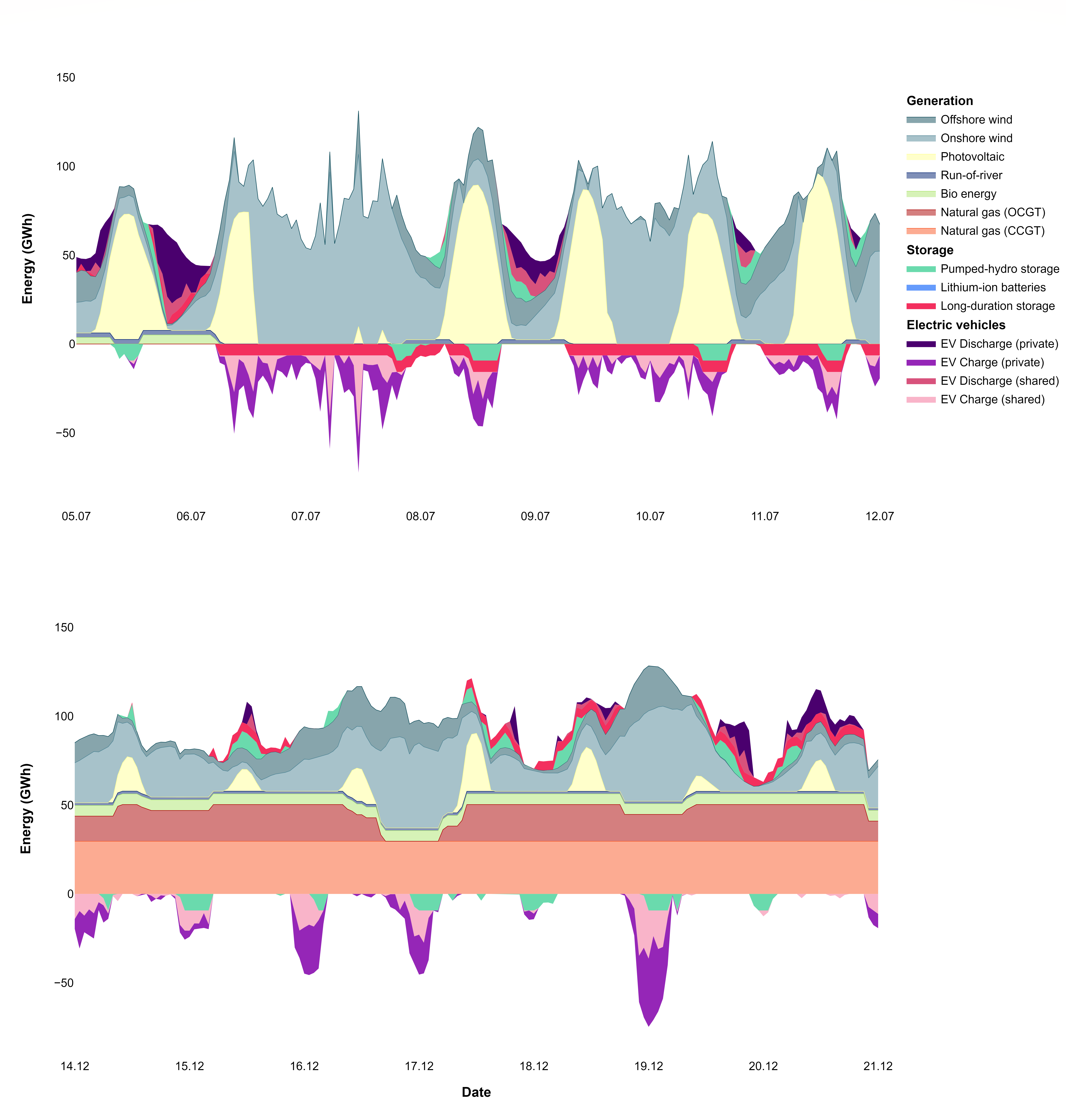}
    \caption{Hourly dispatch for a subset of summer hours (upper panel) and winter hours (lower panel) [``Shared + other BEVs'', bidirectional charging, high uptake]}
    \label{fig:EV_dispatch}
\end{figure}

\clearpage

\begin{figure}[!ht]
    \centering
    \includegraphics[width=16.5cm]{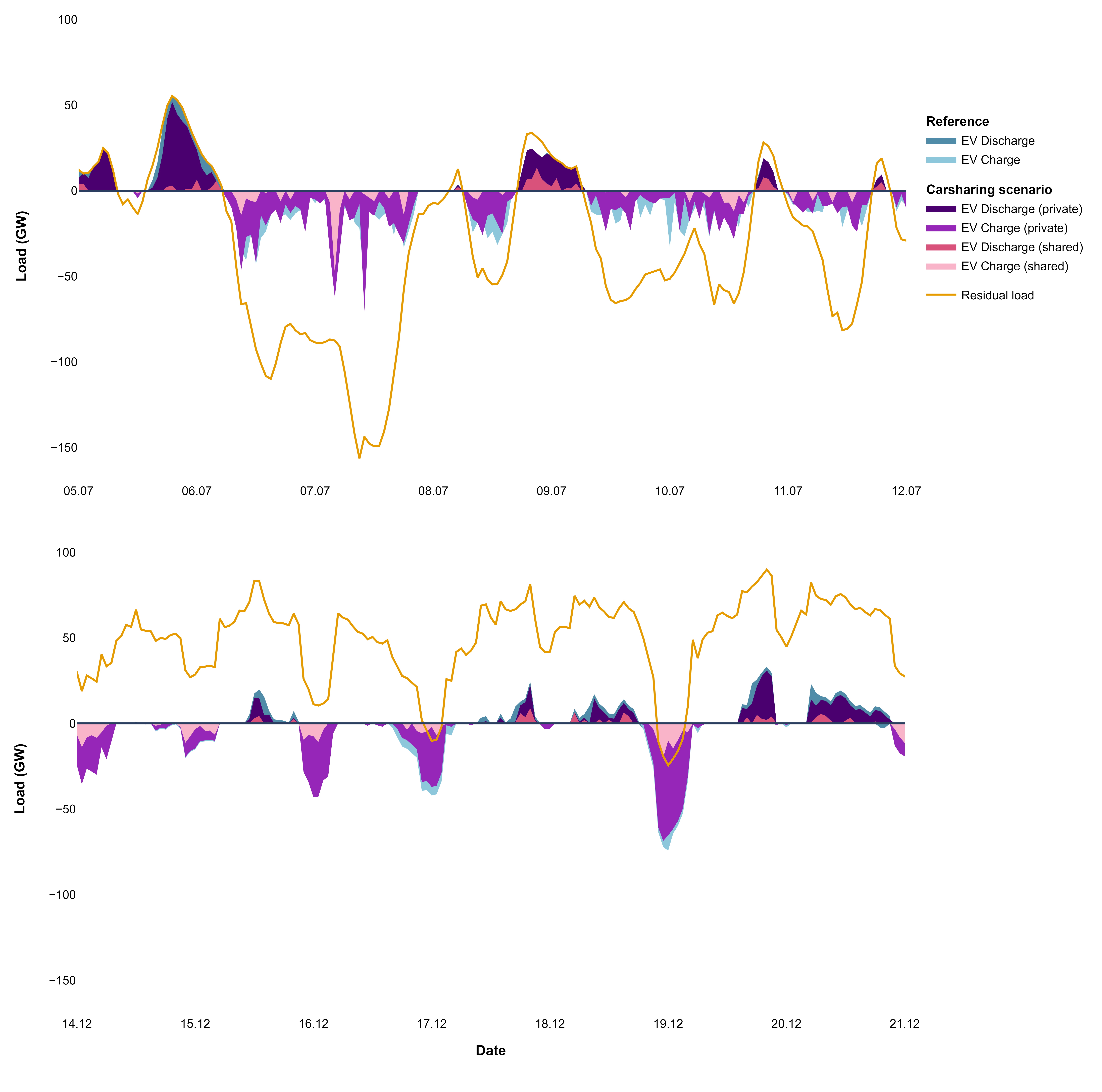}
    \caption{Residual load curve, BEV charging and discharging for a subset of summer hours (upper panel) and winter hours (lower panel) [``Shared + other BEVs'', bidirectional charging, low uptake]}
    \label{fig:EV_dispatch_resload_low}
\end{figure}

\clearpage

\begin{figure}[!ht]
    \centering
    \includegraphics[width=15.5cm]{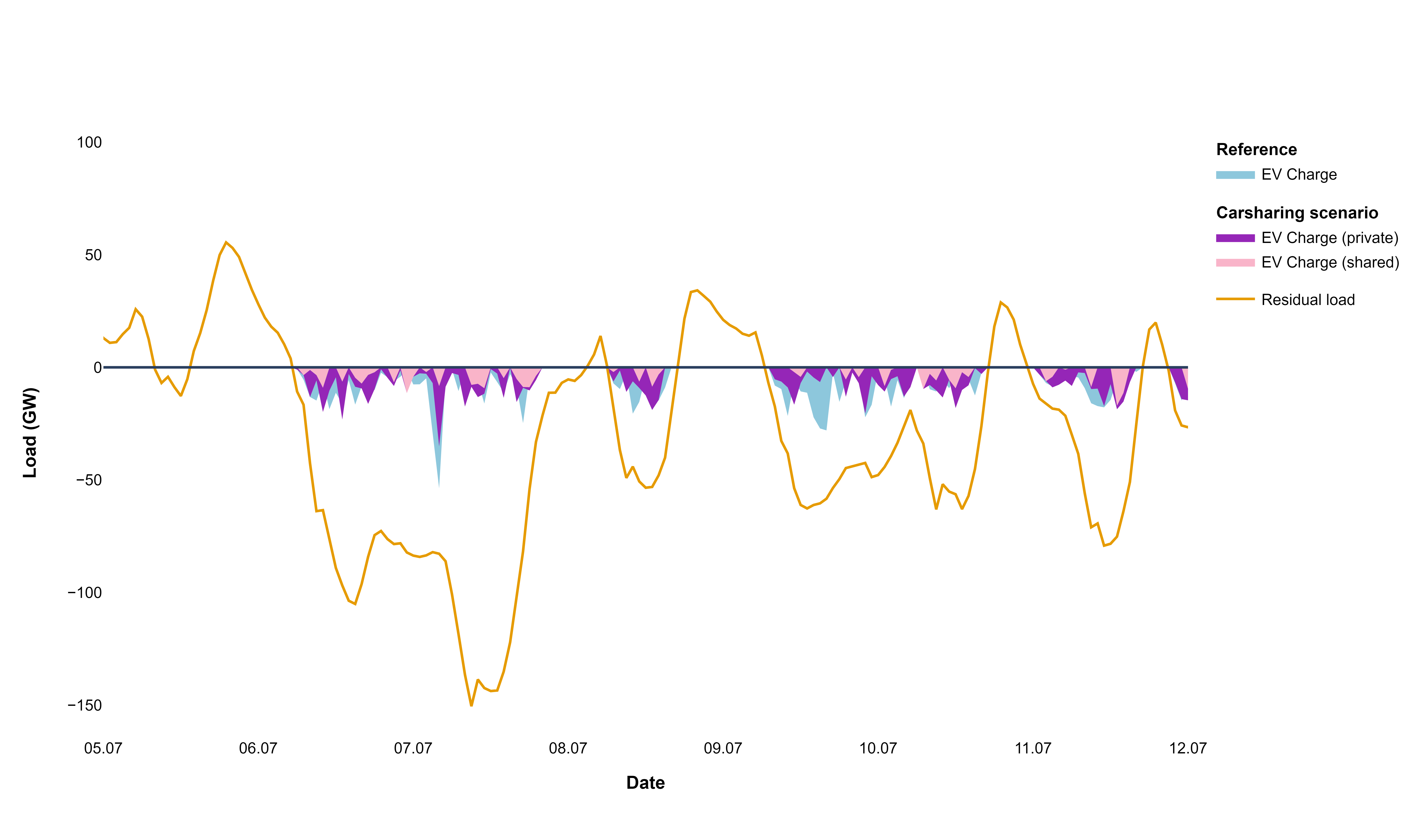}
    \caption{Residual load curve, BEV charging for a subset of summer hours [``Shared + other BEVs'', smart charging, high uptake]}
    \label{fig:EV_dispatch_resload_smart_high}
\end{figure}

\begin{figure}[!ht]
    \centering
    \includegraphics[width=15.5cm]{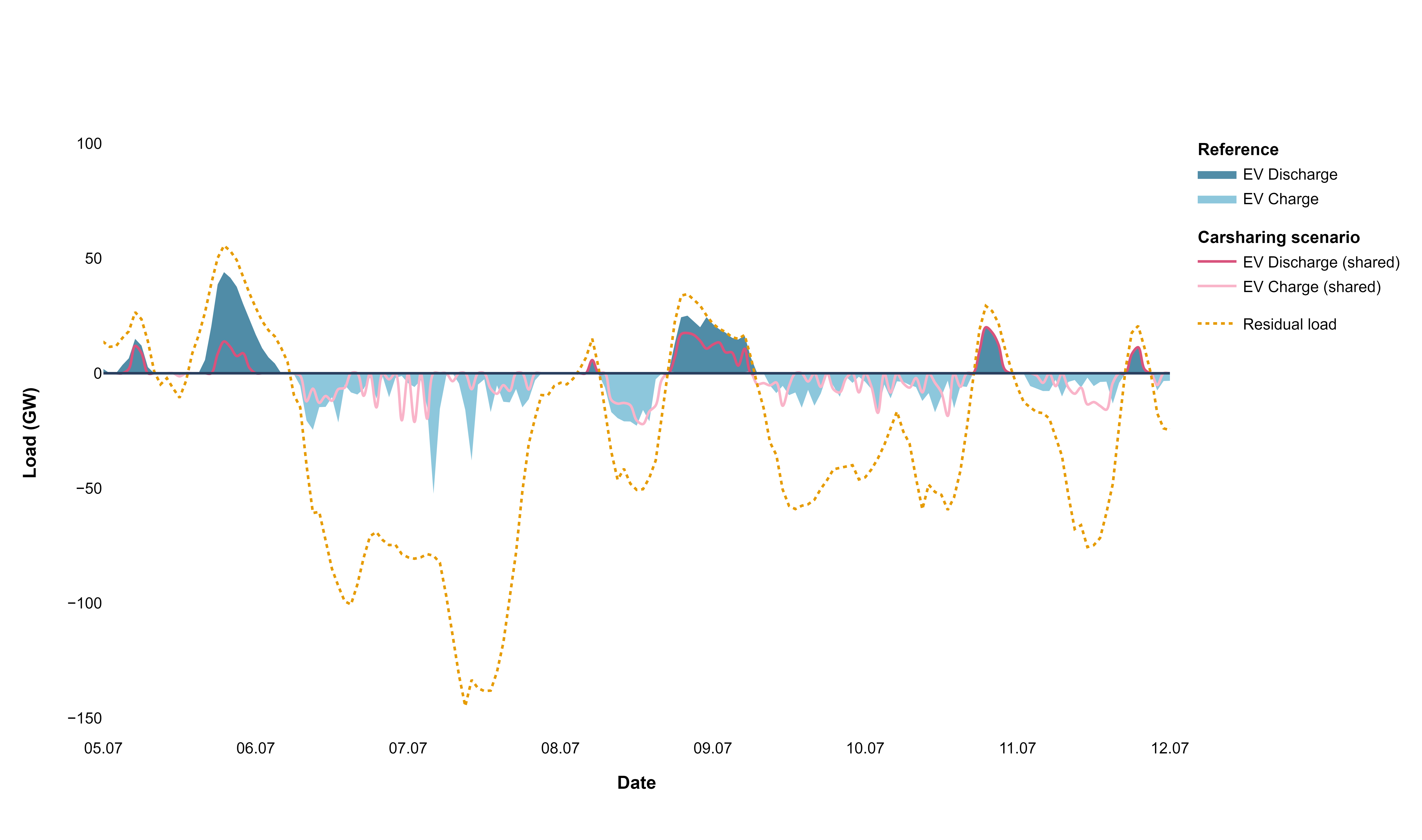}
    \caption{Residual load curve, BEV charging and discharging for a subset of summer hours [``Shared only'', bidirectional charging, high uptake]}
    \label{fig:EV_dispatch_resload_sharedonly_high}
\end{figure}

\clearpage
\newpage

\subsection{Storage capacity and dispatch}
\label{sec:sto}

\begin{figure}[!ht]
    \centering
    \includegraphics[width=15.5cm]{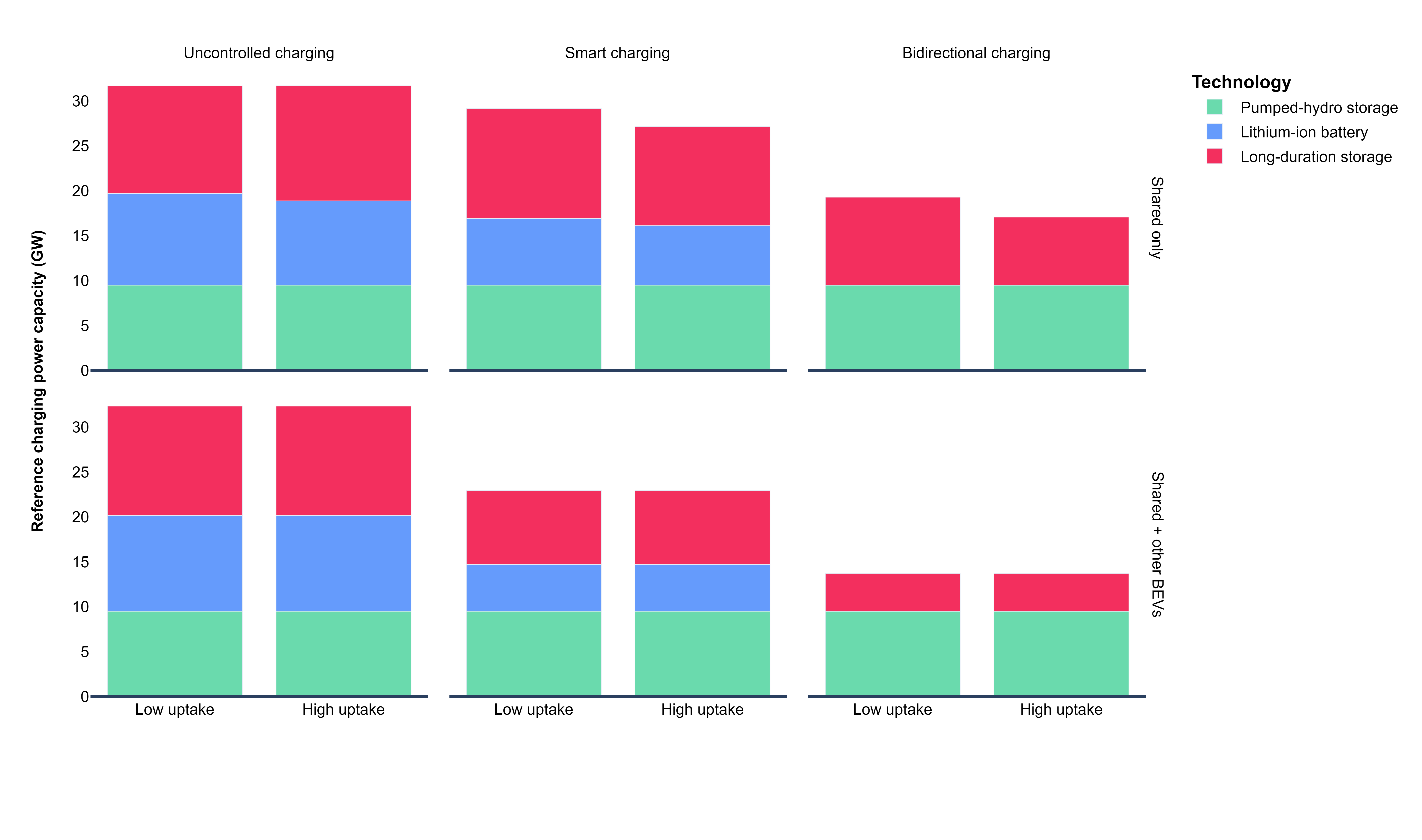}
    \caption{Storage charging power capacity in reference cases}
    \label{fig:sto_charging_ref}
\end{figure}

\begin{figure}[!ht]
    \centering
    \includegraphics[width=15.5cm]{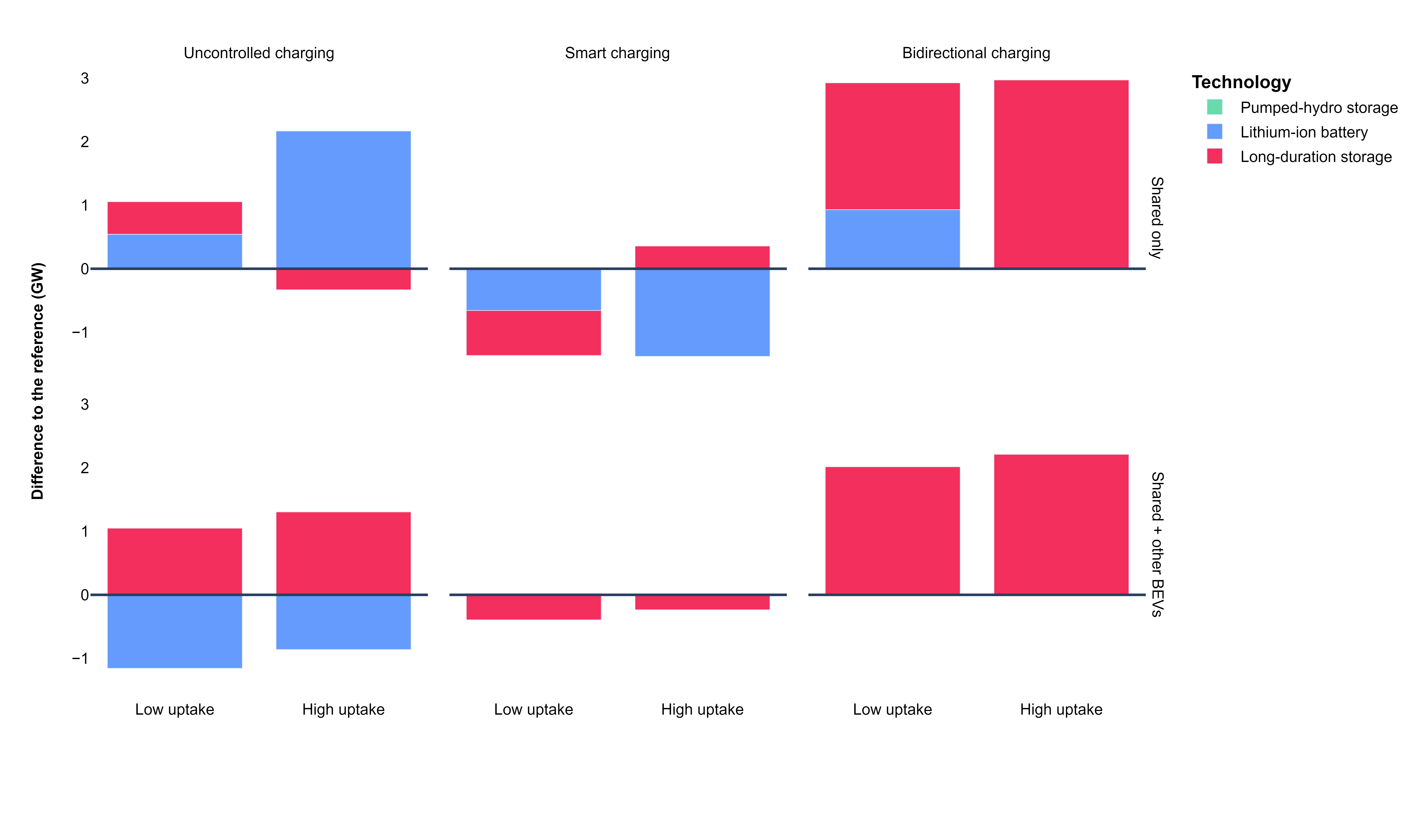}
    \caption{Effects of carsharing on optimal storage charging power capacity}
    \label{fig:sto_charging_scen}
\end{figure}

\clearpage

\begin{figure}[!ht]
    \centering
    \includegraphics[width=16cm]{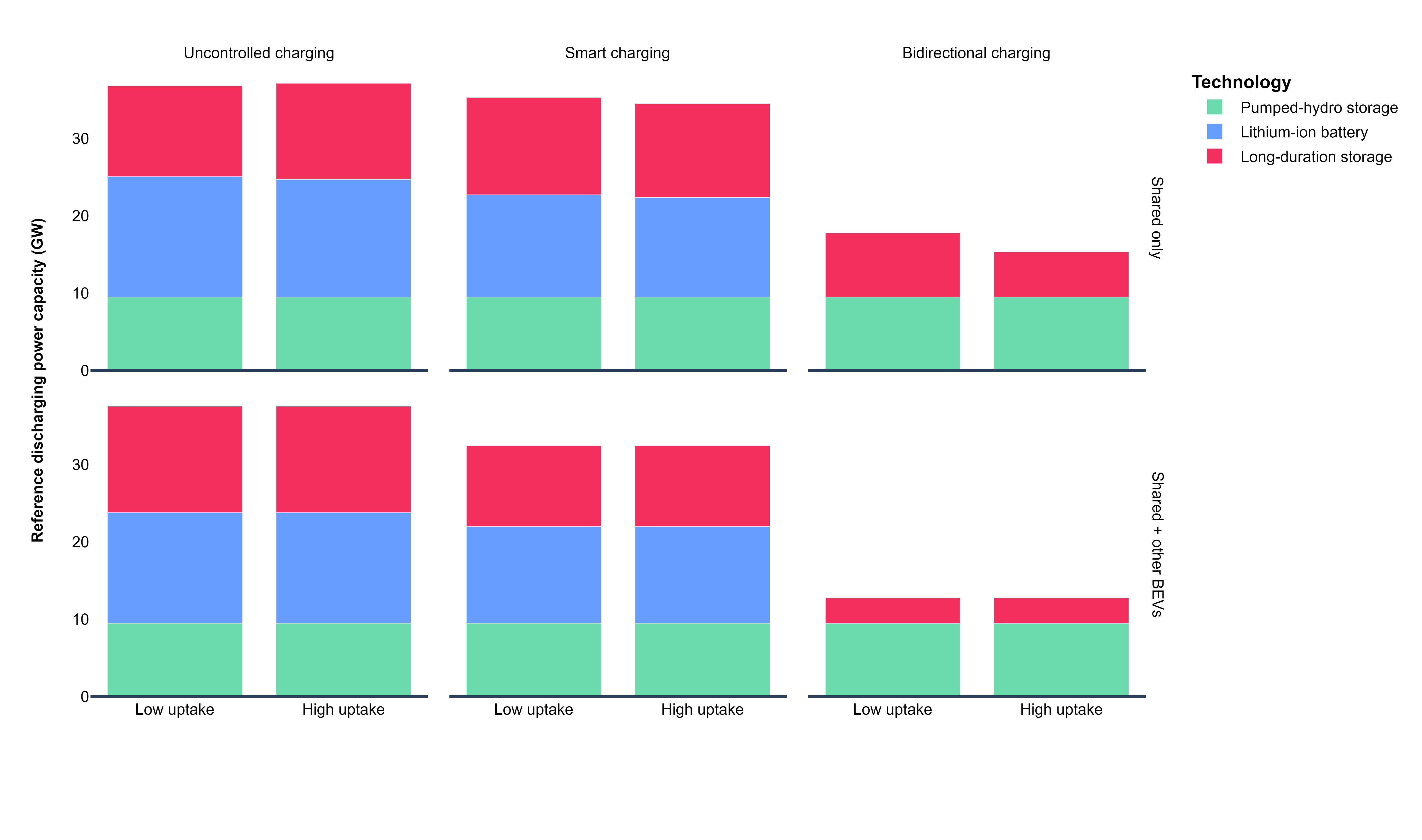}
    \caption{Storage discharging power capacity in reference cases}
    \label{fig:sto_discharging_ref}
\end{figure}

\begin{figure}[!ht]
    \centering
    \includegraphics[width=16cm]{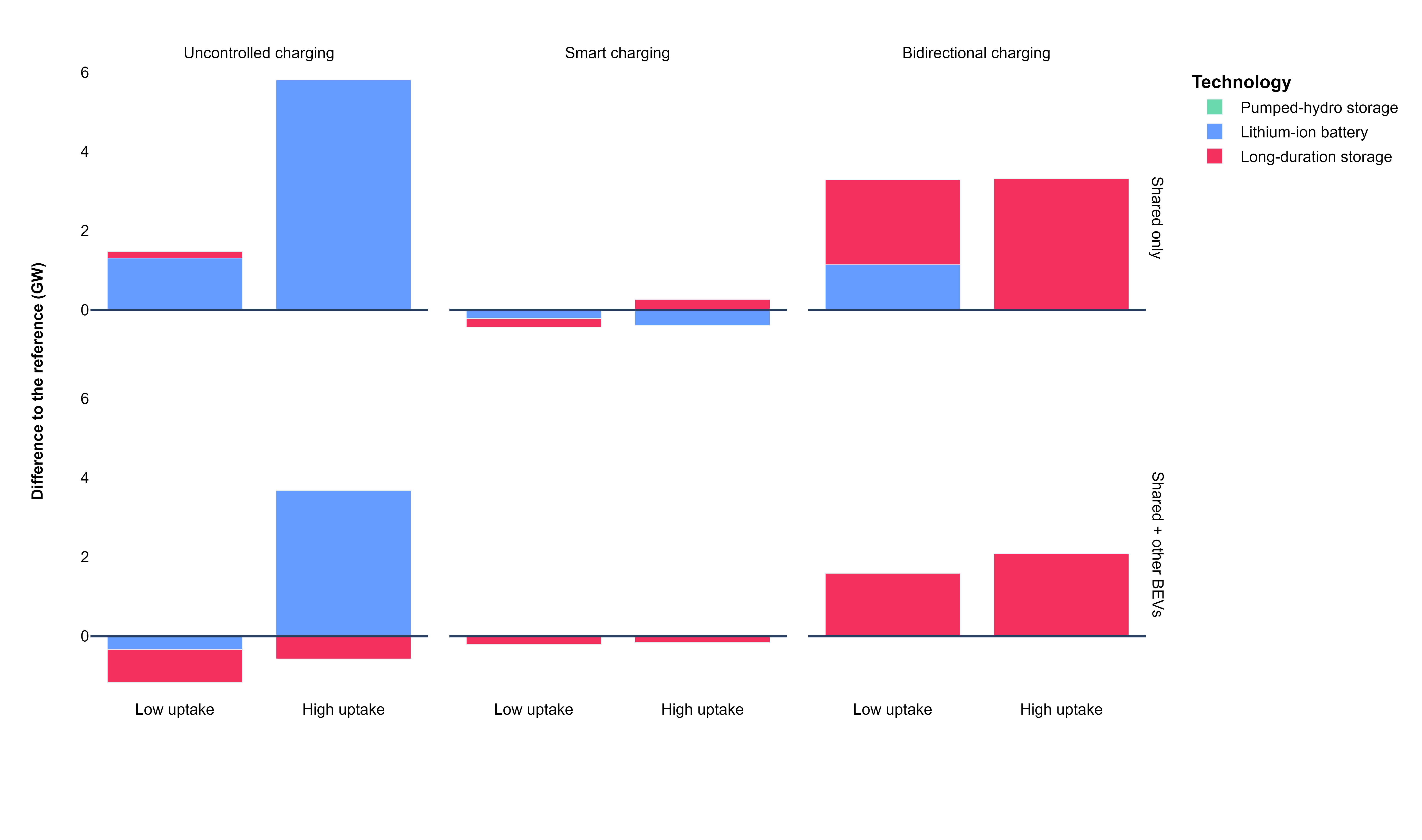}
    \caption{Effects of carsharing on optimal storage discharging power capacity}
    \label{fig:sto_discharging_scen}
\end{figure}

\clearpage

\begin{figure}[!ht]
    \centering
    \includegraphics[width=16cm]{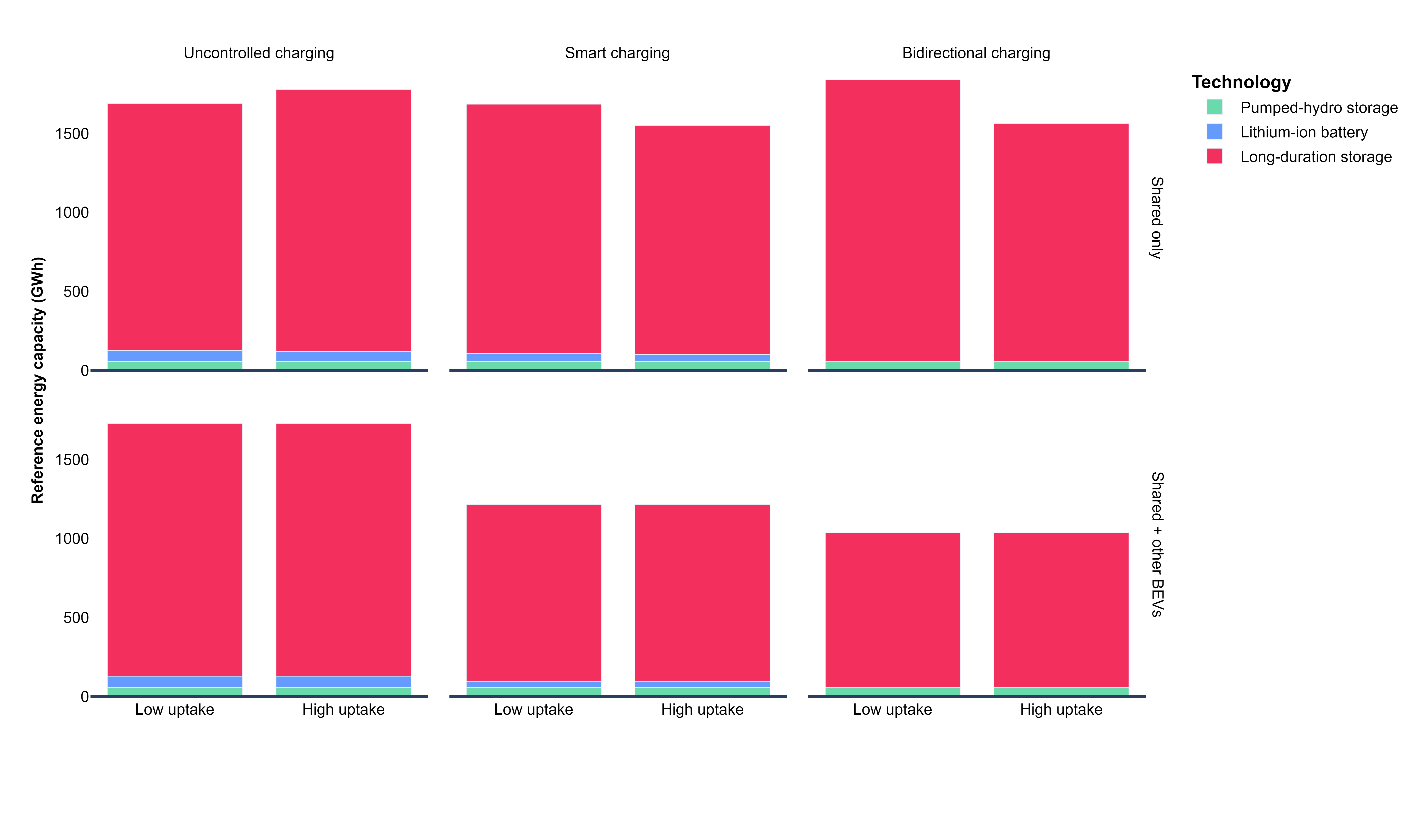}
    \caption{Storage energy capacity in reference cases}
    \label{fig:sto_energy_ref}
\end{figure}

\begin{figure}[!ht]
    \centering
    \includegraphics[width=16cm]{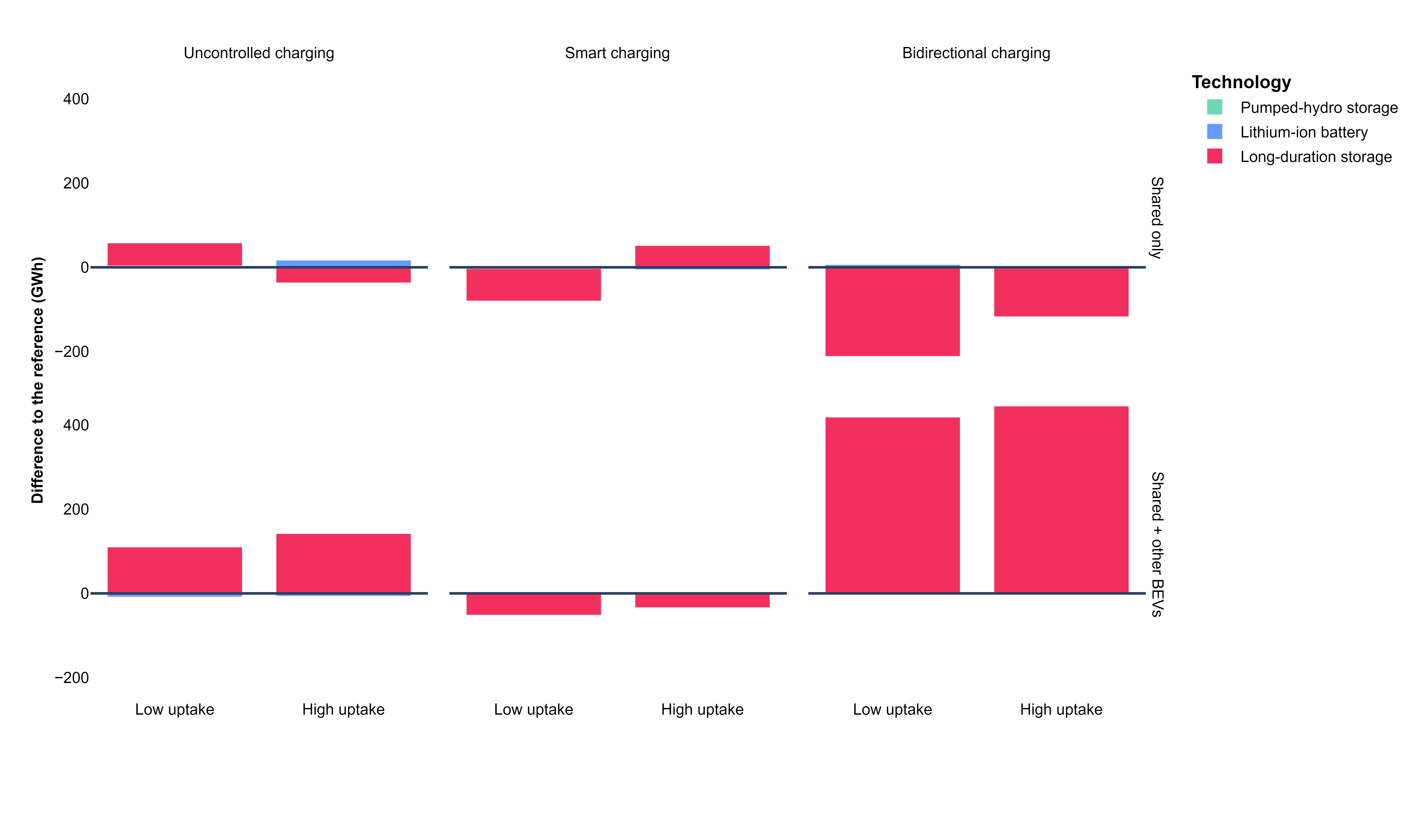}
    \caption{Effects of carsharing on optimal storage energy capacity}
    \label{fig:sto_energy_scen}
\end{figure}

\clearpage

\begin{figure}[!ht]
    \centering
    \includegraphics[width=15.5cm]{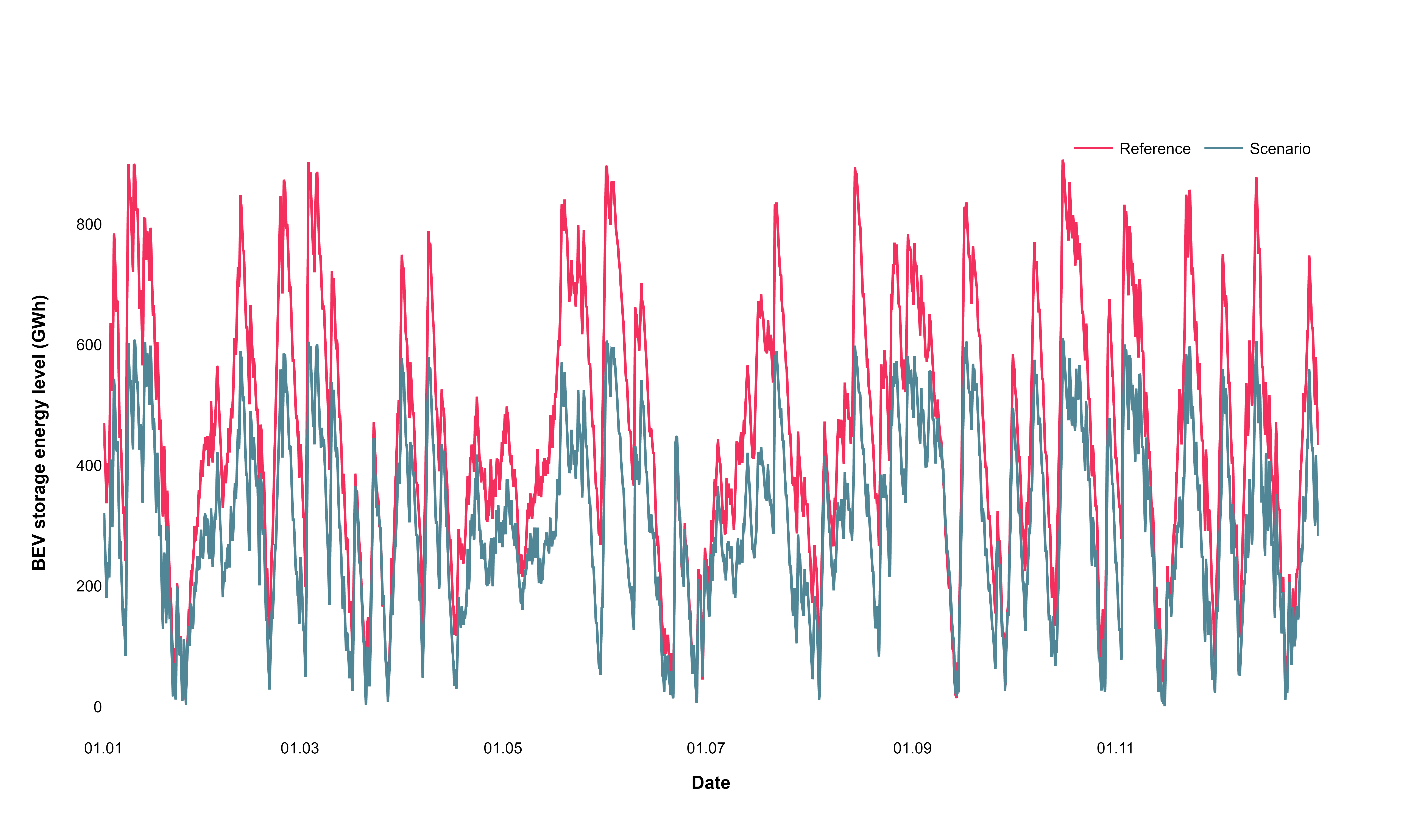}
    \caption{Battery electric vehicles' overall storage level in the reference and the corresponding carsharing scenario [``Shared + other BEVs'', smart charging, high uptake]}
    \label{fig:ev_storagelevel_year}
\end{figure}

\begin{figure}[!ht]
    \centering
    \includegraphics[width=15.5cm]{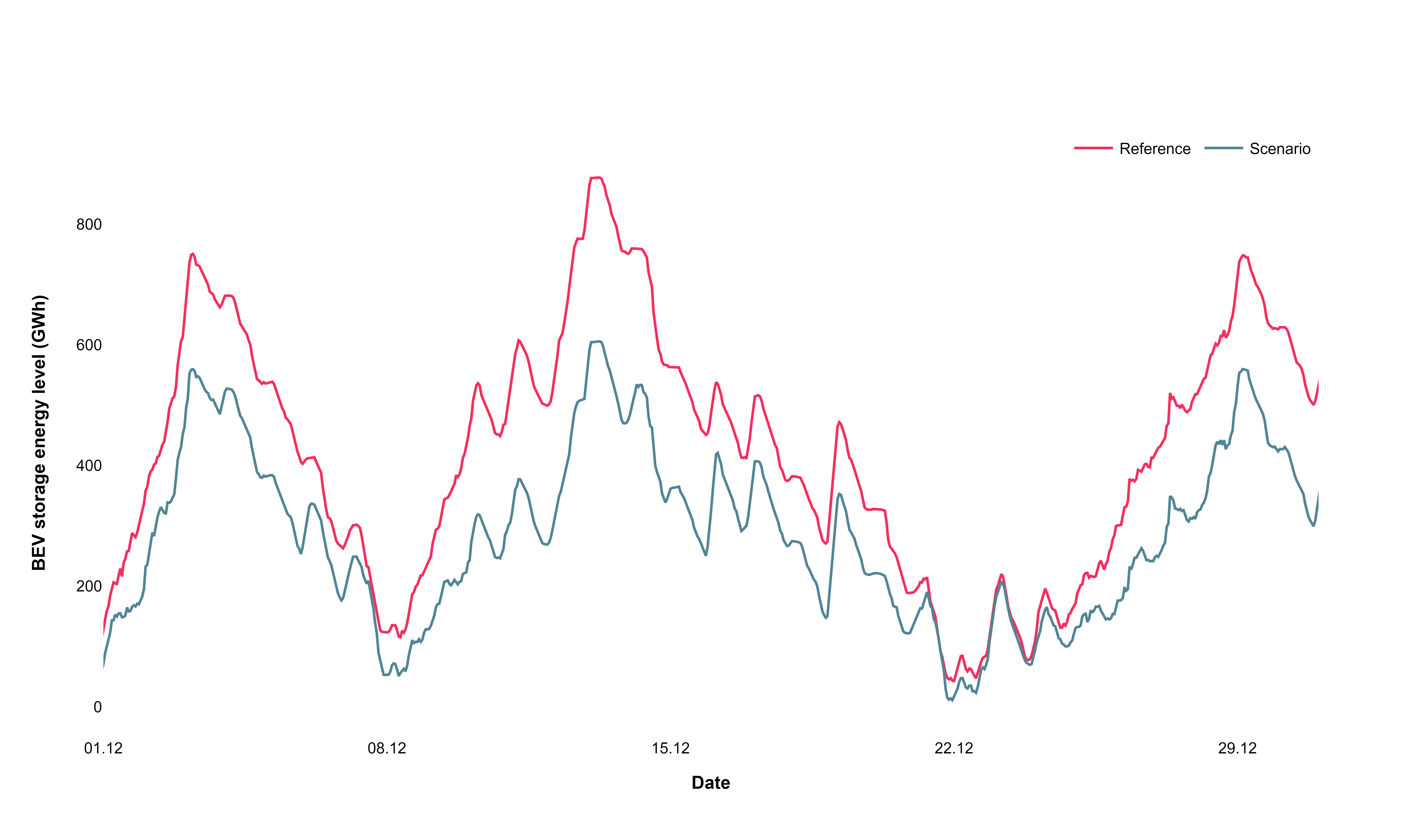}
    \caption{Battery electric vehicles' overall storage level in the reference and the corresponding carsharing scenario, zoomed in for a subset of winter hours [``Shared + other BEVs'', smart charging, high uptake]}
    \label{fig:ev_storagelevel}
\end{figure}

\clearpage
\newpage

\subsection{Sensitivity analysis: additional green hydrogen production}
\label{sec:hydrogen}

\begin{figure}[!ht]
    \centering
    \includegraphics[width=15cm]{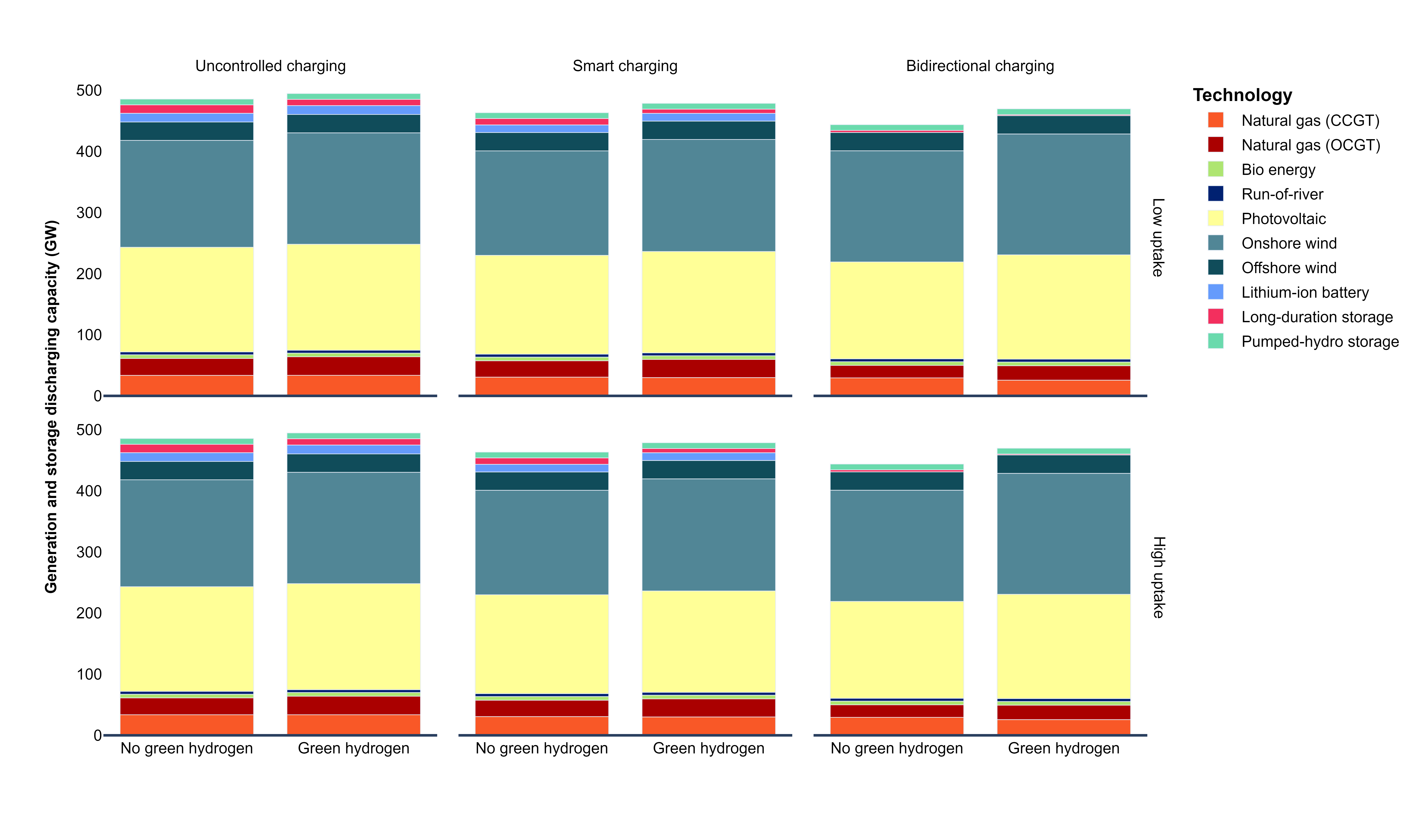}
    \caption{Power generation capacity mixes in reference settings comparing without and with industrial demand for green hydrogen [``Shared + other BEVs'']}
    \label{fig:SensitivityH2_gencap_ref}
\end{figure}

\begin{figure}[!ht]
    \centering
    \includegraphics[width=15cm]{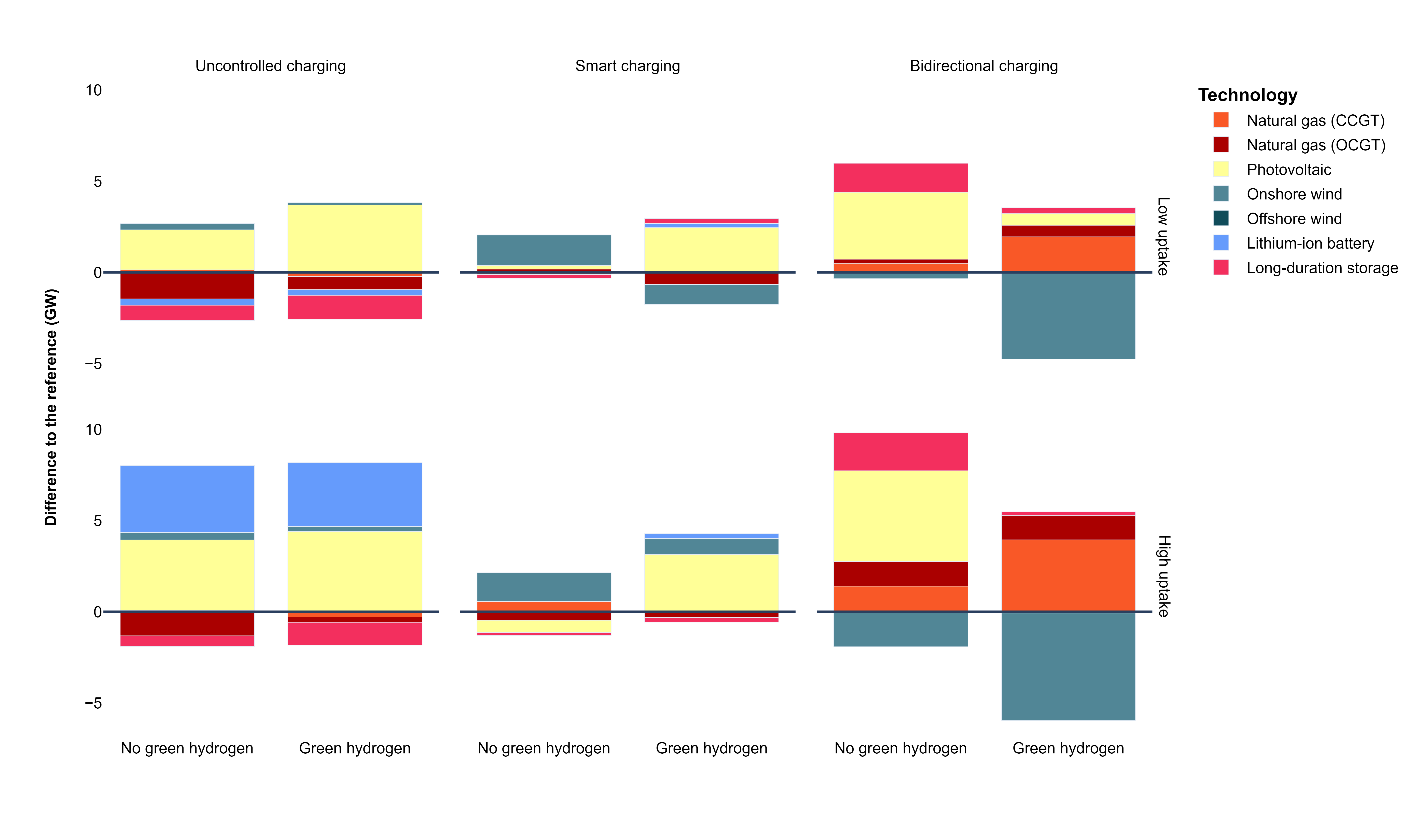}
    \caption{Power generation capacity mixes in carsharing scenarios comparing without and with industrial demand for green hydrogen [``Shared + other BEVs'']}
    \label{fig:SensitivityH2_gencap_scen}
\end{figure}

\clearpage
\newpage

\subsection{Sensitivity analysis: weather years}
\label{sec:sensitivity_weather}

\begin{figure}[!ht]
    \centering
    \includegraphics[width=10cm]{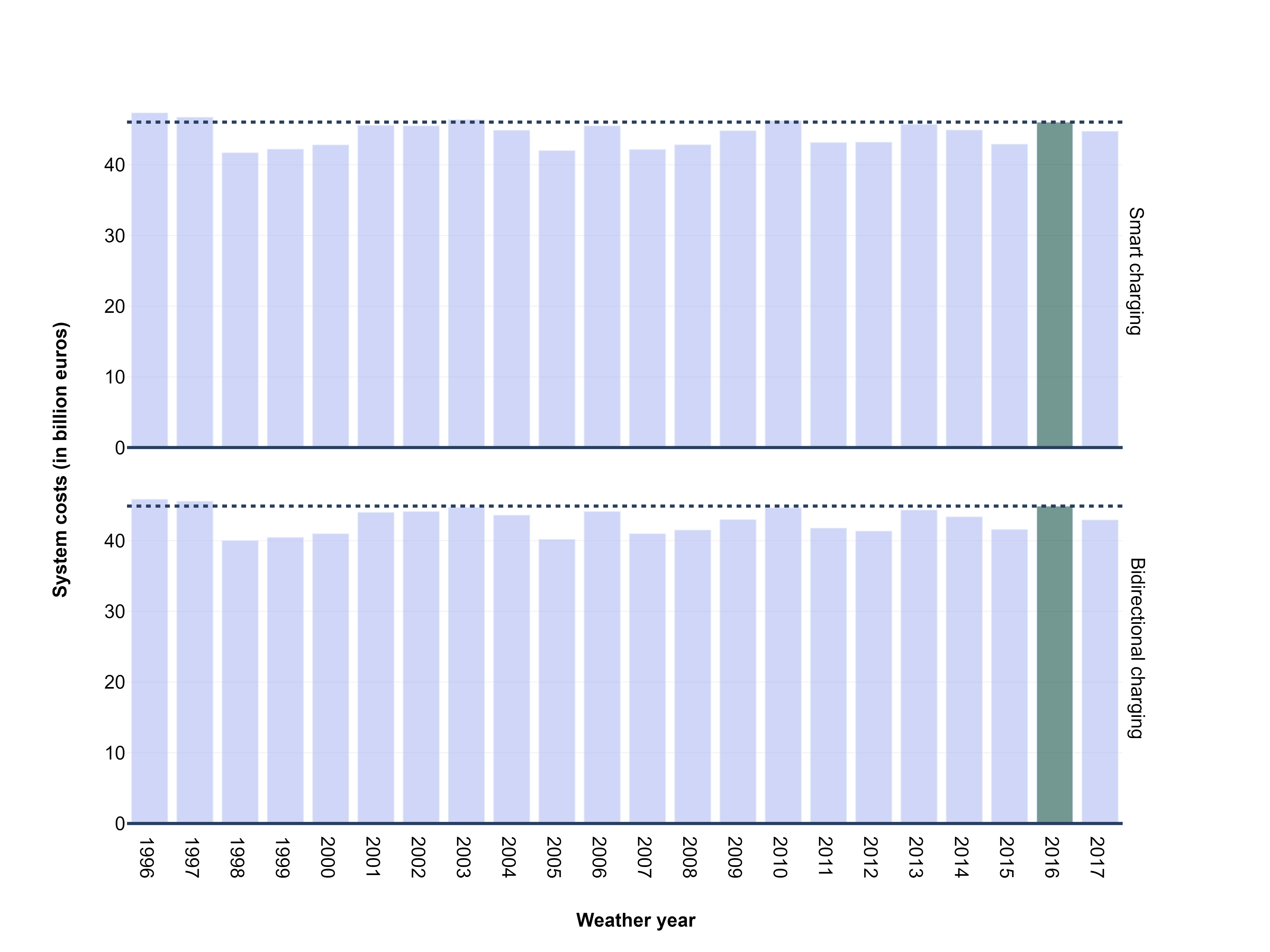}
    \caption{System costs in reference settings for different weather years [``Shared + other BEVs'', high uptake]}
    \label{fig:SensitivityWeather_systemcosts_ref}
\end{figure}

\begin{figure}[!ht]
    \centering
    \includegraphics[width=10cm]{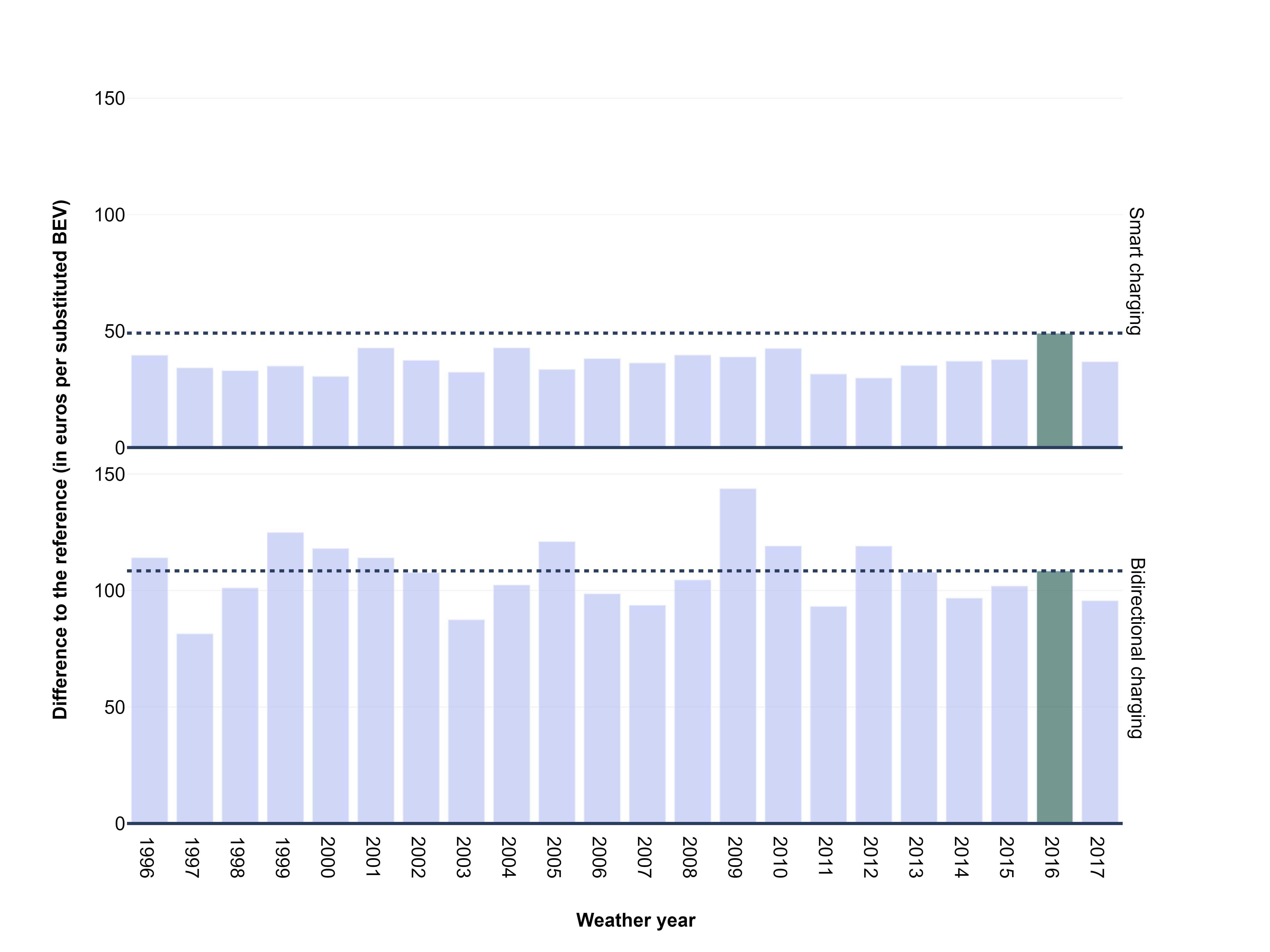}
    \caption{Costs per substituted car in scenario settings for different weather years [``Shared + other BEVs'', high uptake]}
    \label{fig:SensitivityWeather_systemcosts_scen}
\end{figure}

\begin{figure}[!ht]
    \centering
    \includegraphics[width=15cm]{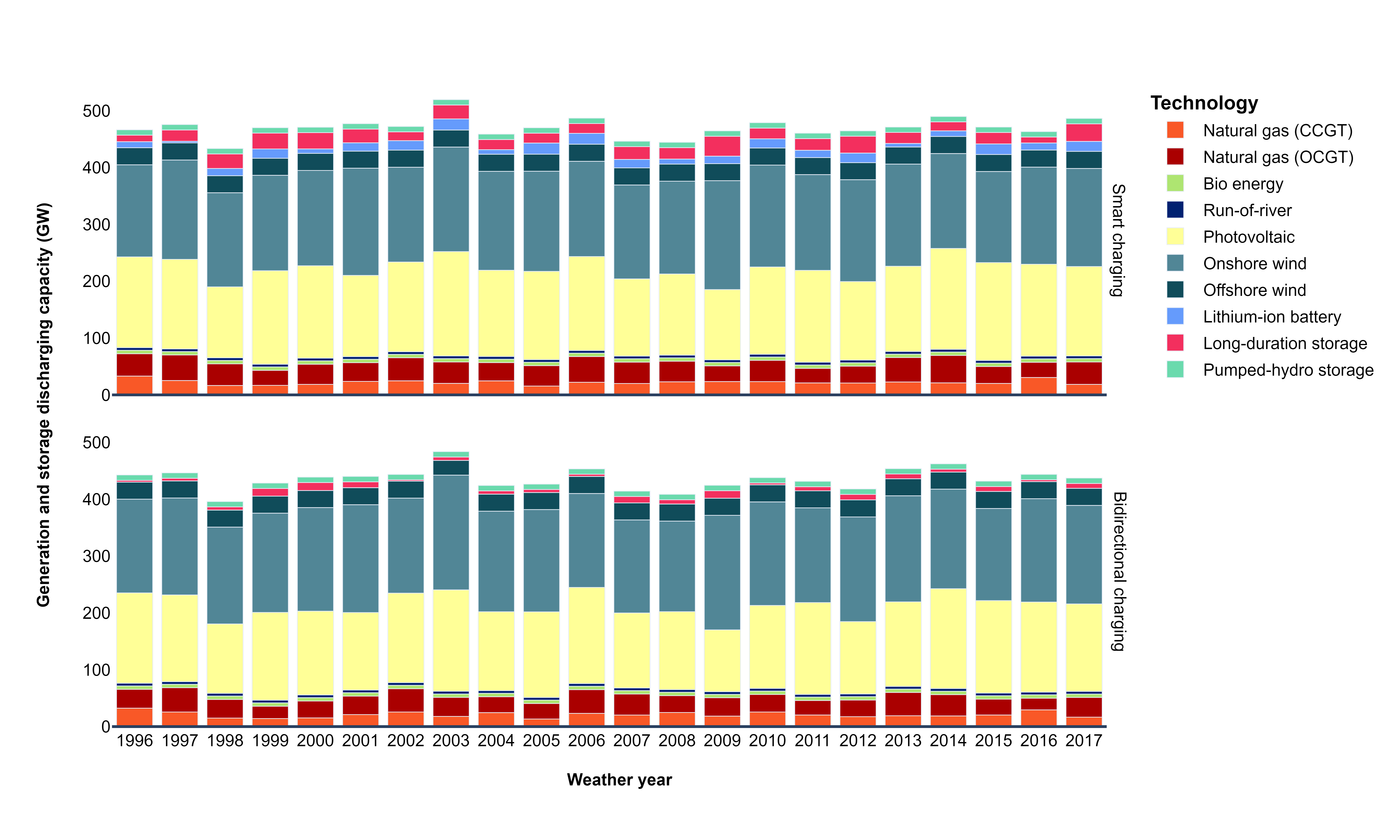}
    \caption{Power generation capacity mixes in reference settings for different weather years [``Shared + other BEVs'', high uptake]}
    \label{fig:SensitivityWeather_capacity_ref}
\end{figure}

\begin{figure}[!ht]
    \centering
    \includegraphics[width=15cm]{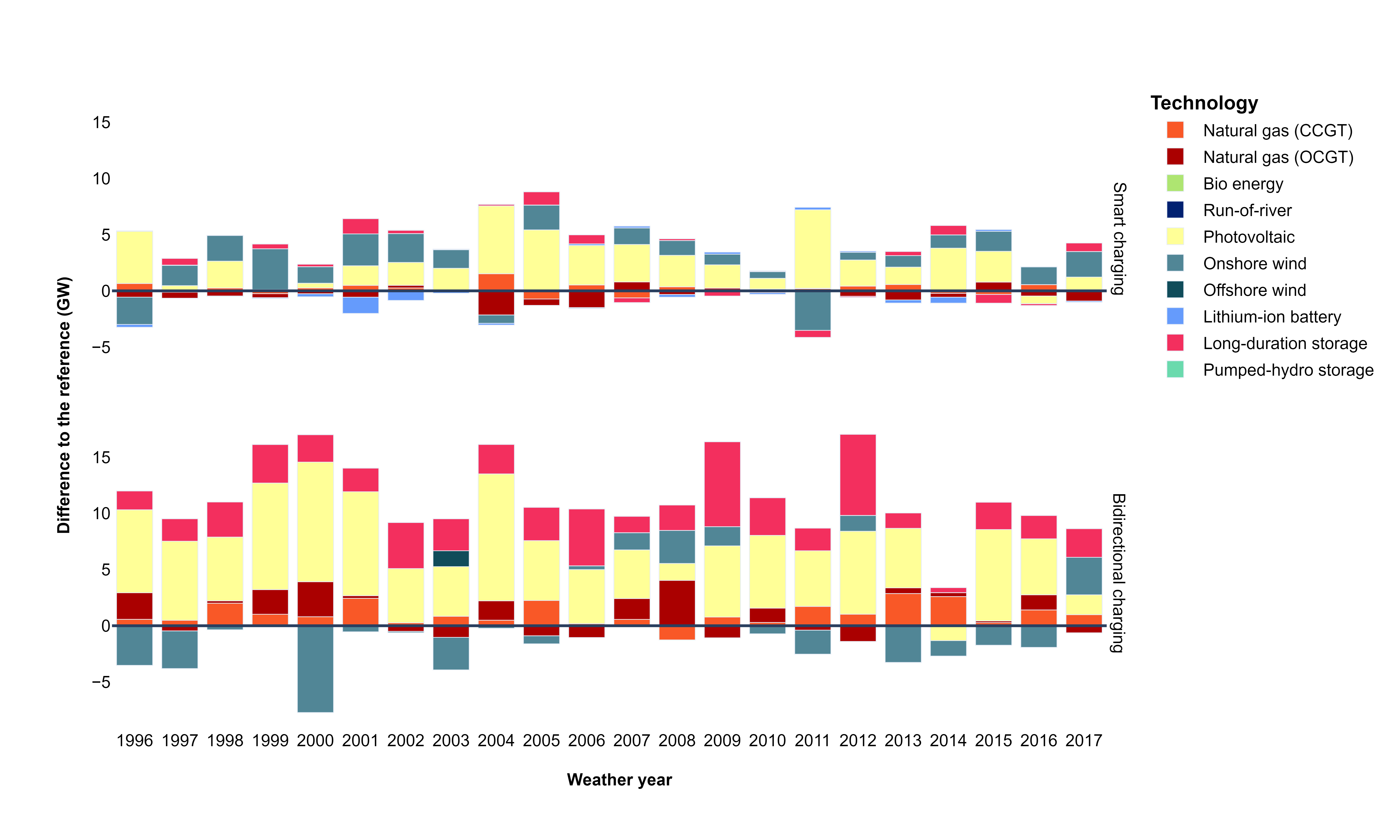}
    \caption{Power generation capacity mixes in scenario settings for different weather years [``Shared + other BEVs'', high uptake]}
    \label{fig:SensitivityWeather_capacity_scen}
\end{figure}

\begin{figure}[!ht]
    \centering
    \includegraphics[width=15cm]{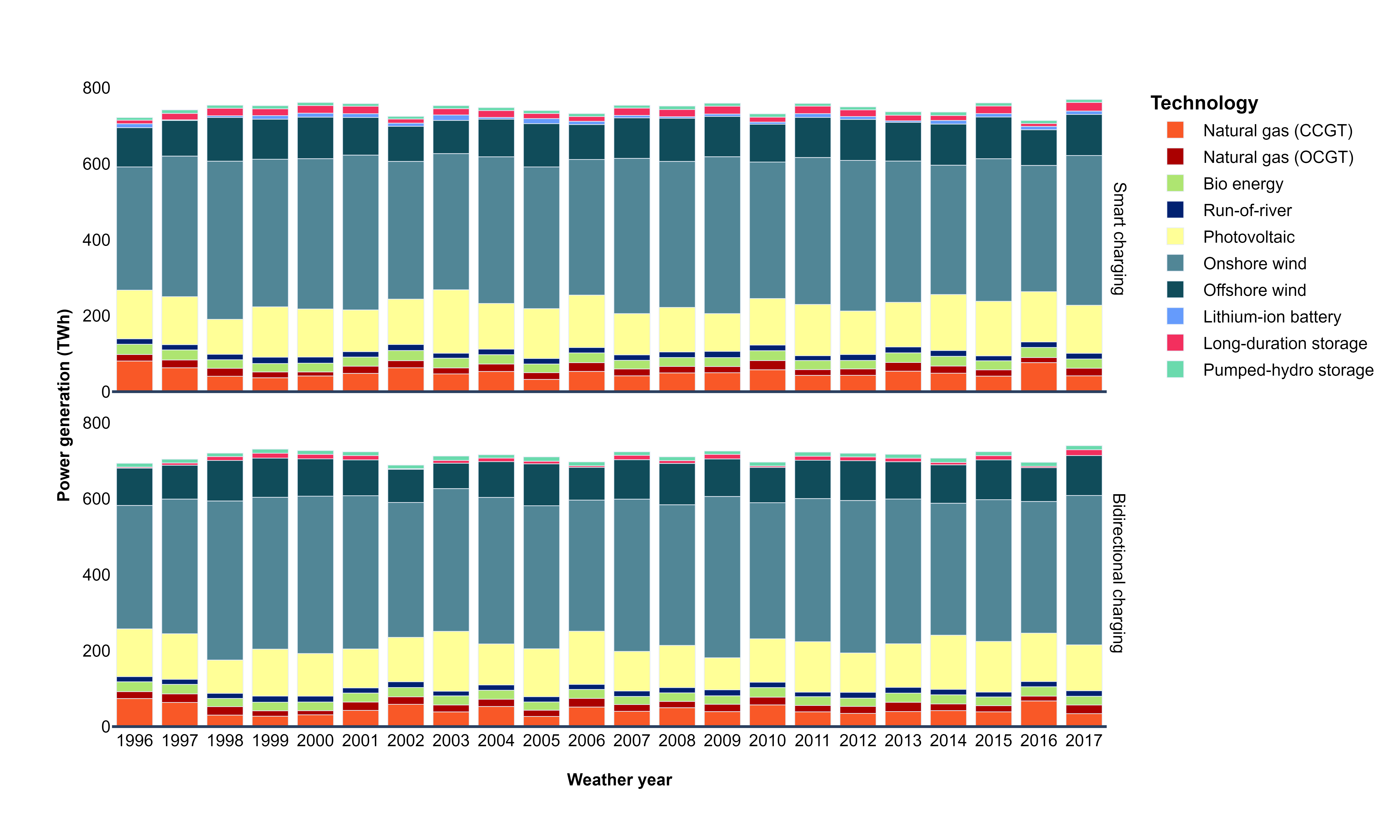}
    \caption{Yearly electricity generation in reference settings for different weather years [``Shared + other BEVs'', high uptake]}
    \label{fig:SensitivityWeather_generation_ref}
\end{figure}

\begin{figure}[!ht]
    \centering
    \includegraphics[width=15cm]{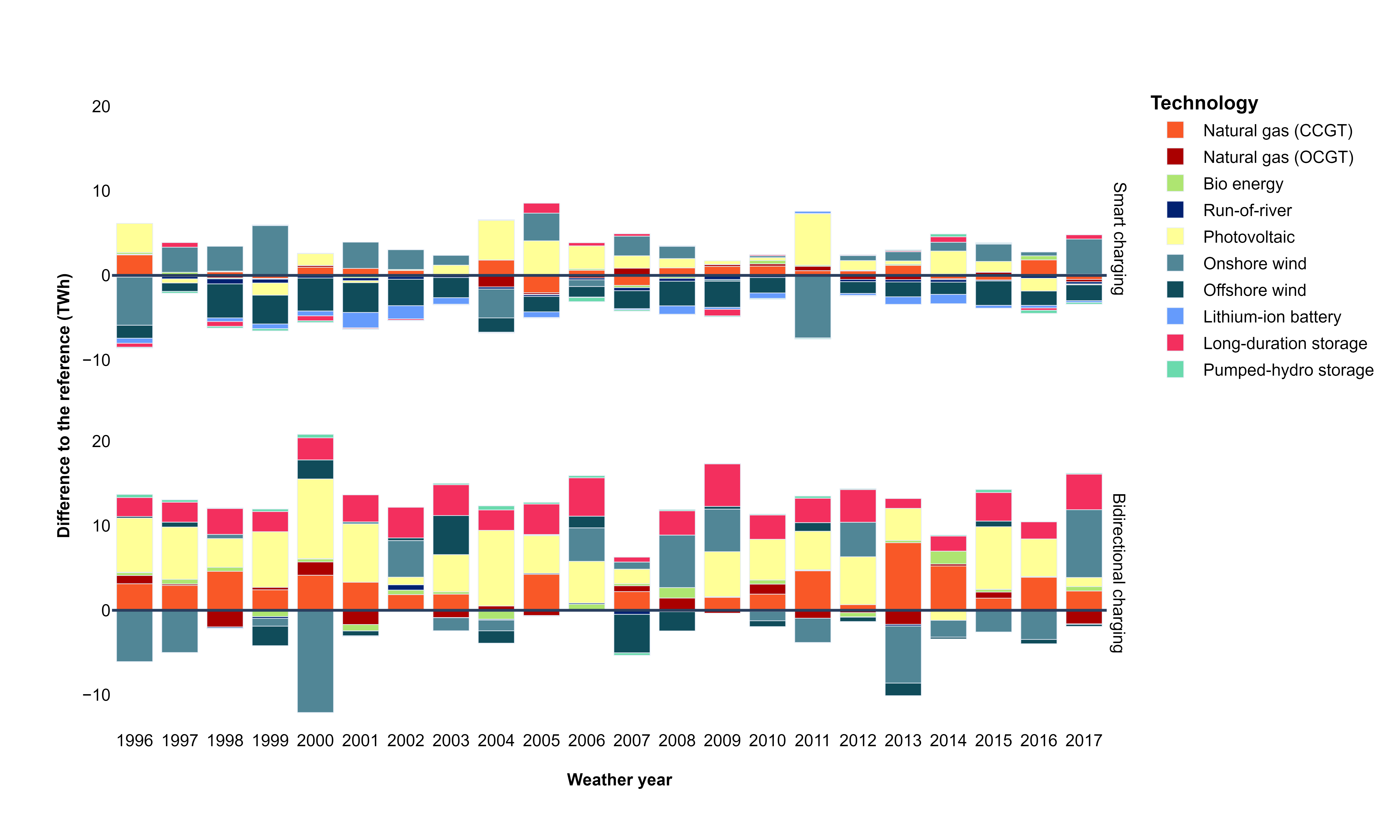}
    \caption{Yearly electricity generation in scenario settings for different weather years [``Shared + other BEVs'', high uptake]}
    \label{fig:SensitivityWeather_generation_scen}
\end{figure}

\begin{figure}[!ht]
    \centering
    \includegraphics[width=12cm]{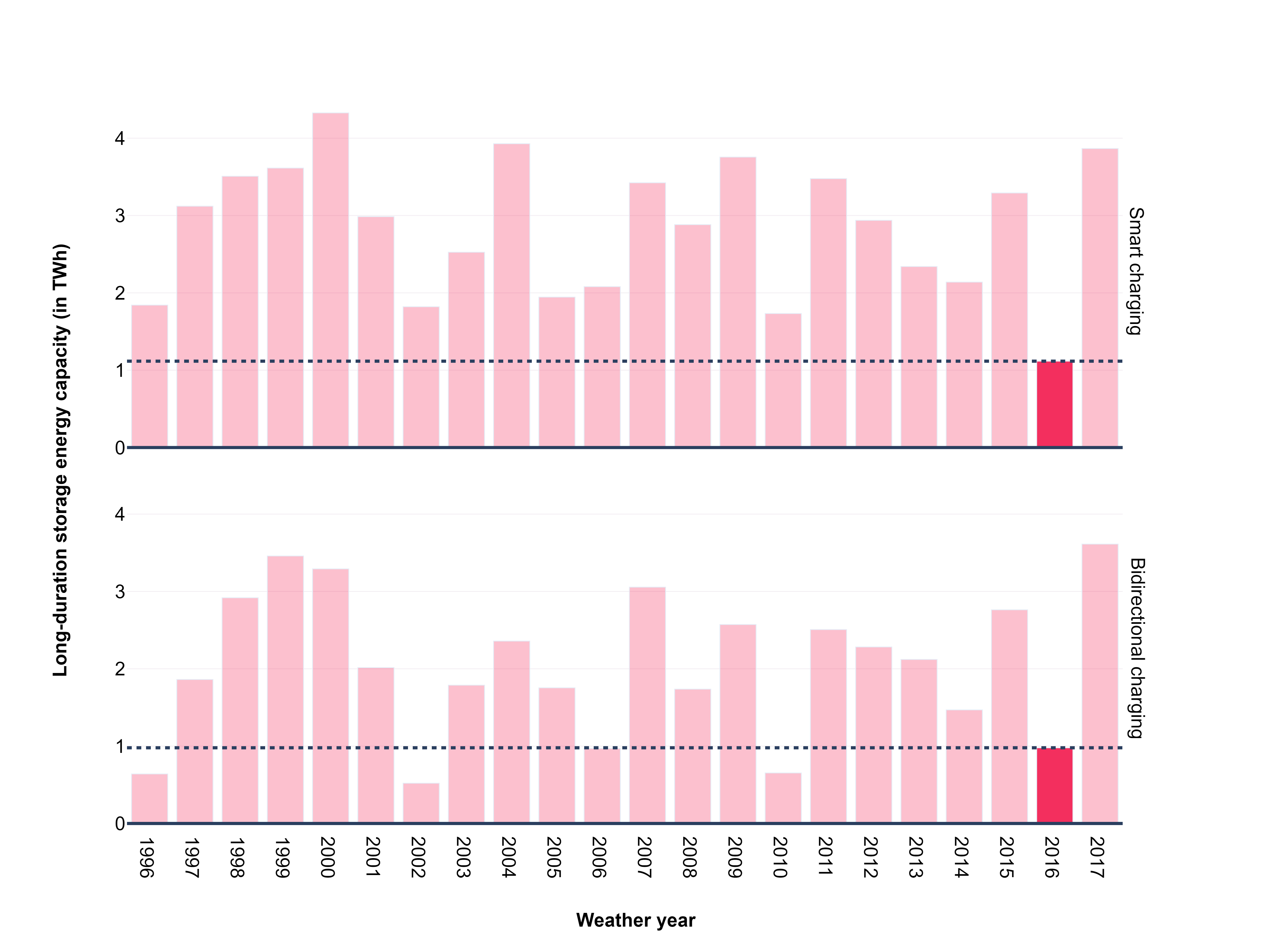}
    \caption{Long-duration storage energy capacity in reference settings for different weather years [``Shared + other BEVs'', high uptake]}
\label{fig:SensitivityWeather_storage_ref}
\end{figure}

\begin{figure}[!ht]
    \centering
    \includegraphics[width=12cm]{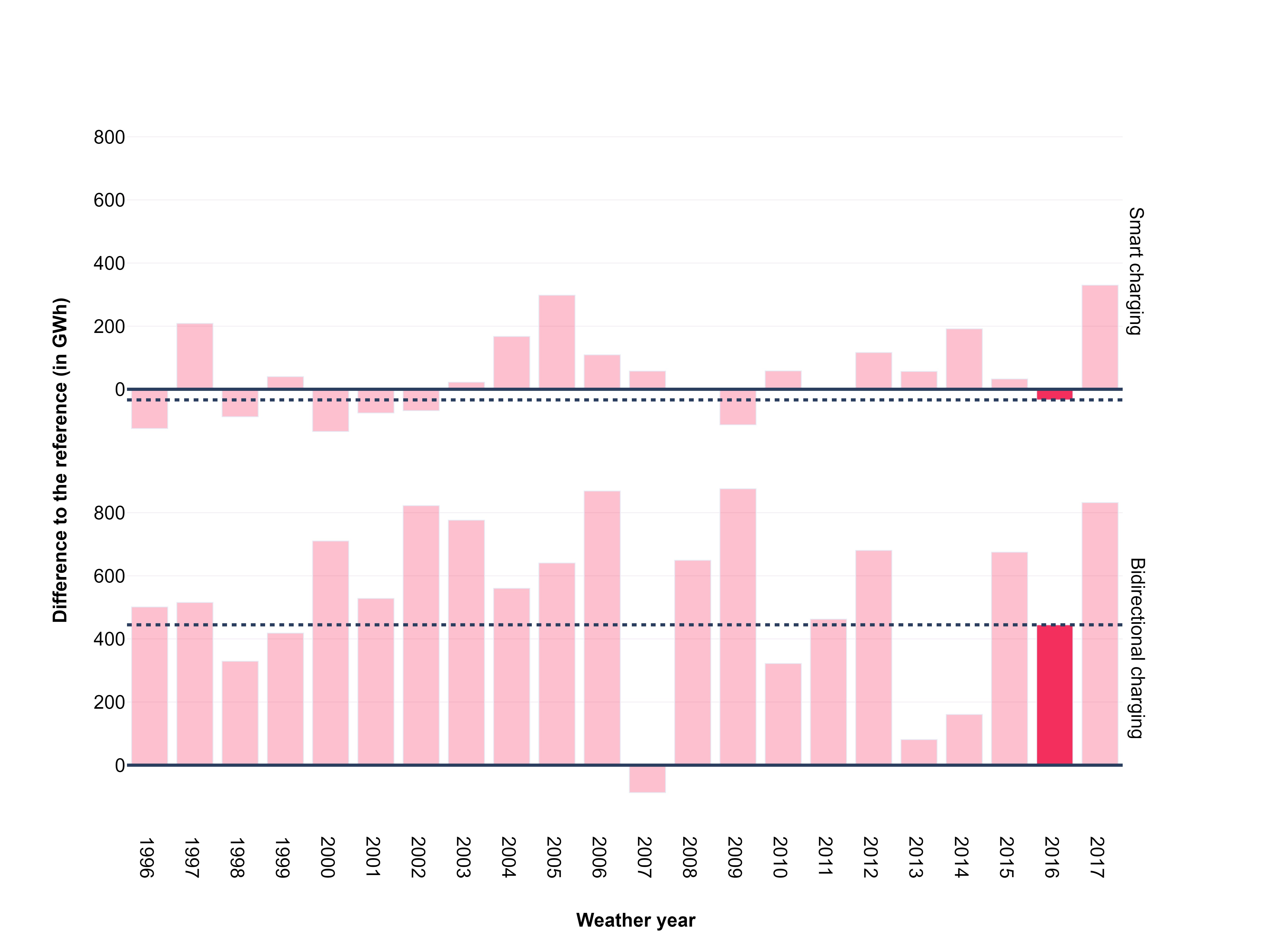}
    \caption{Long-duration storage energy capacity in scenario settings for different weather years [``Shared + other BEVs'', high uptake]}
\label{fig:SensitivityWeather_storage_scen}
\end{figure}

\clearpage
\newpage

\subsection{Sensitivity analysis: technology costs}
\label{sec:sensitivity_costs}

\begin{figure}[!ht]
    \centering
    \includegraphics[width=10cm]{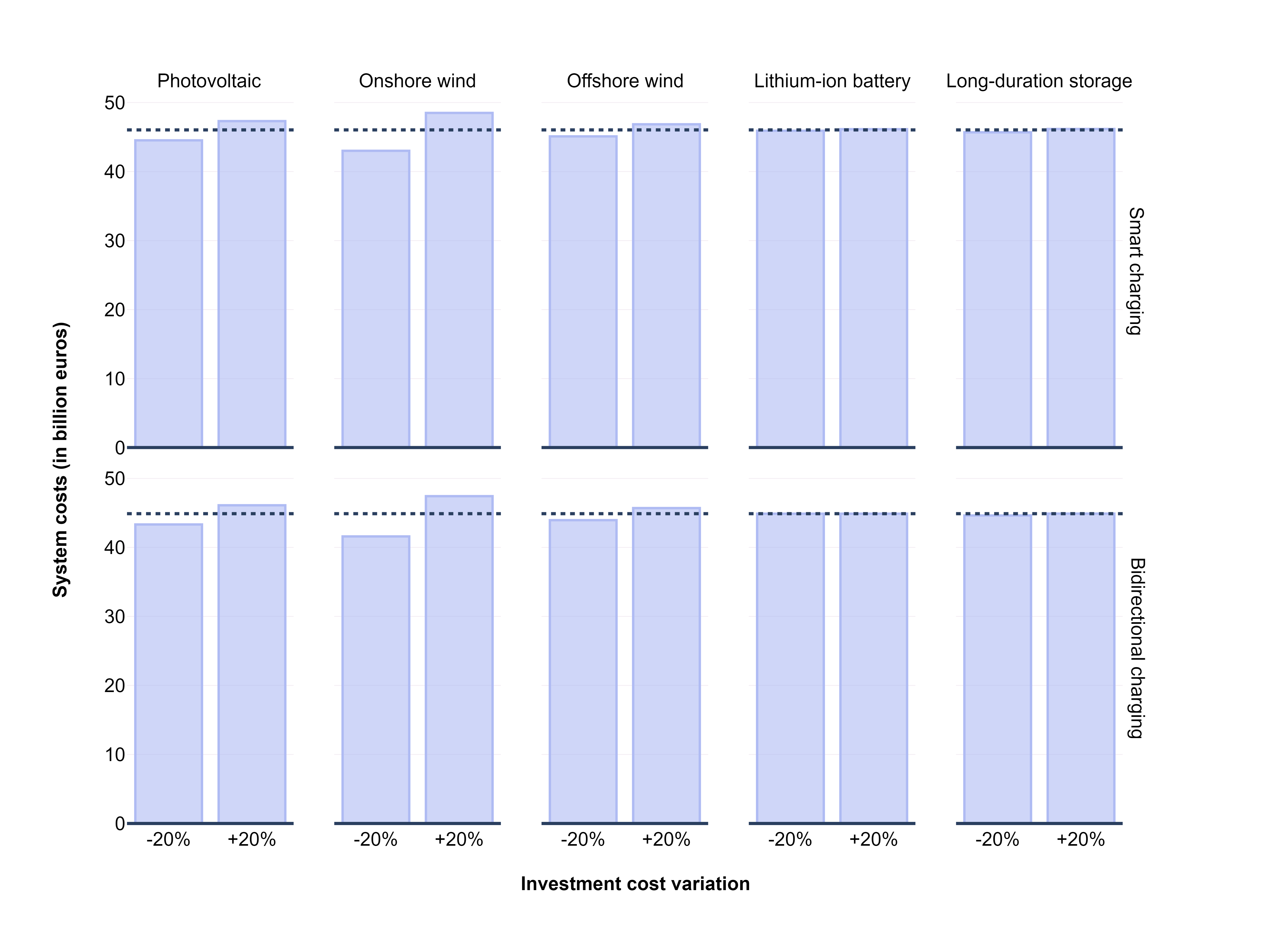}
    \caption{System costs in reference settings for different technology cost assumptions [``Shared + other BEVs'', bidirectional charging, high uptake]}
    \label{fig:SensitivityCosts_systemcosts_ref}
\end{figure}

\begin{figure}[!ht]
    \centering
    \includegraphics[width=10cm]{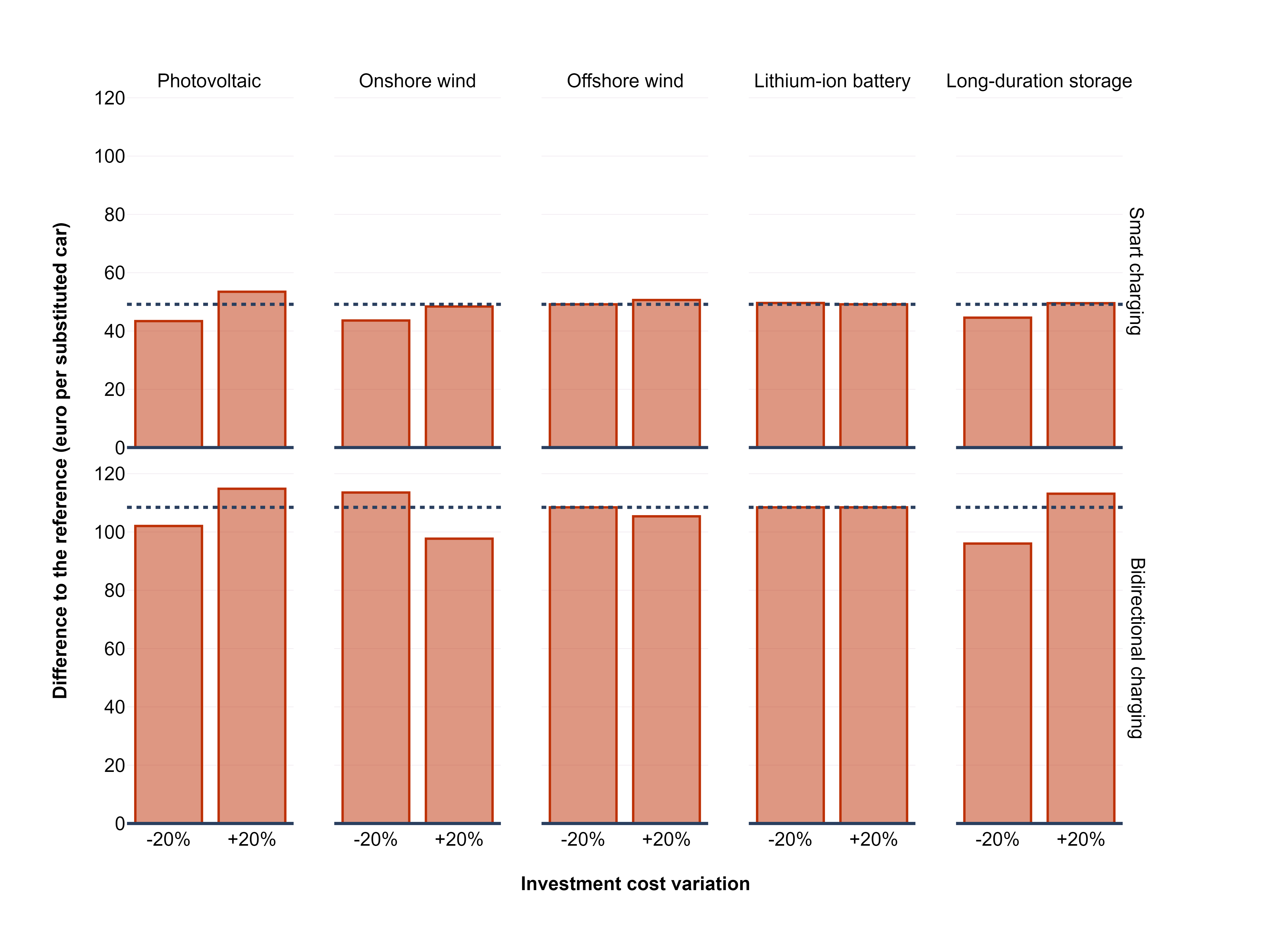}
    \caption{Costs per substituted car in scenario settings for different technology cost assumptions [``Shared + other BEVs'', bidirectional charging, high uptake]}
    \label{fig:SensitivityCosts_systemcosts_scen}
\end{figure}

\begin{figure}[!ht]
    \centering
    \includegraphics[width=15cm]{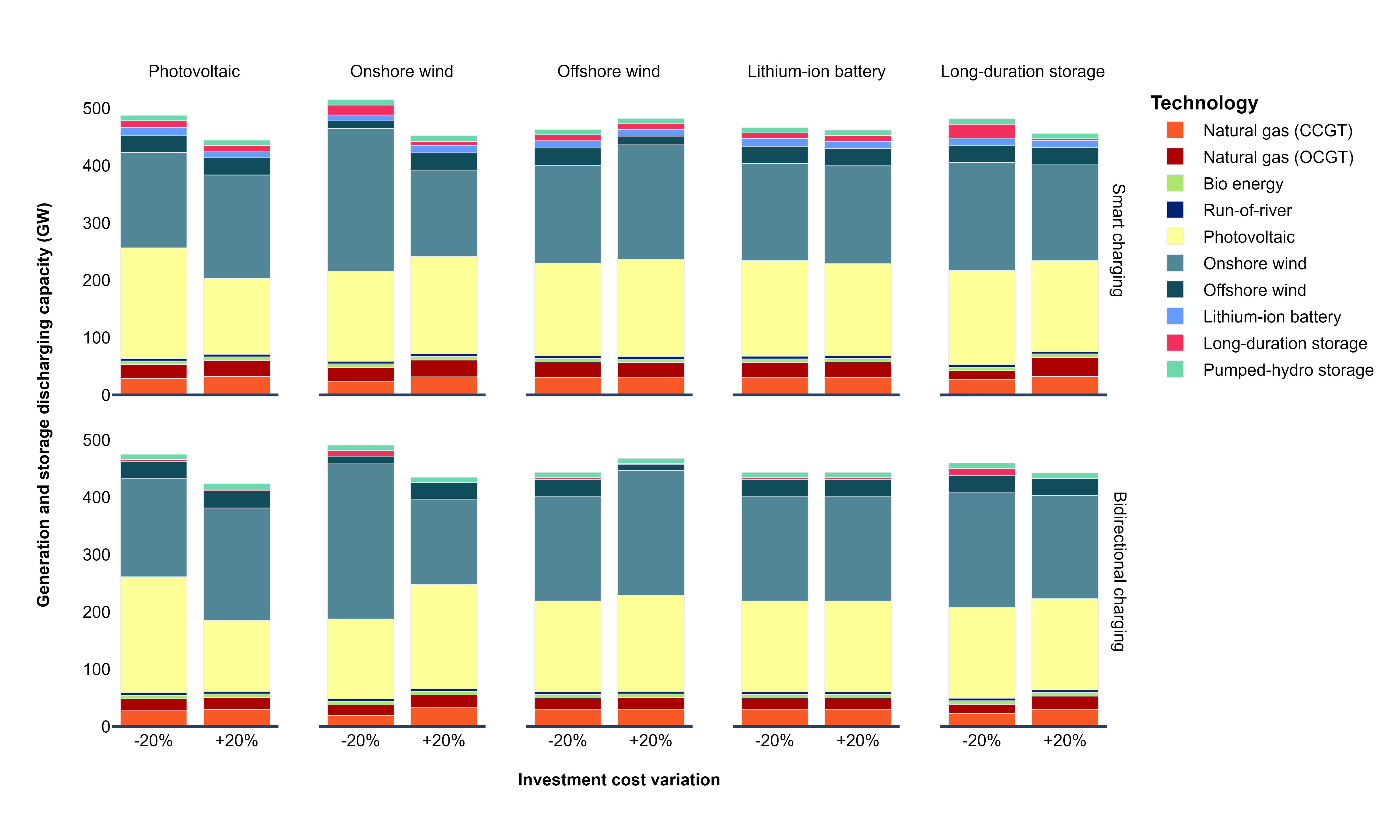}
    \caption{Power generation capacity mixes in reference settings for different technology cost assumptions [``Shared + other BEVs'', bidirectional charging, high uptake]}
    \label{fig:SensitivityCosts_capacity_ref}
\end{figure}

\begin{figure}[!ht]
    \centering
    \includegraphics[width=15cm]{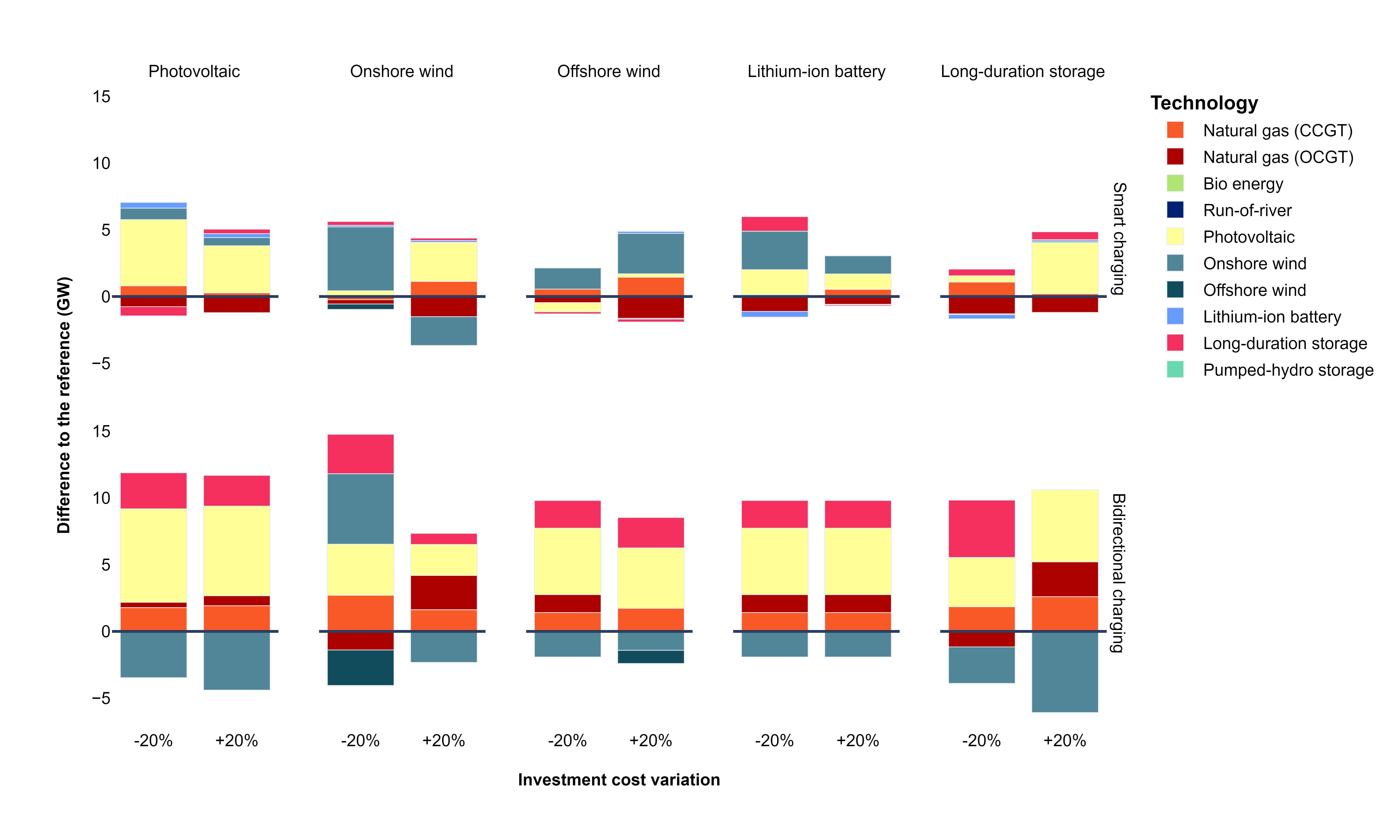}
    \caption{Power generation capacity mixes in scenario settings for different technology cost assumptions [``Shared + other BEVs'', bidirectional charging, high uptake]}
    \label{fig:SensitivityCosts_capacity_scen}
\end{figure}

\begin{figure}[!ht]
    \centering
    \includegraphics[width=14cm]{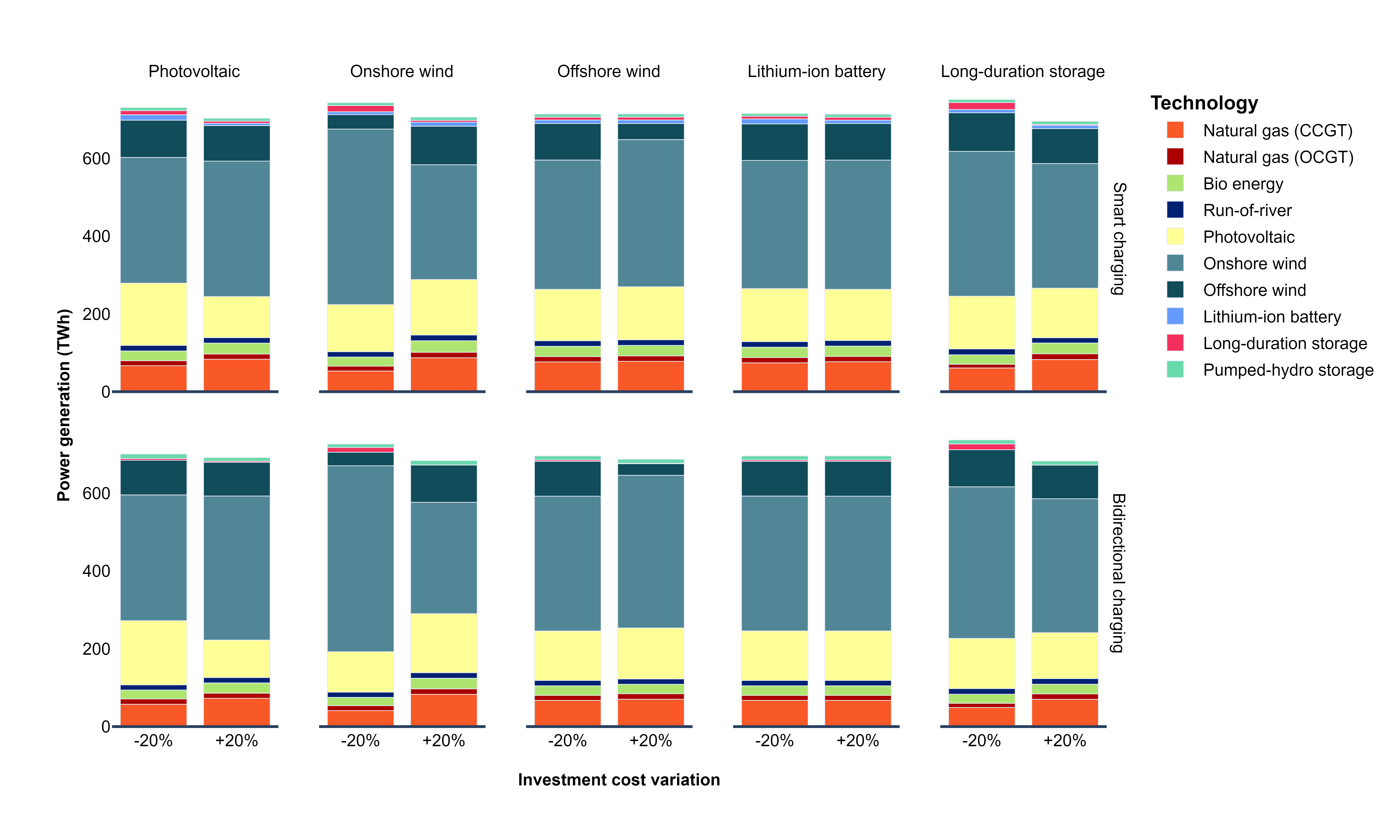}
    \caption{Yearly electricity generation in reference settings for different technology cost assumptions [``Shared + other BEVs'', bidirectional charging, high uptake]}
    \label{fig:SensitivityCosts_generation_ref}
\end{figure}

\begin{figure}[!ht]
    \centering
    \includegraphics[width=14cm]{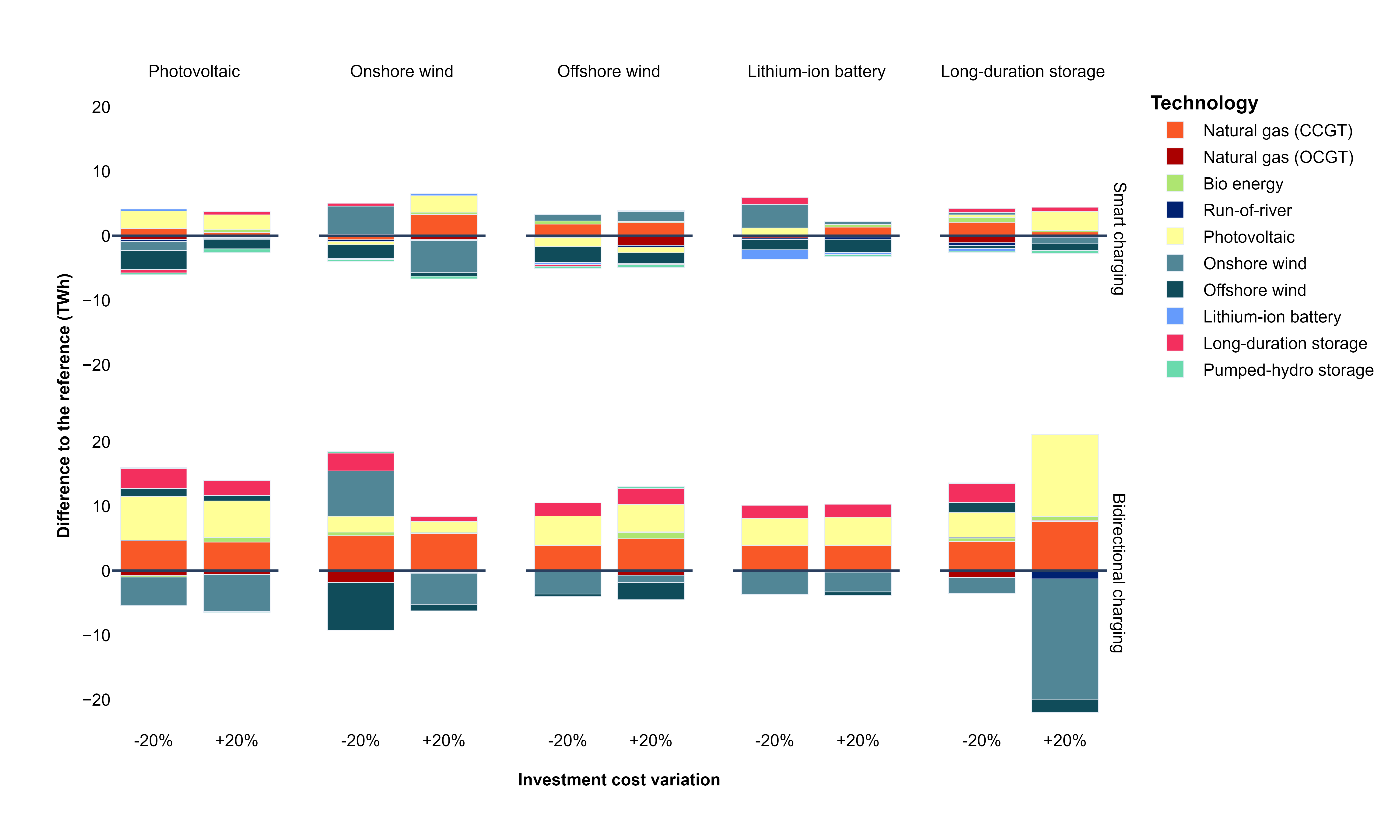}
    \caption{Yearly electricity generation in scenario settings for different technology cost assumptions [``Shared + other BEVs'', bidirectional charging, high uptake]}
    \label{fig:SensitivityCosts_generation_scen}
\end{figure}

\clearpage
\newpage

\subsection{Sensitivity analysis: driving electricity consumption}
\label{sec:sensitivity_drivingelectricityconsumption}

\subfile{sensitivity_eved_systemcosts.tex}

\begin{figure}[!ht]
    \centering
    \includegraphics[width=12cm]{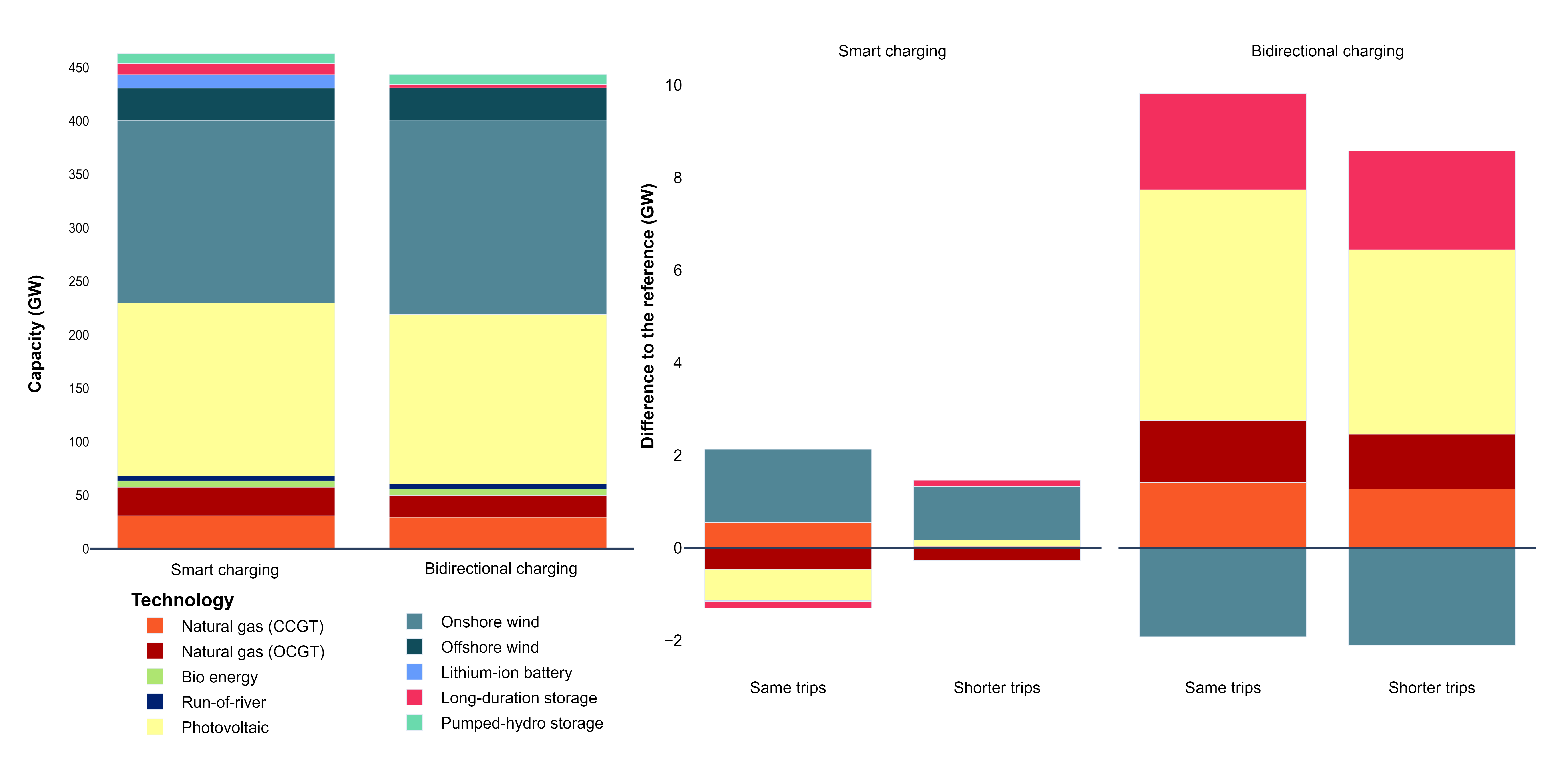}
    \caption{Generation capacity [``Shared + other BEVs'', high uptake]}
    \label{fig:SensitivityEved_capacity}
\end{figure}

\begin{figure}[!ht]
    \centering
    \includegraphics[width=12cm]{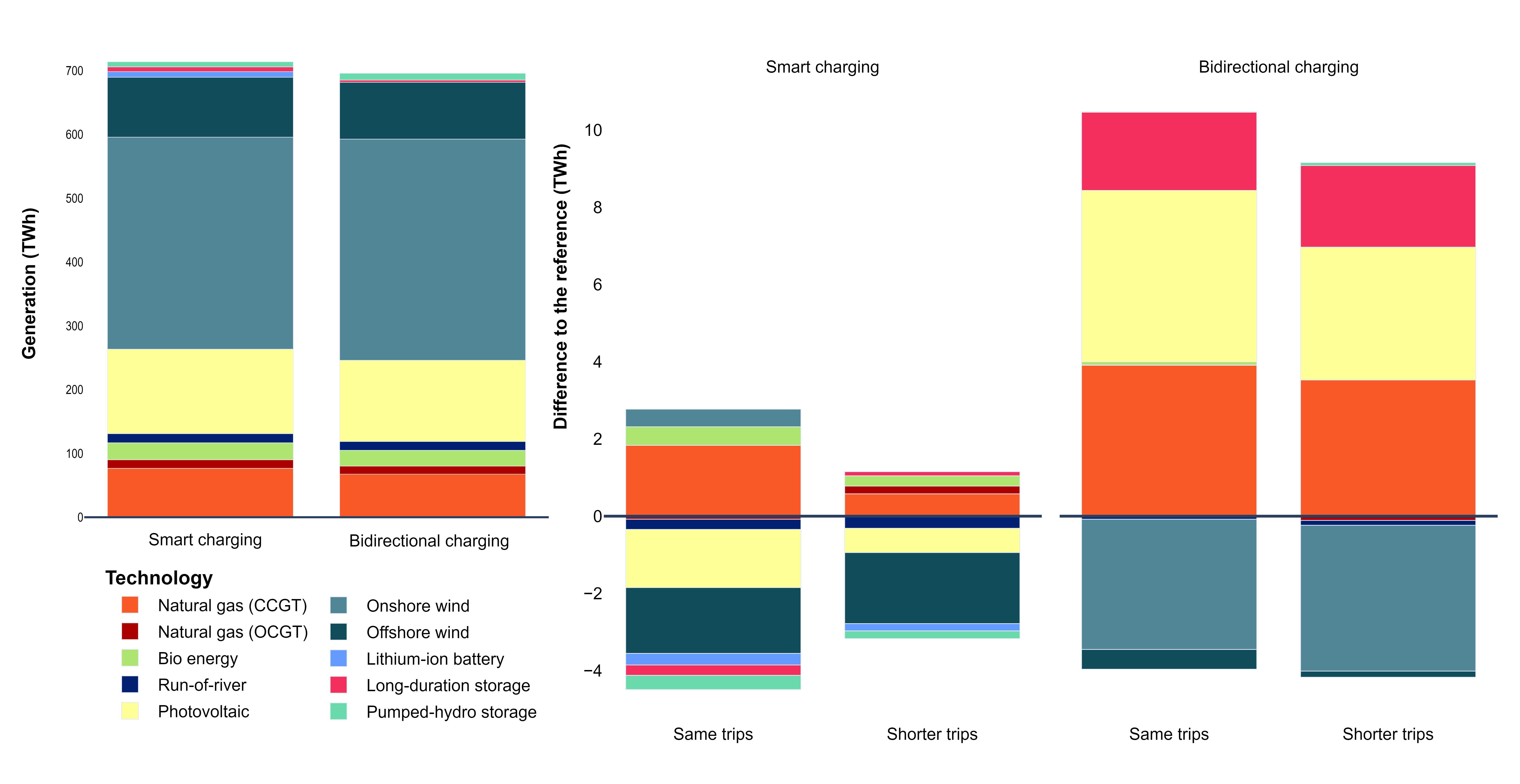}
    \caption{Yearly electricity generation [``Shared + other BEVs'', high uptake]}
    \label{fig:SensitivityEved_generation}
\end{figure}

\end{document}

%% file: car_distribution_cluster_location.tex
\begin{table}[!h]
\centering
\begin{tabular}{@{}
>{\columncolor[HTML]{FFFFFF}}l 
>{\columncolor[HTML]{FFFFFF}}r 
>{\columncolor[HTML]{FFFFFF}}r 
>{\columncolor[HTML]{FFFFFF}}r 
>{\columncolor[HTML]{FFFFFF}}r 
>{\columncolor[HTML]{FFFFFF}}r 
>{\columncolor[HTML]{FFFFFF}}r @{}}
\toprule
\multicolumn{1}{c}{\cellcolor[HTML]{FFFFFF}} &
  \multicolumn{1}{c}{\cellcolor[HTML]{FFFFFF}\textbf{Metropolises}} &
  \multicolumn{1}{c}{\cellcolor[HTML]{FFFFFF}\textbf{Big cities}} &
  \multicolumn{1}{c}{\cellcolor[HTML]{FFFFFF}\textbf{Middle-size cities}} &
  \multicolumn{1}{c}{\cellcolor[HTML]{FFFFFF}\textbf{Small cities}} &
  \multicolumn{1}{c}{\cellcolor[HTML]{FFFFFF}\textbf{Rural areas}} &
  \multicolumn{1}{c}{\cellcolor[HTML]{FFFFFF}\textit{Total}} \\ \midrule
Cluster 3      & 120,308   & 82,729    & 157,940   & 295,736   & 337,853   & 994,566             \\
Cluster 4      & 455,816   & 308,030   & 393,170   & 1,010,433 & 560,125   & 2,727,574           \\
Cluster 5      & 486,317   & 517,490   & 1,112,303 & 907,747   & 1,364,750 & 4,388,607           \\
Cluster 6      & 559,180   & 818,479   & 1,639,890 & 1,840,138 & 2,031,566 & 6,889,253           \\ \midrule
\textit{Total} & 1,621,621 & 1,726,728 & 3,303,303 & 4,054,054 & 4,294,294 & 15,000,000          \\ \bottomrule
\end{tabular}
\caption{Assumed BEV distribution across locations and clusters}
\label{tab:cars_clustloc}
\end{table}

%% file: scenarios.tex
\begin{table}[!h]
\centering
\resizebox{\textwidth}{!}{%
\begin{tabular}{@{}
>{\columncolor[HTML]{FFFFFF}}l 
>{\columncolor[HTML]{FFFFFF}}c 
>{\columncolor[HTML]{FFFFFF}}c 
>{\columncolor[HTML]{FFFFFF}}c 
>{\columncolor[HTML]{FFFFFF}}c |
>{\columncolor[HTML]{FFFFFF}}c 
>{\columncolor[HTML]{FFFFFF}}c 
>{\columncolor[HTML]{FFFFFF}}c 
>{\columncolor[HTML]{FFFFFF}}c @{}}
\toprule
 &
  \multicolumn{4}{c|}{\cellcolor[HTML]{FFFFFF}\textbf{Shared only}} &
  \multicolumn{4}{c}{\cellcolor[HTML]{FFFFFF}\textbf{Shared + other BEVs}} \\
 &
  \multicolumn{2}{c}{\cellcolor[HTML]{FFFFFF}\textit{Low uptake}} &
  \multicolumn{2}{c|}{\cellcolor[HTML]{FFFFFF}\textit{High uptake}} &
  \multicolumn{2}{c}{\cellcolor[HTML]{FFFFFF}\textit{Low uptake}} &
  \multicolumn{2}{c}{\cellcolor[HTML]{FFFFFF}\textit{High uptake}} \\
                                          & Reference & Scenario & Reference & Scenario & Reference & Scenario & Reference & Scenario \\ \midrule
Number of BEVs (in million)               & 4.9       & 1.0      & 7.9       & 1.6      & 15.0      & 11.1     & 15.0      & 8.8      \\
Number of private BEVs (in million)       & 4.9       & 0.0      & 7.9       & 0.0      & 15.0      & 10.2     & 15.0      & 7.2      \\
Number of shared BEVs (in million)        & 0.0       & 1.0      & 0.0       & 1.6      & 0.0       & 1.0      & 0.0       & 1.6      \\
Share of shared BEVs (in \%)              & 0.0       & 100      & 0.0       & 100      & 0.0       & 8.7      & 0.0       & 17.9     \\
Number of substituted BEVs (in million)   & -         & 3.9      & -         & 6.2      & -         & 3.9      & -         & 6.2      \\ \midrule
Overall BEV battery capacity (in GWh)     & 282       & 98       & 461       & 161      & 912       & 729      & 912       & 613      \\
Substituted BEV battery capacity (in GWh) & -         & 183      & -         & 300      & -         & 183      & -         & 299      \\ \midrule
Overall EV power consumption (in TWh)     & 7.7       & 7.7      & 17.7      & 17.7     & 46.7      & 46.7     & 46.7      & 46.7     \\ \bottomrule
\end{tabular}%
}
\caption{BEV units, battery capacity and power consumption for all reference settings and scenarios. \textit{Lecture note}: the table displays the two frameworks defined above. For each framework, we consider several uptake regimes. For each framework and uptake regime, the reference consists in a BEV fleet without carsharing and the scenario consists in a fleet with carsharing. The overall number of BEVs in the fleet and its composition (private vs. shared BEVs) is detailed. The number of substituted BEVs corresponds to the number of BEVs that are removed from the fleet in the carsharing scenario compared to the reference. In the ``Shared + other BEVs'' framework with a high carsharing uptake, the reference fleet amounts to 15 million BEVs. In the scenario, 7.8 million private BEVs are substituted by 1.6 million shared BEVs. 7.2 million private BEVs remain in the fleet as private BEVs. The overall fleet is then composed of 8.8 BEVs in total, which corresponds to a removal of 6.2 million BEVs compared to the reference.}
\label{tab:scenarios}
\end{table}

%% file: cluster_statistics.tex
\begin{table}
\centering
\begin{tabular}{@{}
>{\columncolor[HTML]{FFFFFF}}l 
>{\columncolor[HTML]{FFFFFF}}c 
>{\columncolor[HTML]{FFFFFF}}c 
>{\columncolor[HTML]{FFFFFF}}c 
>{\columncolor[HTML]{FFFFFF}}c 
>{\columncolor[HTML]{FFFFFF}}c @{}}
\toprule
               & \textbf{Metropolises} & \textbf{Big cities} & \textbf{Middle-size cities} & \textbf{Small cities} & \textbf{Rural areas} \\ \midrule
               & \multicolumn{5}{c}{\cellcolor[HTML]{FFFFFF}\textit{Number of sequences}}                                                 \\ \cmidrule(l){2-6} 
Cluster 1      & 107   & 63    & 143   & 67     & 130    \\
Cluster 2      & 363   & 158   & 194   & 298    & 656    \\
Cluster 3      & 790   & 561   & 908   & 1,877  & 1,777  \\
Cluster 4      & 2,984 & 2,072 & 2,277 & 6,437  & 2,931  \\
Cluster 5      & 3,188 & 3,491 & 6,451 & 5,789  & 7,156  \\
Cluster 6      & 3,662 & 5,513 & 9,504 & 11,725 & 10,664 \\
\textit{Total} & 11,094                & 11,858              & 19,477                      & 26,193                & 23,314               \\ \midrule
               & \multicolumn{5}{c}{\cellcolor[HTML]{FFFFFF}\textit{Shares of clusters in each location}}                                 \\ \cmidrule(l){2-6} 
Cluster 1      & 1.0   & 0.5   & 0.7   & 0.3    & 0.6    \\
Cluster 2      & 3.3   & 1.3   & 1.0   & 1.1    & 2.8    \\
Cluster 3      & 7.1   & 4.7   & 4.7   & 7.2    & 7.6    \\
Cluster 4      & 26.9  & 17.5  & 11.7  & 24.6   & 12.6   \\
Cluster 5      & 28.7  & 29.4  & 33.1  & 22.1   & 30.7   \\
Cluster 6      & 33.0  & 46.5  & 48.8  & 44.8   & 45.7   \\
\textit{Total} & 100   & 100   & 100   & 100    & 100    \\ \midrule
               & \multicolumn{5}{c}{\cellcolor[HTML]{FFFFFF}\textit{Average number of trips per day}}                                     \\ \cmidrule(l){2-6} 
Cluster 1      & 2.3   & 2.1   & 2.2   & 2.0    & 2.2    \\
Cluster 2      & 3.5   & 2.8   & 3.0   & 2.9    & 3.5    \\
Cluster 3      & 3.9   & 3.9   & 3.9   & 3.9    & 4.2    \\
Cluster 4      & 3.6   & 3.9   & 4.0   & 3.9    & 3.7    \\
Cluster 5      & 2.8   & 3.4   & 3.6   & 3.2    & 3.4    \\
Cluster 6      & 2.1   & 2.3   & 2.5   & 2.5    & 2.5    \\ \midrule
               & \multicolumn{5}{c}{\cellcolor[HTML]{FFFFFF}\textit{Overall daily travelled distance (in kilometers)}}                    \\ \cmidrule(l){2-6} 
Cluster 1      & 410.3 & 407.2 & 340.6 & 328.5  & 276.3  \\
Cluster 2      & 239.3 & 360.5 & 310.3 & 324.6  & 243.2  \\
Cluster 3      & 112.6 & 156.4 & 160.4 & 150.0  & 123.3  \\
Cluster 4      & 49.7  & 68.3  & 82.2  & 63.4   & 74.2   \\
Cluster 5      & 25.2  & 32.9  & 39.5  & 34.7   & 46.6   \\
Cluster 6      & 10.6  & 13.4  & 13.2  & 14.0   & 17.1   \\ \midrule
               & \multicolumn{5}{c}{\cellcolor[HTML]{FFFFFF}\textit{Average trip distance (in kilometers)}}                               \\ \cmidrule(l){2-6} 
Cluster 1      & 249.2 & 256.2 & 228.5 & 215.2  & 172.9  \\
Cluster 2      & 116.1 & 186.9 & 162.2 & 165.4  & 108.5  \\
Cluster 3      & 43.8  & 65.2  & 65.1  & 60.2   & 41.2   \\
Cluster 4      & 18.5  & 24.4  & 28.7  & 21.5   & 26.5   \\
Cluster 5      & 11.1  & 12.5  & 14.2  & 13.4   & 16.8   \\
Cluster 6      & 5.5   & 6.2   & 6.0   & 6.4    & 7.9    \\ \midrule
               & \multicolumn{5}{c}{\cellcolor[HTML]{FFFFFF}\textit{Average trip duration (in minutes)}}                                  \\ \cmidrule(l){2-6} 
Cluster 1      & 343.7 & 438.7 & 375.0 & 485.6  & 421.6  \\
Cluster 2      & 126.6 & 189.1 & 164.8 & 214.5  & 127.5  \\
Cluster 3      & 57.9  & 79.2  & 79.0  & 73.5   & 52.5   \\
Cluster 4      & 34.2  & 37.8  & 41.1  & 32.6   & 36.7   \\
Cluster 5      & 24.5  & 24.3  & 25.0  & 22.8   & 24.8   \\
Cluster 6      & 14.1  & 14.5  & 13.0  & 12.5   & 13.0   \\ \bottomrule
\end{tabular}
\caption{Cluster statistics by location (working days)}
\label{tab:clust_stat}
\end{table}

%% file: emobpy_condprob.tex
\begin{table}[!h]
\centering
\caption{Conditionality criteria and levels for empirical probability distributions}
\label{tab:emobpy_condprob}
\resizebox{\textwidth}{!}{%
\begin{tabular}{
>{\columncolor[HTML]{FFFFFF}}c 
>{\columncolor[HTML]{FFFFFF}}l 
>{\columncolor[HTML]{FFFFFF}}c 
>{\columncolor[HTML]{FFFFFF}}c 
>{\columncolor[HTML]{FFFFFF}}c 
>{\columncolor[HTML]{FFFFFF}}c 
>{\columncolor[HTML]{FFFFFF}}c 
>{\columncolor[HTML]{FFFFFF}}c }
 &
  \multicolumn{1}{c}{\cellcolor[HTML]{FFFFFF}} &
   &
   &
   &
   &
   &
   \\ \hline
 &
  \multicolumn{1}{c}{\cellcolor[HTML]{FFFFFF}} &
  \multicolumn{2}{c}{\cellcolor[HTML]{FFFFFF}\textbf{Weekday (Mo-Fri)}} &
  \multicolumn{2}{c}{\cellcolor[HTML]{FFFFFF}\textbf{Saturday}} &
  \multicolumn{2}{c}{\cellcolor[HTML]{FFFFFF}\textbf{Sunday}} \\
 &
  \multicolumn{1}{c}{\cellcolor[HTML]{FFFFFF}} &
  Conditionality &
  Level &
  Conditionality &
  Level &
  Conditionality &
  Level \\ \hline
\cellcolor[HTML]{FFFFFF} &
  \textit{Number of trips} &
  no &
  \cellcolor[HTML]{FFFFFF}{\color[HTML]{6200C9} } &
  no &
  \cellcolor[HTML]{FFFFFF}{\color[HTML]{036400} } &
  no &
  \cellcolor[HTML]{FFFFFF}{\color[HTML]{036400} } \\
\cellcolor[HTML]{FFFFFF} &
  \cellcolor[HTML]{EFEFEF}\textit{Destination departure time} &
  \cellcolor[HTML]{EFEFEF}\begin{tabular}[c]{@{}c@{}}number of trips\\      trip rank\end{tabular} &
  \cellcolor[HTML]{FFFFFF}{\color[HTML]{6200C9} } &
  \cellcolor[HTML]{EFEFEF}\begin{tabular}[c]{@{}c@{}}number of trips\\      trip rank\end{tabular} &
  \cellcolor[HTML]{FFFFFF}{\color[HTML]{036400} } &
  \cellcolor[HTML]{EFEFEF}\begin{tabular}[c]{@{}c@{}}number of trips\\      trip rank\end{tabular} &
  \cellcolor[HTML]{FFFFFF}{\color[HTML]{036400} } \\
\multirow{-3}{*}{\cellcolor[HTML]{FFFFFF}\textbf{\begin{tabular}[c]{@{}c@{}}Private \\ BEVs\end{tabular}}} &
  \textit{Joint distance/duration} &
  \begin{tabular}[c]{@{}c@{}}number of trips\\      destination\end{tabular} &
  \multirow{-3}{*}{\cellcolor[HTML]{FFFFFF}{\color[HTML]{6200C9} cluster x location}} &
  \begin{tabular}[c]{@{}c@{}}number of trips\\      destination\end{tabular} &
  \multirow{-3}{*}{\cellcolor[HTML]{FFFFFF}{\color[HTML]{036400} location}} &
  \begin{tabular}[c]{@{}c@{}}number of trips\\      destination\end{tabular} &
  \multirow{-3}{*}{\cellcolor[HTML]{FFFFFF}{\color[HTML]{036400} location}} \\ \hline
\cellcolor[HTML]{FFFFFF} &
  \cellcolor[HTML]{EFEFEF}\textit{Number of trips} &
  \cellcolor[HTML]{EFEFEF}no &
  \cellcolor[HTML]{FFFFFF}{\color[HTML]{6200C9} } &
  \cellcolor[HTML]{EFEFEF}no &
  \cellcolor[HTML]{FFFFFF}{\color[HTML]{036400} } &
  \cellcolor[HTML]{EFEFEF}no &
  \cellcolor[HTML]{FFFFFF}{\color[HTML]{036400} } \\
\cellcolor[HTML]{FFFFFF} &
  \textit{Destination departure time} &
  no &
  \cellcolor[HTML]{FFFFFF}{\color[HTML]{6200C9} } &
  no &
  \cellcolor[HTML]{FFFFFF}{\color[HTML]{036400} } &
  no &
  \cellcolor[HTML]{FFFFFF}{\color[HTML]{036400} } \\
\multirow{-3}{*}{\cellcolor[HTML]{FFFFFF}\textbf{\begin{tabular}[c]{@{}c@{}}Shared \\ BEVs\end{tabular}}} &
  \cellcolor[HTML]{EFEFEF}\textit{Joint distance/duration} &
  \cellcolor[HTML]{EFEFEF}destination &
  \multirow{-3}{*}{\cellcolor[HTML]{FFFFFF}{\color[HTML]{6200C9} cluster x location}} &
  \cellcolor[HTML]{EFEFEF}destination &
  \multirow{-3}{*}{\cellcolor[HTML]{FFFFFF}{\color[HTML]{036400} location}} &
  \cellcolor[HTML]{EFEFEF}destination &
  \multirow{-3}{*}{\cellcolor[HTML]{FFFFFF}{\color[HTML]{036400} location}} \\ \hline
 &
  \multicolumn{1}{c}{\cellcolor[HTML]{FFFFFF}} &
   &
   &
   &
   &
   &
   \\
 &
  \multicolumn{1}{c}{\cellcolor[HTML]{FFFFFF}} &
   &
   &
   &
   &
   &
  
\end{tabular}%
}
\end{table}

%% file: emobpy_assumptions.tex
\begin{table}[!h]
\centering
\caption{emobpy assumptions for generating grid availability time series}
\label{tab:emobpy_assumptions}
\resizebox{\textwidth}{!}{%
\begin{tabular}{@{}cccccccccc@{}}
\toprule
\multicolumn{1}{l}{} &
  \multicolumn{1}{l}{} &
  Charging station &
  \multicolumn{5}{c}{Power rating} &
   &
  Battery capacity \\
\multicolumn{1}{l}{} &
  destination &
  availability &
  \multicolumn{1}{l}{0 kW} &
  3.7 kW &
  \multicolumn{1}{l}{11 kW} &
  \multicolumn{1}{l}{22 kW} &
  \multicolumn{1}{l}{75kW} &
  \multicolumn{1}{l}{150 kW} &
  (kWh) \\ \midrule
\multirow{6}{*}{Private BEVs}         & home           & 0.9  & 0.1  & 0.6 & 0   & 0.3 & 0     & 0     & \multirow{5}{*}{58}  \\
                                      & work/school    & 0.9  & 0.1  & 0   & 0.3 & 0.5 & 0.1   & 0     &                      \\
                                      & errands        & 0.8  & 0.2  & 0   & 0.2 & 0.5 & 0.1   & 0     &                      \\
                                      & leisure        & 0.8  & 0.2  & 0   & 0.2 & 0.5 & 0.1   & 0     &                      \\
                                      & driving        & 0.01 & 0.99 & 0   & 0   & 0   & 0.005 & 0.005 &                      \\ \midrule
\multirow{6}{*}{Shared BEVs}          & home           & 1    & 0    & 0   & 0   & 0   & 1     & 0     & \multirow{5}{*}{100} \\
                                      & work/school    & 1    & 0    & 0   & 0   & 0   & 1     & 0     &                      \\
                                      & errands        & 1    & 0    & 0   & 0   & 0   & 1     & 0     &                      \\
                                      & leisure        & 1    & 0    & 0   & 0   & 0   & 1     & 0     &                      \\
                                      & driving & 0.01 & 0.99 & 0   & 0   & 0   & 0.005 & 0.005 &                      \\ \bottomrule
\end{tabular}%
}
\end{table}

%% file: dieter_cardistribution.tex
\begin{table}[!h]
\centering
\resizebox{\textwidth}{!}{%
\begin{tabular}{@{}
>{\columncolor[HTML]{FFFFFF}}c 
>{\columncolor[HTML]{FFFFFF}}c 
>{\columncolor[HTML]{FFFFFF}}c 
>{\columncolor[HTML]{FFFFFF}}c 
>{\columncolor[HTML]{FFFFFF}}c 
>{\columncolor[HTML]{FFFFFF}}c 
>{\columncolor[HTML]{FFFFFF}}c @{}}
\toprule
\multicolumn{1}{l}{\cellcolor[HTML]{FFFFFF}} & Metropolises & Big cities & Middle-sized cities & Small-sized cities & Rural areas & Total         \\ \midrule
\textit{Cars (\%)}                           & 11           & 12         & 22                  & 27                 & 29          & 100           \\
\textit{Cars (million   units)}              & 1.62         & 1.73       & 3.30                & 4.05               & 4.29        & 15 \\ \bottomrule
\end{tabular}%
}
\caption{Distribution of cars across location types}
\label{tab:dieter_cardistribution}
\end{table}

%% file: dieter_parametersgeneration.tex
\begin{table}[!h]
\centering
\resizebox{\textwidth}{!}{%
\begin{tabular}{@{}
>{\columncolor[HTML]{FFFFFF}}l 
>{\columncolor[HTML]{FFFFFF}}c 
>{\columncolor[HTML]{FFFFFF}}c 
>{\columncolor[HTML]{FFFFFF}}c 
>{\columncolor[HTML]{FFFFFF}}c 
>{\columncolor[HTML]{FFFFFF}}c 
>{\columncolor[HTML]{FFFFFF}}c 
>{\columncolor[HTML]{FFFFFF}}c 
>{\columncolor[HTML]{FFFFFF}}c 
>{\columncolor[HTML]{FFFFFF}}c 
>{\columncolor[HTML]{FFFFFF}}c @{}}
\toprule
 &
  \multicolumn{2}{c}{\cellcolor[HTML]{FFFFFF}\textbf{Capacity bounds}} &
  \cellcolor[HTML]{FFFFFF} &
  \cellcolor[HTML]{FFFFFF} &
  \cellcolor[HTML]{FFFFFF} &
  \cellcolor[HTML]{FFFFFF} &
  \cellcolor[HTML]{FFFFFF} &
  \cellcolor[HTML]{FFFFFF} &
  \cellcolor[HTML]{FFFFFF} &
  \cellcolor[HTML]{FFFFFF} \\ \cmidrule(lr){2-3}
\textbf{} &
  \textbf{Lower} &
  \textbf{Upper} &
  \multirow{-2}{*}{\cellcolor[HTML]{FFFFFF}\textbf{Interest rates}} &
  \multirow{-2}{*}{\cellcolor[HTML]{FFFFFF}\textbf{Lifetime}} &
  \multirow{-2}{*}{\cellcolor[HTML]{FFFFFF}\textbf{Availability}} &
  \multirow{-2}{*}{\cellcolor[HTML]{FFFFFF}\textbf{Overnight costs}} &
  \multirow{-2}{*}{\cellcolor[HTML]{FFFFFF}\textbf{Fixed costs}} &
  \multirow{-2}{*}{\cellcolor[HTML]{FFFFFF}\textbf{Efficiency}} &
  \multirow{-2}{*}{\cellcolor[HTML]{FFFFFF}\textbf{Carbon content}} &
  \multirow{-2}{*}{\cellcolor[HTML]{FFFFFF}\textbf{Fuel costs}} \\
\textbf{} &
  {[}in GW{]} &
  {[}in GW{]} &
  \multicolumn{1}{l}{\cellcolor[HTML]{FFFFFF}\textbf{}} &
  {[}years{]} &
   &
  {[}1,000 EUR/MW{]} &
  {[}1,000 EUR/MW{]} &
   &
  {[}t/MWh{]} &
  {[}EUR/MWh{]} \\ \midrule
Hard coal &
  0.0 &
  9.8 &
  0.04 &
  40 &
  0.955 &
  3845 &
  61.520 &
  0.430 &
  0.337 &
  8.27 \\
Lignite &
  0.0 &
  9.3 &
  0.04 &
  40 &
  0.953 &
  3845 &
  61.520 &
  0.380 &
  0.399 &
  5.50 \\
Natural gas (OCGT) &
  0.0 &
  +Inf &
  0.04 &
  25 &
  0.948 &
  435 &
  7.830 &
  0.400 &
  0.201 &
  30.00 \\
Natural gas (CCGT) &
  0.0 &
  +Inf &
  0.04 &
  25 &
  0.960 &
  830 &
  27.390 &
  0.542 &
  0.201 &
  30.00 \\
Bio energy &
  6.0 &
  6.0 &
  0.04 &
  25 &
  1.0 &
  3300 &
  118.80 &
  0.487 &
  0.000 &
  32.50 \\
Run-of-river &
  4.73 &
  4.73 &
  0.04 &
  80 &
  1.0 &
  3312 &
  66.240 &
  0.900 &
  0.000 &
  0.00 \\
Photovoltaic &
  0.0 &
  +Inf &
  0.04 &
  30 &
  1.0 &
  519 &
  12.975 &
  1.000 &
  0.000 &
  0.00 \\
Onshore wind &
  0.0 &
  +Inf &
  0.04 &
  27 &
  1.0 &
  1035 &
  13.455 &
  1.000 &
  0.000 &
  0.00 \\
Offshore wind &
  0.0 &
  +Inf &
  0.04 &
  27 &
  1.0 &
  1934 &
  36.746 &
  1.000 &
  0.000 &
  0.00 \\ \bottomrule
\end{tabular}%
}
\caption{ Cost and technology parameters for electricity generation technologies}
\label{tab:costs_powergeneration}
\end{table}

%% file: dieter_parametersstorage.tex
\begin{table}[!h]
\centering
\resizebox{\textwidth}{!}{%
\begin{tabular}{@{}
>{\columncolor[HTML]{FFFFFF}}l 
>{\columncolor[HTML]{FFFFFF}}c 
>{\columncolor[HTML]{FFFFFF}}c 
>{\columncolor[HTML]{FFFFFF}}c 
>{\columncolor[HTML]{FFFFFF}}c 
>{\columncolor[HTML]{FFFFFF}}c 
>{\columncolor[HTML]{FFFFFF}}c 
>{\columncolor[HTML]{FFFFFF}}c 
>{\columncolor[HTML]{FFFFFF}}c 
>{\columncolor[HTML]{FFFFFF}}c 
>{\columncolor[HTML]{FFFFFF}}c 
>{\columncolor[HTML]{FFFFFF}}c 
>{\columncolor[HTML]{FFFFFF}}c 
>{\columncolor[HTML]{FFFFFF}}c 
>{\columncolor[HTML]{FFFFFF}}c @{}}
\toprule
 &
  \multicolumn{4}{c}{\cellcolor[HTML]{FFFFFF}\textbf{Capacity bounds}} &
  \multicolumn{1}{l}{\cellcolor[HTML]{FFFFFF}} &
  \multicolumn{1}{l}{\cellcolor[HTML]{FFFFFF}} &
  \multicolumn{1}{l}{\cellcolor[HTML]{FFFFFF}} &
  \multicolumn{3}{c}{\cellcolor[HTML]{FFFFFF}\textbf{Overnight costs}} &
  \multicolumn{2}{c}{\cellcolor[HTML]{FFFFFF}\textbf{Efficiency}} &
  \multicolumn{2}{c}{\cellcolor[HTML]{FFFFFF}\textbf{Marginal costs}} \\ \cmidrule(lr){2-5} \cmidrule(l){9-15} 
 &
  \multicolumn{2}{c}{\cellcolor[HTML]{FFFFFF}\textbf{energy}} &
  \multicolumn{2}{c}{\cellcolor[HTML]{FFFFFF}\textbf{power in/out}} &
  \textbf{Interest rate} &
  \textbf{Lifetime} &
  \textbf{Availability} &
  \textbf{energy} &
  \textbf{charging power} &
  \textbf{discharging power} &
  \textbf{charging} &
  \textbf{discharging} &
  \textbf{charging} &
  \textbf{dicharging} \\
 &
  \textit{lower} &
  \textit{upper} &
  \textit{lower} &
  \textit{upper} &
  \textbf{} &
  \textbf{} &
  \textbf{} &
  \textbf{} &
  \textbf{} &
  \textbf{} &
  \textbf{} &
  \textbf{} &
  \textbf{} &
  \textbf{} \\
 &
  \multicolumn{2}{c}{\cellcolor[HTML]{FFFFFF}{[}GWh{]}} &
  \multicolumn{2}{c}{\cellcolor[HTML]{FFFFFF}{[}GW{]}} &
  \multicolumn{1}{l}{\cellcolor[HTML]{FFFFFF}} &
  {[}years{]} &
   &
  {[}1,000 EUR/MWh{]} &
  {[}1,000 EUR/MW{]} &
  {[}1,000 EUR/MW{]} &
   &
   &
  {[}EUR/MW{]} &
  {[}EUR/MW{]} \\ \midrule
Lithium-ion batteries &
  0 &
  +Inf &
  0 &
  +Inf &
  0.04 &
  20 &
  0.98 &
  142 &
  80 &
  80 &
  0.96 &
  0.96 &
  0.5 &
  0.5 \\
Long-duration storage &
   &
   &
   &
   &
   &
   &
   &
   &
   &
   &
  \multicolumn{1}{l}{\cellcolor[HTML]{FFFFFF}} &
  \multicolumn{1}{l}{\cellcolor[HTML]{FFFFFF}} &
   &
   \\
\multicolumn{1}{r}{\cellcolor[HTML]{FFFFFF}\textit{Electrolysers}} &
  0 &
  +Inf &
  0 &
  +Inf &
  0.04 &
  25 &
  1.0 &
  - &
  650 &
  - &
  0.71 &
  - &
  0.5 &
  - \\
\multicolumn{1}{r}{\cellcolor[HTML]{FFFFFF}\textit{Compressor}} &
  0 &
  +Inf &
  0 &
  +Inf &
  0.04 &
  15 &
  1.0 &
  - &
  80.166 &
  - &
  0.995 &
  - &
  0.0 &
  - \\
\multicolumn{1}{r}{\cellcolor[HTML]{FFFFFF}\textit{Cavern storage}} &
  0 &
  +Inf &
  0 &
  +Inf &
  0.04 &
  100 &
  0.95 &
  2 &
  - &
  - &
  - &
  - &
  - &
  - \\
\multicolumn{1}{r}{\cellcolor[HTML]{FFFFFF}\textit{Reconversion}} &
  0 &
  +Inf &
  0 &
  +Inf &
  0.04 &
  25 &
  0.95 &
  - &
  - &
  435 &
  \multicolumn{1}{l}{\cellcolor[HTML]{FFFFFF}-} &
  0.4 &
  - &
  0.5 \\
Pumped-hydro storage &
  57 &
  57 &
  9.5 &
  9.5 &
  0.04 &
  80 &
  0.98 &
  10 &
  550 &
  550 &
  0.97 &
  0.91 &
  0.5 &
  0.5 \\ \bottomrule
\end{tabular}%
}
\caption{Cost and technology parameters for electricity storage technologies}
\label{tab:parameters_storage}
\end{table}

%% file: system_costs.tex
\begin{table}[!h]
\centering
\resizebox{\columnwidth}{!}{%
\begin{tabular}{@{}
>{\columncolor[HTML]{FFFFFF}}l 
>{\columncolor[HTML]{FFFFFF}}l 
>{\columncolor[HTML]{FFFFFF}}c 
>{\columncolor[HTML]{FFFFFF}}c 
>{\columncolor[HTML]{FFFFFF}}c 
>{\columncolor[HTML]{FFFFFF}}c 
>{\columncolor[HTML]{FFFFFF}}c 
>{\columncolor[HTML]{FFFFFF}}c @{}}
\toprule
\multicolumn{1}{c}{\cellcolor[HTML]{FFFFFF}} &
  \multicolumn{1}{c}{\cellcolor[HTML]{FFFFFF}} &
  \multicolumn{6}{c}{\cellcolor[HTML]{FFFFFF}\textbf{Shared only}} \\ \midrule
\multicolumn{1}{c}{\cellcolor[HTML]{FFFFFF}} &
  \multicolumn{1}{c}{\cellcolor[HTML]{FFFFFF}} &
  \multicolumn{2}{c}{\cellcolor[HTML]{FFFFFF}Uncontrolled} &
  \multicolumn{2}{c}{\cellcolor[HTML]{FFFFFF}Smart charging} &
  \multicolumn{2}{c}{\cellcolor[HTML]{FFFFFF}Bidirectional} \\
\multicolumn{1}{c}{\cellcolor[HTML]{FFFFFF}} &
  \multicolumn{1}{c}{\cellcolor[HTML]{FFFFFF}} &
  \textit{Low} &
  \textit{High} &
  \textit{Low} &
  \textit{High} &
  \textit{Low} &
  \textit{High} \\ \midrule
\cellcolor[HTML]{FFFFFF} &
  Reference (in billion euros) &
  44.7 &
  45.7 &
  44.1 &
  44.5 &
  43.2 &
  43.5 \\
\cellcolor[HTML]{FFFFFF} &
  Difference (in million euros) &
  27.0 &
  81.6 &
  105.0 &
  282.5 &
  672.8 &
  939.3 \\
\multirow{-3}{*}{\cellcolor[HTML]{FFFFFF}\textit{No H$_2$ demand}} &
  Difference (in euros per   substituted car) &
  6.7 &
  13.1 &
  27.1 &
  45.2 &
  173.6 &
  150.3 \\ \midrule
\cellcolor[HTML]{FFFFFF} &
  Reference (in billion euros) &
  46.0 &
  47.0 &
  45.4 &
  45.9 &
  44.5 &
  44.8 \\
\cellcolor[HTML]{FFFFFF} &
  Difference (in million euros) &
  25.4 &
  80.1 &
  108.5 &
  284.5 &
  674.4 &
  928.4 \\
\multirow{-3}{*}{\cellcolor[HTML]{FFFFFF}\textit{Green H$_2$ demand}} &
  Difference (in euros per substituted car) &
  6.5 &
  12.8 &
  28.0 &
  45.5 &
  174.1 &
  148.6 \\ \midrule
\multicolumn{1}{c}{\cellcolor[HTML]{FFFFFF}} &
  \multicolumn{1}{c}{\cellcolor[HTML]{FFFFFF}} &
  \multicolumn{6}{c}{\cellcolor[HTML]{FFFFFF}\textbf{Shared + other BEVs}} \\ \cmidrule(l){3-8} 
\multicolumn{1}{c}{\cellcolor[HTML]{FFFFFF}} &
  \multicolumn{1}{c}{\cellcolor[HTML]{FFFFFF}} &
  \multicolumn{2}{c}{\cellcolor[HTML]{FFFFFF}Uncontrolled} &
  \multicolumn{2}{c}{\cellcolor[HTML]{FFFFFF}Smart charging} &
  \multicolumn{2}{c}{\cellcolor[HTML]{FFFFFF}Bidirectional} \\
\multicolumn{1}{c}{\cellcolor[HTML]{FFFFFF}} &
  \multicolumn{1}{c}{\cellcolor[HTML]{FFFFFF}} &
  \textit{Low} &
  \textit{High} &
  \textit{Low} &
  \textit{High} &
  \textit{Low} &
  \textit{High} \\ \midrule
\cellcolor[HTML]{FFFFFF} &
  Reference (in billion euros) &
  48.9 &
  48.9 &
  46.0 &
  46.0 &
  44.9 &
  44.9 \\
\cellcolor[HTML]{FFFFFF} &
  Difference (in million euros) &
  -86.0 &
  -45.3 &
  134.4 &
  306.6 &
  382.5 &
  676.4 \\
\multirow{-3}{*}{\cellcolor[HTML]{FFFFFF}\textit{No H$_2$ demand}} &
  Difference (in euros per   substituted car) &
  -22.3 &
  -7.3 &
  34.7 &
  49.2 &
  98.9 &
  108.4 \\ \midrule
\cellcolor[HTML]{FFFFFF} &
  Reference (in billion euros) &
  50.2 &
  50.2 &
  47.4 &
  47.4 &
  46.3 &
  46.3 \\
\cellcolor[HTML]{FFFFFF} &
  Difference (in million euros) &
  -85.9 &
  -46.6 &
  126.2 &
  293.2 &
  357.8 &
  640.7 \\
\multirow{-3}{*}{\cellcolor[HTML]{FFFFFF}\textit{Green H$_2$ demand}} &
  Difference (in euros per   substituted car) &
  -22.2 &
  -7.5 &
  32.6 &
  47.0 &
  92.5 &
  102.7 \\ \bottomrule
\end{tabular}%
}
\caption{Power sector costs for reference and scenarios across frameworks, uptake regimes and charging strategies}
\label{tab:system_costs}
\end{table}

%% file: res_shares.tex
\begin{table}[!h]
\centering
\begin{tabular}{@{}
>{\columncolor[HTML]{FFFFFF}}l 
>{\columncolor[HTML]{FFFFFF}}l 
>{\columncolor[HTML]{FFFFFF}}c 
>{\columncolor[HTML]{FFFFFF}}c 
>{\columncolor[HTML]{FFFFFF}}c 
>{\columncolor[HTML]{FFFFFF}}c 
>{\columncolor[HTML]{FFFFFF}}c 
>{\columncolor[HTML]{FFFFFF}}c @{}}
\toprule
\multicolumn{1}{c}{\cellcolor[HTML]{FFFFFF}} &
  \multicolumn{1}{c}{\cellcolor[HTML]{FFFFFF}} &
  \multicolumn{6}{c}{\cellcolor[HTML]{FFFFFF}\textbf{Shared only}} \\ \cmidrule(l){3-8} 
\multicolumn{1}{c}{\cellcolor[HTML]{FFFFFF}} &
  \multicolumn{1}{c}{\cellcolor[HTML]{FFFFFF}} &
  \multicolumn{2}{c}{\cellcolor[HTML]{FFFFFF}Uncontrolled} &
  \multicolumn{2}{c}{\cellcolor[HTML]{FFFFFF}Smart charging} &
  \multicolumn{2}{c}{\cellcolor[HTML]{FFFFFF}Bidirectional} \\
\multicolumn{1}{c}{\cellcolor[HTML]{FFFFFF}} &
  \multicolumn{1}{c}{\cellcolor[HTML]{FFFFFF}} &
  \textit{Low} &
  \textit{High} &
  \textit{Low} &
  \textit{High} &
  \textit{Low} &
  \textit{High} \\ \midrule
\cellcolor[HTML]{FFFFFF} &
  Reference (in \%) &
  85.5 &
  85.5 &
  85.5 &
  85.6 &
  86.6 &
  87.2 \\
\multirow{-2}{*}{\cellcolor[HTML]{FFFFFF}\textit{No H$_2$ demand}} &
  Difference (in \%) &
  0.1 &
  0.1 &
  -0.1 &
  0.0 &
  -0.9 &
  -1.3 \\ \midrule
\cellcolor[HTML]{FFFFFF} &
  Reference (in \%) &
  85.2 &
  85.0 &
  85.2 &
  85.6 &
  86.9 &
  87.8 \\
\multirow{-2}{*}{\cellcolor[HTML]{FFFFFF}\textit{Green H$_2$ demand}} &
  Difference (in \%) &
  0.1 &
  0.3 &
  0.0 &
  -0.1 &
  -1.5 &
  -1.8 \\ \midrule
\multicolumn{1}{c}{\cellcolor[HTML]{FFFFFF}} &
  \multicolumn{1}{c}{\cellcolor[HTML]{FFFFFF}} &
  \multicolumn{6}{c}{\cellcolor[HTML]{FFFFFF}\textbf{Shared + other BEVs}} \\ \cmidrule(l){3-8} 
\multicolumn{1}{c}{\cellcolor[HTML]{FFFFFF}} &
  \multicolumn{1}{c}{\cellcolor[HTML]{FFFFFF}} &
  \multicolumn{2}{c}{\cellcolor[HTML]{FFFFFF}Uncontrolled} &
  \multicolumn{2}{c}{\cellcolor[HTML]{FFFFFF}Smart charging} &
  \multicolumn{2}{c}{\cellcolor[HTML]{FFFFFF}Bidirectional} \\
\multicolumn{1}{c}{\cellcolor[HTML]{FFFFFF}} &
  \multicolumn{1}{c}{\cellcolor[HTML]{FFFFFF}} &
  \textit{Low} &
  \textit{High} &
  \textit{Low} &
  \textit{High} &
  \textit{Low} &
  \textit{High} \\ \midrule
\cellcolor[HTML]{FFFFFF} &
  Reference (in \%) &
  85.2 &
  85.2 &
  86.5 &
  86.5 &
  88.0 &
  88.0 \\
\multirow{-2}{*}{\cellcolor[HTML]{FFFFFF}\textit{No H$_2$ demand}} &
  Difference (in \%) &
  0.1 &
  0.3 &
  0.0 &
  -0.3 &
  -0.2 &
  -0.6 \\ \midrule
\cellcolor[HTML]{FFFFFF} &
  Reference (in \%) &
  84.8 &
  84.8 &
  86.9 &
  86.9 &
  89.5 &
  89.5 \\
\multirow{-2}{*}{\cellcolor[HTML]{FFFFFF}\textit{Green H$_2$ demand}} &
  Difference (in \%) &
  0.2 &
  0.2 &
  0.0 &
  0.0 &
  -0.8 &
  -1.4 \\ \bottomrule
\end{tabular}%
\caption{Share of renewable energy in electricity consumption for reference and scenarios across frameworks, uptake regimes and charging strategies}
\label{tab:res_shares}
\end{table}

%% file: sensitivity_eved_systemcosts.tex
\begin{table}[!h]
\centering
\resizebox{\columnwidth}{!}{%
\begin{tabular}{@{}
>{\columncolor[HTML]{FFFFFF}}c 
>{\columncolor[HTML]{FFFFFF}}l 
>{\columncolor[HTML]{FFFFFF}}c 
>{\columncolor[HTML]{FFFFFF}}c @{}}
\toprule
\multicolumn{1}{l}{\cellcolor[HTML]{FFFFFF}}                     &                                             & Smart charging & Bidirectional charging \\ \midrule
\cellcolor[HTML]{FFFFFF} & Reference (in billion euros)  & 46.0  & 44.9  \\
\cellcolor[HTML]{FFFFFF} & Difference (in million euros) & 306.6 & 676.4 \\
\multirow{-3}{*}{\cellcolor[HTML]{FFFFFF}\textit{Same trips}}    & Difference (in euros per   substituted car) & 49.2           & 108.4                  \\ \midrule
\cellcolor[HTML]{FFFFFF} & Reference (in billion euros)  & 46.0  & 44.9  \\
\cellcolor[HTML]{FFFFFF} & Difference (in million euros) & 160.6 & 518.4 \\
\multirow{-3}{*}{\cellcolor[HTML]{FFFFFF}\textit{Shorter trips}} & Difference (in euros per   substituted car) & 25.8           & 83.1                   \\ \bottomrule
\end{tabular}%
}
\caption{Power sector costs for reference and scenarios {[}``Shared + other BEVs'', high uptake{]}. The ``shorter trips'' setting refers to a sensitivity analysis where trips undertaken with shared cars correspond to a 10\% smaller driving electricity consumption compared to the main setting where mobility needs are assumed constant.}
\label{tab:sensitivity_eved_systemcosts}
\end{table}